\documentclass[11pt,A4paper,twoside,openright,review]{elsarticle}
\usepackage{epsfig}
\usepackage{hyperref}
\usepackage{amsmath}
\usepackage{amssymb}
\usepackage[english]{babel}
\usepackage{fancyhdr}
\usepackage{psfrag}
\usepackage{slashed}
\usepackage{axodraw}
\def\beq{\begin{equation}}
\def\eeq{\end{equation}}
\def\bea{\begin{eqnarray}}
\def\eea{\end{eqnarray}}

	\newcommand{\abs}[1]{ \mathopen{}\left| {#1}\right| }

\def\beq{\begin{equation}}
\def\eeq{\end{equation}}
\def\bea{\begin{eqnarray}}
\def\eea{\end{eqnarray}}
\def\bean{\begin{eqnarray*}}
\def\eean{\end{eqnarray*}}

\newcommand{\be}{\begin{eqnarray}}
\newcommand{\ee}{\end{eqnarray}}

\hypersetup{
    pdfnewwindow=true,      
    colorlinks=true,        
    linkcolor=black,        
    citecolor=blue,         
    filecolor=blue,         
    urlcolor=blue           
}

\begin{document}

\begin{frontmatter}

\title{\Large{\bf{New Paradigm for Baryon and Lepton Number Violation \\}}}

\vspace{13.0cm}
\author{\large{Pavel Fileviez P\'erez}}
\address{Particle and Astroparticle Physics Division \\ Max Planck Institut fuer Kernphysik, 69117 Heidelberg, Germany}

\begin{abstract}
The possible discovery of proton decay, neutron-antineutron oscillation, neutrinoless double beta decay in low energy experiments, 
and exotic signals related to the violation of the baryon and lepton numbers at collider experiments will change our 
understanding of the conservation of fundamental symmetries in nature. In this review we discuss the rare 
processes due to the existence of baryon and lepton number violating interactions. The simplest grand unified theories and the 
neutrino mass generation mechanisms are discussed. The theories where the baryon and lepton numbers are defined as local 
gauge symmetries spontaneously broken at the low scale are discussed in detail. The simplest supersymmetric gauge theory which 
predicts the existence of lepton number violating processes at the low scale is investigated. The main goal of this review is to 
discuss the main implications of baryon and lepton number violation in physics beyond the Standard Model.   
\end{abstract}

\end{frontmatter}

\clearpage
\newpage

\small
\tableofcontents
\normalsize

\newpage
\section{Introduction}
The discovery of the Higgs boson at the Large Hadron Collider did close a chapter in the history of particle physics. Now, the Standard Model of particle physics 
is completed and describes most of the experiments with a great precision. There are many open questions in particle physics and cosmology which we cannot answer 
in the context of the Standard Model. Some of these questions are related to the understanding of the baryon asymmetry and the dark matter in the Universe. 
There are even more fundamental questions related to the origin of all interactions in nature. One of the most appealing scenarios for high energy physics 
correspond to the case where a new theory is realized at the TeV scale and one could solve some of these enigmas in this context.

The great desert hypothesis is very often considered as a well-defined scenario for physics beyond the Standard Model. In the great desert picture one 
has the Standard Model describing physics at the electroweak scale and a more fundamental theory at the high scale. This high scale is often related to the 
scale where there is a grand unified theory or even a theory where the unification of the gauge interactions and gravity is possible. From the low-energy point 
of view  or bottom-up approach it is important to have an effective field theory which is not sensitive to the unknown physics at the high scale. 

It is well-known that in the Standard Model baryon and lepton numbers are conserved symmetries 
at the classical level. However, in our modern view the Standard Model is just an effective theory which 
describes physics at the electroweak scale and one should think about the impact of 
all possible higher-dimensional operators which could modify the Standard Model predictions. For example, 
in the Standard Model one can have the following higher-dimensional operators 
\begin{eqnarray}
{\cal L} & \supset & \frac{c_L}{\Lambda_L} \ell_L \ell_L H^2  +  \frac{c_B}{\Lambda_B^2} q_L q_L q_L \ell_L 
 +  \frac{c_{F}}{\Lambda_F^2} (\bar{q}_L \gamma^\mu q_L) ( \bar{q}_L \gamma_\mu q_L),
\end{eqnarray}       
where the first operator violates lepton number, the second violates baryon and lepton numbers, and the third breaks the flavour symmetry of the Standard Model gauge sector. 
The experimental bounds demand $\Lambda_L < 10^{14}$ GeV, $\Lambda_B > 10^{15}$ GeV~\cite{Nath:2006ut}, and $\Lambda_F > 10^{3-4}$ TeV~\cite{Isidori:2014rba}, when $c_L$, $c_B$ and $c_F$ 
are of order one. Since in any generic grand unified theory one generates the second operator due to new gauge interactions, it is impossible to achieve unification at the low scale without 
predicting an unstable proton and therefore one needs the great desert. One could imagine a theoretical framework where the second operator is absent, the proton is stable and 
the unification scale is only constrained by flavour violating processes. In this case the unification scale must be above $10^4$ TeV in order to satisfy all experimental constraints.
In this review we present new theories for physics beyond the Standard Model where 
the effective theory at the TeV scale predicts a stable proton and  the unification scale could be very low.

In the first part of this review we present new extensions of the Standard Model where the baryon and lepton 
numbers are local gauge symmetries spontaneously broken at the low scale. In this context one can define a simple 
anomaly free theory adding a simple set of fields with baryon and lepton numbers which we call ``lepto-baryons".
We show that in the spectrum of these theories one has a candidate to describe the cold dark matter in the Universe.
We discuss the predictions for direct detection and the constraints from the allowed relic density. Using the 
relic density constraints we find an upper bound on the symmetry breaking scale which 
tells us that these theories can be tested or ruled out at current or future collider experiments. The relation between 
the baryon asymmetry and dark matter density is investigated. We show that even if the local baryon number 
is broken at the low scale one can have a consistent relation between the baryon asymmetry and the $B-L$ 
asymmetry generated through a mechanism such as leptogenesis. Finally, we discuss the possibility to achieve 
the unification of gauge interactions at the low scale in agreement with the experiment. 
 
In the second part of this review we discuss the main issues in supersymmetric theories related 
to the existence of baryon and lepton number violating interactions. We discuss the possible origin 
of the $R-$parity violating interactions and show that the simplest theories based on the local $B-L$ gauge 
symmetry predicts that $R-$parity must be spontaneously broken at the supersymmetric scale. 
We discuss the most striking signatures at the Large Hadron Collider which one can use to test the theory 
in the near future. The predictions for the lepton number violating channels with multi-leptons are discussed 
in detail. We discuss the main features of this generic mechanism for spontaneous $R-$parity breaking 
and the consequences for cosmology. The theories presented in this review can be considered 
as appealing extensions of the Standard Model of particle physics which could be tested in the near future. 
%
\section{Particle Physics and the Great Desert Hypothesis}
%
The Standard Model of particle physics describes the properties of quarks, leptons and gauge bosons with great precision at the electroweak scale, $\Lambda_{EW} \sim 100-200$ GeV.
Unfortunately, very often when we try to understand the origin of the Standard Model interactions one postulates the existence of a grand unified theory or another fundamental theory which 
describes physics at the very high scale, $M_{GUT} \sim 10^{15-17}$ GeV. Therefore, typically one postulates a large energy gap or desert between the electroweak scale and the new 
scale $M_{GUT}$. See Fig.~1 for a simple representation of the desert hypothesis in particle physics. Notice that in the desert one could have some ``animals" such as the 
fields needed for the seesaw mechanism of neutrino masses. However, in the simplest picture there are only two main energy scales where new theories are defined, the electroweak scale and the unification scale. 

Unfortunately, if the desert hypothesis is true there is no hope to test the unified theories at colliders or low-energy experiments. 
One of the main reasons to postulate the great desert is that in the Standard Model one can have new dimension six operators which mediate proton decay. 
For example, one can write down the following baryon and lepton number violating operator with three quarks and one lepton field,
\begin{equation}
{\cal{L}}_{SM} \supset c_B \frac{q_L q_L q_L \ell_L}{\Lambda_B^2},
\label{d6}
\end{equation}
and using the experimental bounds on the proton decay lifetime, $\tau_p > 10^{32-34}$ years, one obtains a very strong bound on the scale 
$\Lambda_B$, i.e. $\Lambda_B > 10^{15-16}$ GeV. One can understand the origin of these interactions in the context of grand unified theories.
\begin{figure}[t] 
\begin{center}
	\includegraphics[scale=0.4]{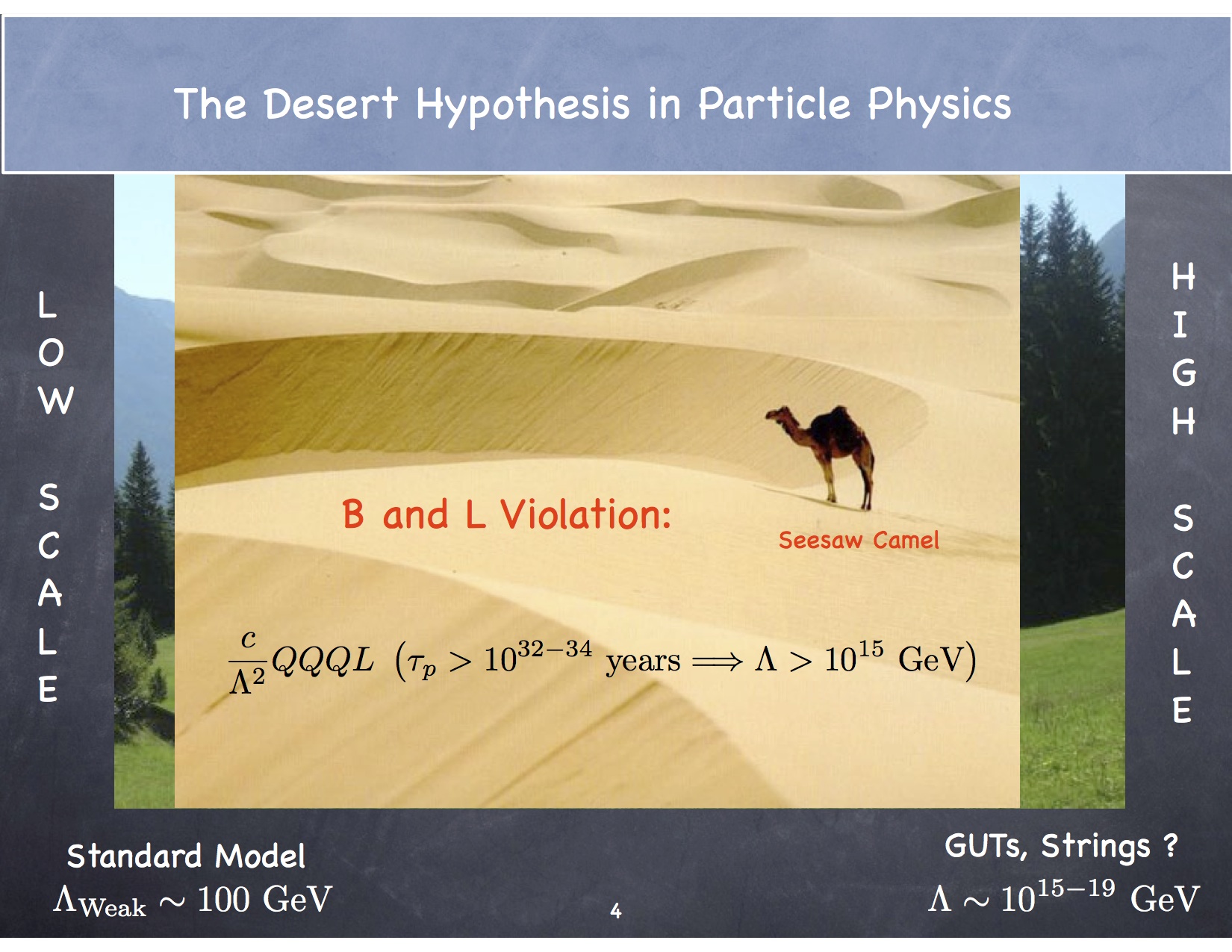}
	\caption{The Desert Hypothesis in Particle Physics and proton decay mediated by dimension six operators. 
	The Standard Model describes physics at the electroweak scale and at the high scale one has some speculative ideas related to the unification of fundamental forces. 
	The grand unified theories could be realized at the $10^{15-16}$ GeV scale, while string theories could be realized at the Planck scale. }
\end{center}
\end{figure}
\begin{figure}[t] 
\begin{center}
	\includegraphics[scale=0.4]{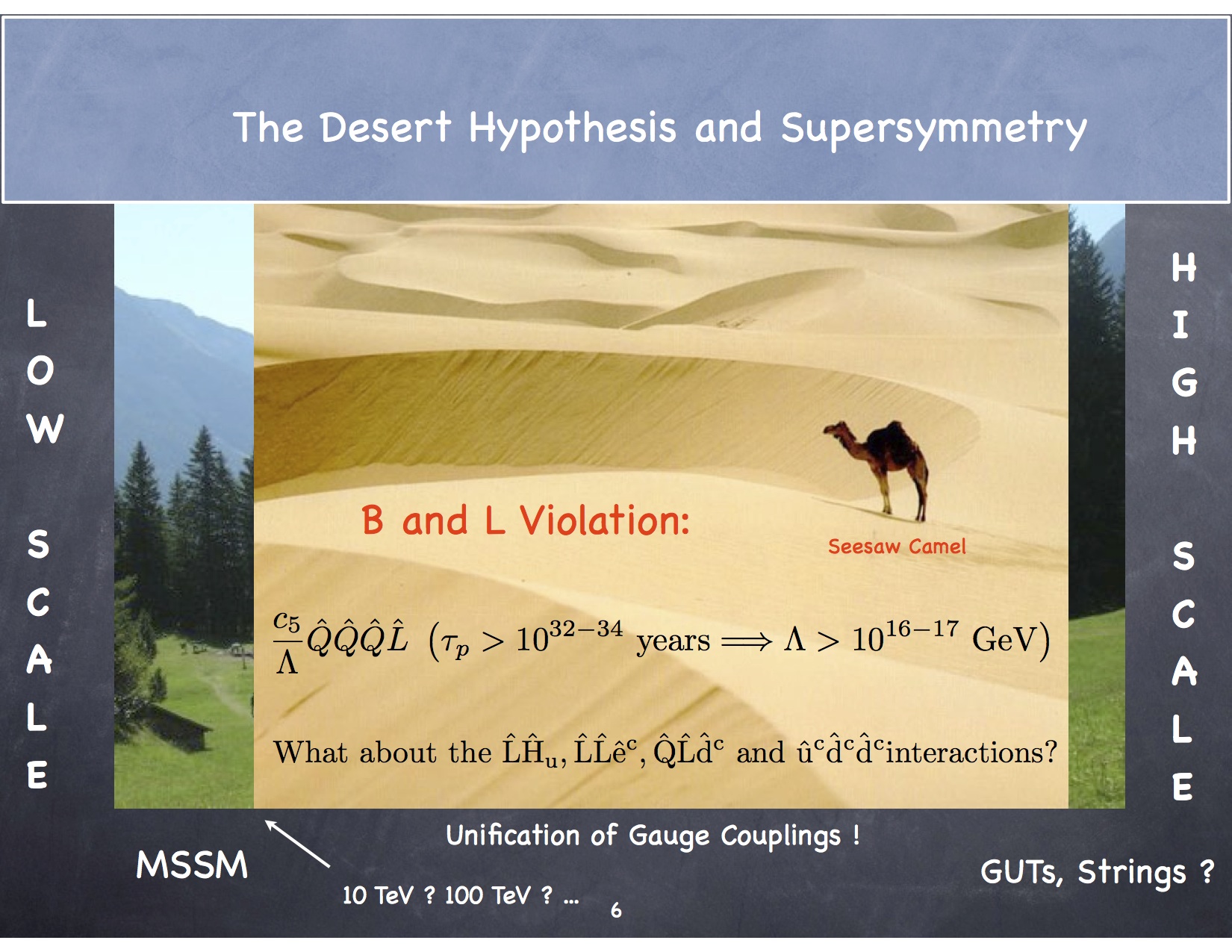}
	\caption{The Desert Hypothesis in the case when we assume Low Energy Supersymmetry. 
	We list the $R-$parity violating interactions and the dimension five operators mediating proton decay. 
	In this case one could describe physics at the TeV scale in the context of the Minimal Supersymmetric Standard Model 
	and at the high scale a possible supersymmetric grand unified theories or superstring theories could play a role.}
	\label{Desert-2}
\end{center}
\end{figure}

The simplest grand unified theory is based on the gauge group $SU(5)$ and it was proposed by H. Georgi and S. Glashow in 1974~\cite{Georgi:1974sy}. 
In this context one can understand the origin of the Standard Model interactions since they are just different manifestations of the same fundamental 
force defined at the high scale. The desert hypothesis was postulated in the same article~\cite{Georgi:1974sy} when the authors realized that the new gauge bosons present 
in $SU(5)$ mediate proton decay and generate the dimension six baryon number violating operator in Eq.~(\ref{d6}). In the context of grand unified 
theories the Standard Model quarks and leptons are unified in the same multiplets. Therefore, the baryon number is explicitly broken at the high 
scale. One must mention that the desert hypothesis is also in agreement with the extrapolation of the gauge couplings up to the high scale.
H. Georgi, H. R. Quinn and S. Weinberg~\cite{Georgi:1974yf} have shown also in 1974 that the Standard Model gauge couplings could be unified at the high scale if we 
assume no new physics between the electroweak and high scales. In 1974 they did not know the values of the gauge couplings with good 
precision but the main idea was correct. 

The minimal supersymmetric Standard Model (MSSM) has been considered as one of the most appealing theories to describe physics at the TeV scale. 
In this context the desert hypothesis plays a main role because the unification of the Standard Model gauge couplings is realized with good precision.
The authors in Refs.~\cite{Dimopoulos:1981yj,Ibanez:1981yh,Marciano:1981un,Einhorn:1981sx} studied the extrapolation of the gauge couplings 
assuming the great desert, showing that they meet without the need of large threshold effects. This result has been quite influential in the particle physics community 
and very often is used as one of the main motivations for having supersymmetry at the low scale. Unfortunately, in the context of the minimal 
supersymmetric Standard Model one finds new interactions which violate the baryon and lepton numbers. In the MSSM superpotential one has 
the terms
\begin{equation}
{\cal{W}}_{MSSM} \supset \epsilon  \hat{L}  \hat{H}_u \ + \  \lambda  \hat{L}  \hat{L}  \hat{e}^c \ + \ \lambda^{'}   \hat{Q}  \hat{L}  \hat{d}^c 
\ + \  \lambda^{''}  \hat{u}^c  \hat{d}^c  \hat{d}^c,
\label{RpV}
\end{equation} 
where the first three terms violate the total lepton number and the last term breaks the baryon number. Here we use the standard superfield notation for all MSSM multiplets.
It is well-known that the last two interactions together mediate the dimension four contributions to proton decay and one needs to impose a discrete symmetry by hand to forbid these interactions. 
Imposing the discrete symmetry called matter parity defined as
\begin{equation}
M=(-1)^{3(B-L)},
\label{d4}
\end{equation} 
one can forbid these interactions. Here, $B$ and $L$ are the baryon and lepton numbers, respectively. Matter parity is defined in such way that it is equal to $-1$ 
for any matter superfield and $+1$ for any gauge or Higgs superfield present in the MSSM. There is a very simple relation between $R$-parity and $M$-parity,
\begin{equation}
R=(-1)^{2S} M,
\label{d4}
\end{equation} 
where $S$ is the spin of the particle. Notice that the interactions in Eq.~(\ref{RpV}) break the $R$-parity discrete symmetry as well and often we refer to these terms as 
$R$-parity violating interactions. It is important to mention that even if we impose the conservation of $M(R)$-parity, still there are interactions which 
mediate proton decay. These are the dimension five operators such as
\begin{equation}
{\cal{W}}_{MSSM} \supset c_5 \frac{\hat{Q} \hat{Q} \hat{Q} \hat{L}}{\Lambda_B}.
\label{d4}
\end{equation} 
In the context of grand unified theories these interactions are mediated by colored fermions and one needs an extra suppression mechanism to satisfy the experimental 
bounds on the proton decay lifetime. One can say that in the context of supersymmetric theories these interactions are problematic and even assuming the great desert 
we need a suppression mechanism. See Fig.~\ref{Desert-2} for a naive representation of the desert hypothesis in the case when we have low energy supersymmetry.

As we have discussed before, the desert hypothesis plays a major role in the particle physics community, and it is often assumed as true.
This hypothesis is very naive and maybe we never will know if it is true because there is no way to test the theories at the high scale. 
The only theory we know which describes physics at the electroweak scale is the Standard Model of particle physics. Then, if we only worry about the theory defined 
at the low scale, the need to assume a large cutoff is uncomfortable. This issue tells us that one should think about effective theories defined 
at the low scale which are free of these baryon number violating operators mediating proton decay. 
%
\section{Proton Stability}
%
We have discussed in the previous section that the main reason to assume the great 
desert between the weak and Planck scales is the proton stability. In the renormalizable 
Standard Model of particle physics the proton is stable since the baryon number is broken 
in three units by the $SU(2)$ instantons. However, there are higher-dimensional operators~\cite{Weinberg:1979sa} 
of dimension six which are allowed by the Standard Model gauge symmetry and violate 
baryon number in one unit. Those operators mediate proton decay and are highly constrained 
by the experimental bounds. The simplest dimension six baryon number violating operators 
are given by
\begin{figure}[t] 
\begin{center}
\vspace{-2.0cm}
	\includegraphics[scale=0.6]{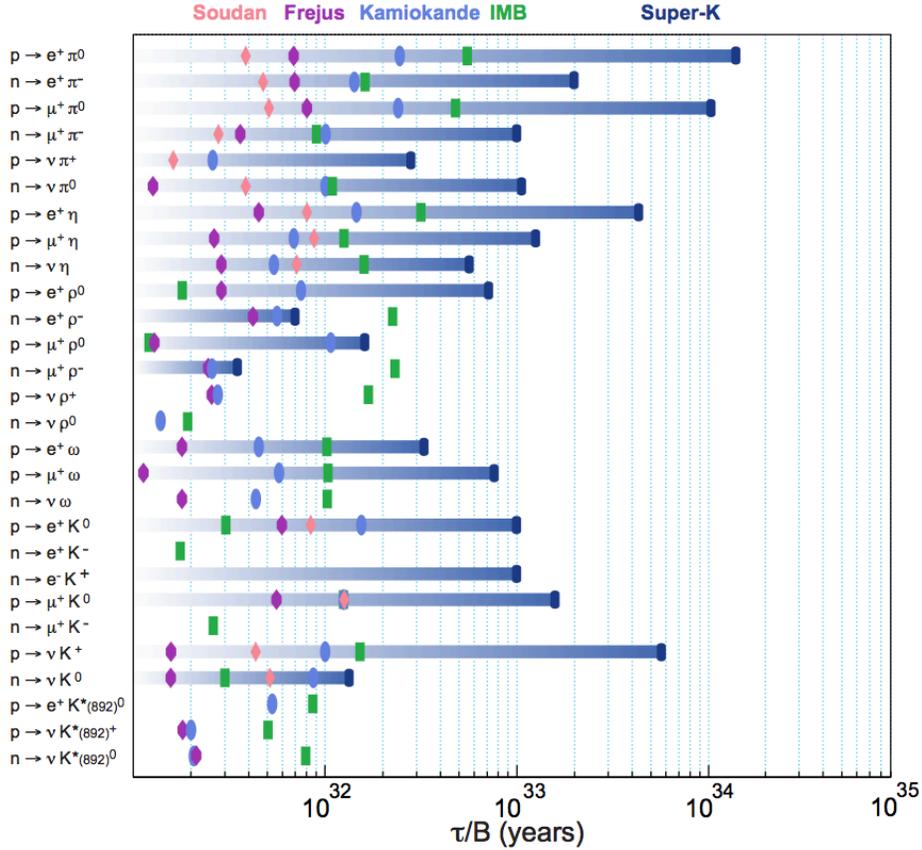}
	\vspace{-2.0cm}
	\caption{Experimental bounds on the proton decay lifetimes~\cite{Raaf}. Here one can see the bounds from different experimental collaborations.}
	\label{protonbounds}
\end{center}
\end{figure} 
\begin{eqnarray}
{\cal{O}}_{1} &=& c_{1} \ \overline{(u^c)}_L \gamma^\mu q_L \  \overline{(e^c)}_L \gamma_\mu q_L,\\
{\cal{O}}_{2} &=& c_{2} \ \overline{(u^c)}_L \gamma^\mu q_L \  \overline{(d^c)}_L \gamma_\mu \ell_L,\\
{\cal{O}}_{3} &=& c_{3} \ \overline{(d^c)}_L \gamma^\mu q_L \  \overline{(u^c)}_L \gamma_\mu \ell_L.
\end{eqnarray}
Here $q_L \sim (3,2,1/6)$, $\ell_L \sim (1,2,-1/2)$, $(e^c)_L \sim (1,1,1)$, $(u^c)_L \sim (\bar{3},1,-2/3)$ and 
$(d^c)_L \sim (\bar{3},1,1/3)$ are the Standard Model quarks and leptons.
These operators are generated once we integrate out superheavy gauge bosons present in grand unified theories.
The first two operators are mediated by the gauge bosons $(X_\mu,Y_\mu) \sim (3,2,-5/6)$, while the last one is mediated 
by $(X_\mu^{'},Y_{\mu}^{'}) \sim (3,2,1/6)$. These gauge bosons are present in $SU(5)$ and flipped $SU(5)$ grand 
unified theories, respectively.

The coefficients $c_i$ in the above equations are proportional to the inverse of the gauge boson masses squared, $M_V^{-2}$. Therefore, using the experimental 
bounds listed in Fig.~\ref{protonbounds} one finds naively a very strong lower bound on the mass of the gauge bosons, i.e. $M_V > 10^{15-16}$ GeV.
Now, since these gauge bosons acquire mass once the grand unified symmetry is broken, the unified scale has to be very large. 
This simple result tells us that there is no way to test these theories at collider experiments. Even if proton decay is found in current and future experiments it will be very difficult to realize the test of a grand unified theory. See Ref.~\cite{Nath:2006ut} for a review on grand unified theories and proton decay. 
For the possibility to find an upper bound on the total proton decay lifetime see Ref.~\cite{Dorsner:2004xa}.

As one can appreciate, these bounds have dramatic implications for particle physics because the 
simplest unified theories must describe physics at the high scale and the great desert is needed. A grand unified theory could be defined at the 
low scale. However, in this case baryon number should be a local symmetry spontaneously broken at the low scale. In the next section we will 
discuss the theories with gauged baryon number which predict that the proton is stable and define a new path to low scale unification. 

\section{Neutron-antineutron oscillations}
In this section we discuss the simplified models for $n\bar{n}$ oscillations studied in Ref.~\cite{Arnold:2012sd}.
In Table~\ref{table1} we list the different scalar di-quarks relevant for this study.
 \begin{table}[h]
\begin{center}
    \begin{tabular}{| c | c |}
    \hline
       Interactions & \ \ \ \ \ $ SU(3) \otimes SU(2) \otimes U(1)_Y$\ \ \ \ \   \\ \hline\hline
    $\ \ X q_L q_L, X u_R d_R \ \ $ & $\left(\bar{6},1, -1/3\right)$  \\ \hline
    $X q_L q_L$ & $\left(\bar{6},3, -1/3\right)$  \\ \hline
    $X d_R d_R$ & $\left(3,1, 2/3\right)$, $\left(\bar{6},1, 2/3\right)$   \\ \hline
    $X u_R u_R$ & $\left(\bar{6},1, -4/3\right)$  \\ \hline  
    \hline
    \end{tabular}
\end{center}
\caption{\footnotesize{Possible interactions between the scalar di-quarks and the Standard Model fermions~\cite{Arnold:2012sd}.}}
\label{table1}
\end{table}
Now, we list the simplest models which give rise to processes with $\Delta B = 2$ and $\Delta L = 0$, 
but only the first three models contribute to $n\bar{n}$ oscillations at tree-level due to the symmetry 
properties of the Yukawa interactions. The sextet field is defined as
\bea
X^{\alpha \beta} = \left(
      \begin{array}{ccc}
        \tilde{X}^{11} & \tilde{X}^{12}/\sqrt{2} & \tilde{X}^{13}/\sqrt{2} \\
        \tilde{X}^{12}/\sqrt{2} & \tilde{X}^{22} & \tilde{X}^{23}/\sqrt{2} \\
        \tilde{X}^{13}/\sqrt{2} & \tilde{X}^{23}/\sqrt{2} & \tilde{X}^{33} \\
      \end{array}
    \right),\
\eea
and the simplest models are discussed below.
\\
\\
\noindent $\textbf{Model 1:}$ In this case the extra fields are $X_{1} \in \left(\bar{6}, 1, -1/3\right)$ and  $X_{2} \in \left(\bar{6}, 1, 2/3\right)$, and the relevant 
Lagrangian reads as
\bea\label{Lag1}
\mathcal{L}_1 & = &  -\ h_{1}^{a b} X_{1}^{\alpha \beta} \left(q_{L \alpha}^a \epsilon \,q_{L \beta}^b\right)- {h}_{2}^{a b}  X_{2}^{\alpha \beta} (d_{R \alpha}^a d_{R \beta}^b)\nonumber\\
& & \hspace{-8mm} - \ {h}_{1}^{\prime a b}  X_{1}^{\alpha \beta} (u_{R \alpha}^a d_{R \beta}^b) + \lambda \ X_{1}^{\alpha \alpha'} X_{1}^{\beta \beta'} X_2^{\gamma \gamma'}\epsilon_{\alpha \beta \gamma} \epsilon_{\alpha'\beta'\gamma'}.\
\eea
\noindent
Here the Greek indices $\alpha$, $\beta$ and $\gamma$ are the color indices, while $a$ and $b$ are family indices. $h_1$ must be antisymmetric in flavor space but this antisymmetry is not retained upon rotation into the mass eigenstate basis. The coupling $h_2$ must be symmetric because of the symmetric color structure in the second term.   
\\ \\
\noindent$\textbf{Model 2:}$ In the second model one has the fields $X_{1} \in \left(\bar{6}, 3, -1/3\right)$ and  $X_{2} \in \left(\bar{6}, 1, 2/3\right)$ with
\bea
\mathcal{L}_2 &\!\! =\!\! &  -\ h_{1}^{a b} X_{1}^{\alpha \beta A} (q_{L \alpha}^a \epsilon \,\tau^A \,q_{L \beta}^b)  - {h}_{2}^{a b}  X_{2}^{\alpha \beta} (d_{R \alpha}^a d_{R \beta}^b) 
 + \ \lambda \ X_{1}^{\alpha \alpha' A} X_{1}^{\beta \beta' A} X_2^{\gamma \gamma'}\epsilon_{\alpha \beta \gamma} \epsilon_{\alpha'\beta'\gamma'}.\
\eea
\noindent
Here the matrix $\epsilon \, \tau^A$ is symmetric, while the first and second terms have symmetric color structures, $h_1$ and $h_2$ must be symmetric in flavor.  
\\ \\
\noindent$\textbf{Model 3:}$ One can have a simple scenario with the fields  $X_{1} \in \left(\bar{6}, 1, 2/3\right)$ and $X_{2} \in \left(\bar{6}, 1, -4/3\right)$ 
and the Lagrangian in this case is given by 
\bea
\mathcal{L}_3 & = &  - \ h_{1}^{a b} X_{1}^{\alpha \beta} (d_{R \alpha}^a d_{R \beta}^b)  - \ {h}_{2}^{a b}  X_{2}^{\alpha \beta} (u_{R \alpha}^a u_{R \beta}^b) 
+ \ \lambda \ X_{1}^{\alpha \alpha'} X_{1}^{\beta \beta'} X_{2}^{\gamma \gamma'}\epsilon_{\alpha \beta \gamma} \epsilon_{\alpha'\beta'\gamma'}.
\eea
\noindent
Both terms have symmetric color structures and no weak structure, so $h_1$ and $h_2$ must be symmetric in flavor. 
\\ \\
\noindent$\textbf{Model 4:}$ In the last scenario one has $X_{1} \in \left(3, 1, 2/3\right)$ and $X_{2} \in \left(\bar{6}, 1, -4/3\right)$ with
\bea
\mathcal{L}_4 & = &  -\ h_{1}^{a b} X_{1\alpha} \left(d_{R \beta}^a \,d_{R \gamma}^b\right)\epsilon^{\alpha\beta\gamma}  -  {h}_{2}^{a b}  X_{2}^{\alpha \beta} (u_{R \alpha}^a u_{R \beta}^b) + \ \lambda \ X_{1 \alpha} X_{1 \beta} X_{2}^{\alpha \beta}. \
\eea
\noindent
Because of the antisymmetric color structure in the first term, $h_1$ must be antisymmetric in flavor which prevents it from introducing meson-antimeson mixing.  The antisymmetric structure of $h_1$ also prevents the existence of six-quark operators involving all first-generation quarks, and one prevents $n\bar{n}$ oscillations.   

\begin{figure}[t!]
\centering \hspace{5mm}
\includegraphics[scale=.9]{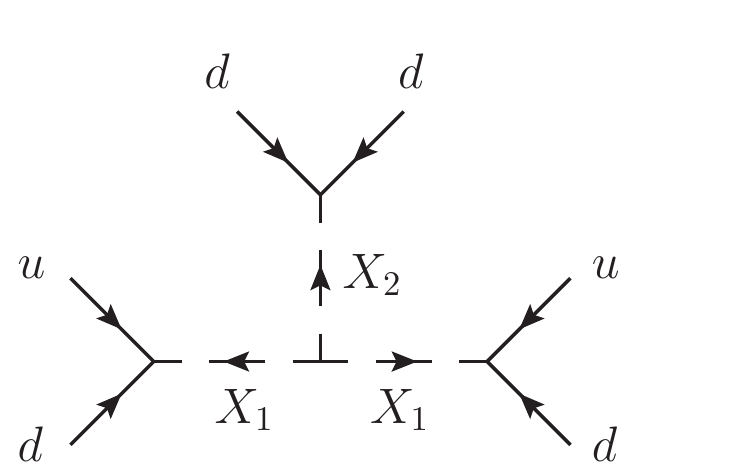}
\caption{Feynman Graph for neutron-antineutron oscillations~\cite{Mohapatra:1980qe,Arnold:2012sd}.}
\label{fig1b}
\end{figure}

In order to understand the constraints from the $n-\bar{n}$ oscillation experiments the authors in Ref.~\cite{Arnold:2012sd} studied model 1. 
The transition matrix element 
\bea
\Delta m = \langle \bar{n}| \mathcal{H}_{\rm eff} |n\rangle,
\label{Heff}
\eea
leads to a transition probability for a neutron at rest to change into an antineutron after time $t$ equal to $P_{n\rightarrow \bar{n}}(t) = \sin^2\!\left(|\Delta m |\, t\right)$.
Neglecting the coupling $h_1$ in the Lagrangian given by Eq.(\ref{Lag1})  the effective $|\Delta B|=2$ Hamiltonian that causes $n\bar{n}$ oscillations reads as~\cite{Arnold:2012sd}
\bea
\mathcal{H_{\rm eff}} &\!\!\!=\!\!\!& - \frac{(h_1^{\prime 11})^2 h_2^{11} \lambda}{4 M_{1}^4 M_{2}^2} d_{R i}^{{\alpha}} d_{R i'}^{{\beta}} u_{R j}^{{\gamma}} d_{R j'}^{{\delta}} u_{R k}^{{\lambda}} d_{R k'}^{{\chi}} \epsilon_{{\alpha} {\beta}} \epsilon_{{\gamma} {\delta}} \epsilon_{{\lambda} {\chi}}\nonumber\\
& &\hspace{-10mm}\times \Big(\epsilon_{ijk} \epsilon_{i'j'k'} + \epsilon_{i'jk} \epsilon_{ij'k'}+\epsilon_{ij'k} \epsilon_{i'j k'}+\epsilon_{ijk'} \epsilon_{i'j'k}\Big) +{\rm h.c.}
\eea
where Latin indices are color and Greek indices are spinor. It arises from the tree-level diagram in Fig.\ \ref{fig1b}.  
In the vacuum insertion approximation to Eq.\ (\ref{Heff}) one finds~\cite{Arnold:2012sd,Bart} that
\bea\label{deltam}
|\Delta m |= 2 \lambda \beta^2\frac{|  (h_1^{\prime 11})^2 h_2^{11}| }{3 M_{1}^4 M_{2}^2}\ .
\eea
Here the constant $\beta \simeq 0.01~{\rm GeV}^3$. Using the current experimental limit on $\Delta m$~\cite{Abe:2011ky},
\bea
|\Delta m| < 2\times 10^{-33} \ {\rm GeV},
\eea
assuming the relation $M_1 = M_2 = M$, the values of the couplings $h_1^{\prime 11} = h_2^{11} = 1$, and $\lambda = M$ one obtains
\bea
M \gtrsim 500 \ {\rm TeV}.
\eea
The effect on $n\bar{n}$ oscillations is maximized if we choose $M_2 > M_1$ because $|\Delta m |$ can be larger.  
Assuming $M_1 = 5 \ {\rm TeV}$ and $\lambda = M_2$ experiments in the future may be able to probe $n\bar{n}$ oscillations with increased sensitivity of $|\Delta m| \simeq 7\times 10^{-35} \ \rm GeV$. 
If no oscillations are observed the new lower bound in the case of equal  masses will be
\bea
M \gtrsim  1000 \ {\rm TeV}\ .
\eea
In this way one can see the possible bounds that can be achieved if $n\bar{n}$ oscillations are not discovered.
It is important to say that $n\bar{n}$ experiments cannot probe the grand unified scale 
but the discovery of this process will change the way we think about physics beyond the Standard Model.
More experiments are badly needed in this area. See Ref.~\cite{Phillips:2014fgb} for a recent review 
on $n\bar{n}$ oscillations.

\section{Mechanisms for Neutrino Masses}
The discovery of massive neutrinos have motivated the experts in the field to think 
about the possible mechanisms to generate neutrino masses. The neutrinos 
could be Majorana or Dirac particles. In the Majorana case the total lepton number 
must be broken in two units, $\Delta L = 2$, as in the seesaw mechanisms. In this section    
we discuss the simplest mechanisms for generating neutrino masses 
at tree and one-loop level. The simplest mechanisms at tree level 
are the following:

{\bf{\underline{Type I Seesaw}}}~\cite{TypeI-1,TypeI-2,TypeI-3,TypeI-4,TypeI-5}: 
This is perhaps the simplest mechanism for generating neutrino masses. 
In this case one adds a SM singlet, $\nu^c \sim (1,1,0)$, and using the interactions   
\begin{equation}
- {\cal L}_\nu^I = Y_\nu \ l \ H \ \nu^c \ + \ \frac{1}{2} M \ \nu^c \ \nu^c \ + \ \textrm{h.c.}, 
\end{equation}
in the limit $M \gg Y_\nu \ v_0 $ one finds
\begin{equation}
{\cal M}_\nu^I =  \frac{1}{2} \ Y_\nu \ M^{-1} \ Y_\nu^T \ v^2_0,
\end{equation}
where M is typically defined by the $B-L$ breaking scale.
Then, one understands the smallness of the neutrino masses 
due to the existence of a mass scale, $M \gg Y_\nu v_0 \gg m_\nu$. 
Here, if we assume $Y_\nu \sim 1$ the scale $M \sim 10^{14-15}$ GeV.
Now, in general it is not possible to make predictions for the neutrino 
masses and mixing in this framework since we do not know the 
matrices $Y_\nu$ and $M$. Then, one should look for a theory 
where one could predict these quantities. In the context of 
$SO(10)$ grand unified theories one can relate the charged 
fermion and neutrino masses in a consistent way.

Once the right-handed neutrinos are present in the theory the 
$B-L$ symmetry can be defined as a local symmetry. In this 
scenario the same Higgs breaking $B-L$ can generate Majorana 
masses for the right-handed neutrinos. Therefore, the scale $M$ is 
defined by the $B-L$ breaking scale. In this case if the scale is in the 
TeV range one can produce the right-handed neutrinos with 
large cross sections through the $B-L$ gauge boson
$$pp \to  Z_{B-L}^*  \to N_1 N_1  \to e^{\pm}_i  e^{\pm}_j W^{\mp} W^{\mp},$$ 
and one can have signatures with two same-sign dileptons at the LHC.
Here $e_i=e,\mu,\tau$ and $N_1$ is the lightest physical $\nu^c$-like state. 
See Ref.~\cite{Perez:2009mu} for a recent study of these collider signals.
For the original idea of looking for lepton number violation at colliders see Ref.~\cite{Keung:1983uu}.

{\bf {\underline{Type II Seesaw}}}~\cite{TypeII-1,TypeII-2,TypeII-3,TypeII-4,TypeII-5}: 
In this scenario one introduces a new Higgs boson, 
$\Delta \sim (1,3,1)$, which couples to the leptonic 
doublets and the SM Higgs boson
\begin{equation}
- {\cal L}_\nu^{II}= Y_\nu \ \ell_L \ \Delta \ \ell_L \ + \ \mu \ H \  \Delta^\dagger \ H \ + \ \textrm{h.c.},
\end{equation}
and when the neutral component in $\Delta=(\delta^0, \delta^+, \delta^{++})$ acquires 
a vacuum expectation value, $v_\Delta$, one finds
\begin{equation}
\label{TypeII}
{\cal M}_\nu^{II}= \sqrt{2} \ Y_\nu \ v_{\Delta} = \mu \ Y_\nu \ v_0^2 / M_{\Delta}^2. 
\end{equation}
Notice that if $\mu \sim M_{\Delta}$ and $M_{\Delta} \sim 10^{14-15}$ GeV 
the vev $v_{\Delta}$ should be of order $1$ eV. However, 
in general the triplet mass can be around the TeV scale 
and $\mu$ can be small. Unfortunately, in this context one cannot 
make predictions for neutrino masses because in general the matrix 
$Y_\nu$ and $v_\Delta$ are unknown. 
Then, as in the previous case, one should look for a theory 
where one can predict these quantities. We would like to mention 
that the field $\Delta$ can be light in agreement with the unification constraints 
in a simple grand unified theory based on the $SU(5)$ gauge symmetry~\cite{Dorsner:2005fq,Dorsner:2005ii,Dorsner:2006hw}.

In this scenario one can have spectacular signatures at the LHC 
if the Higgs triplet is light. When $M_{\Delta} \leq 1$ TeV one could produce 
at the LHC the doubly and singly charged Higgses present in the model. 
When $v_{\Delta} < 10^{-4}$ GeV the dominant decays are $H^{++} \to e_i^+ e^+_j$ 
and $H^{+} \to e_i^+ \bar{\nu}$, and we could learn about the neutrino spectrum. 
These signatures have been investigated in great detail in Refs.~\cite{TypeII-LHC-1,TypeII-LHC-2,TypeII-LHC-3,TypeII-LHC-4,TypeII-LHC-5}.

{\bf {\underline{Type III Seesaw}}}~\cite{TypeIII,Ma,Goran,Adjoint-SU(5),SUSY-Adjoint-SU(5),LR-TypeIII}:
In the case of Type III seesaw one adds new fermions, $\rho_L \sim (1,3,0)$, and the neutrino 
masses are generated using the following interactions
\begin{equation}
- {\cal L}_\nu^{III}= Y_\nu \ \ell_L \ \rho_L \ H \ + \ M_{\rho} \ \textrm{Tr} \  \rho^2_L \ + \ \textrm{h.c.},
\end{equation}
where $\rho_L=(\rho^0, \rho^+, \rho^-)$. Integrating 
out the neutral component of the fermionic triplet 
one finds 
\begin{equation}
{\cal M}_\nu^{III}= \frac{1}{2} Y_\nu \ M_{\rho}^{-1} \ Y_\nu^T \ v_0^2.
\end{equation}
Here, as in the case of Type I seesaw, if $Y_\nu \sim 1$ one needs 
$M_{\rho} \sim 10^{14-15}$ GeV. One faces the same problem, if we 
want to make predictions for neutrinos masses and mixings, 
a theory where $Y_\nu$ and $M_{\rho}$ can be predicted is needed. 
The existence of a light fermionic triplet is consistent with unification 
of gauge interactions in the context of $SU(5)$ theories. 
See Refs.~\cite{Goran,Dorsner:2006fx,Bajc:2007zf}  for details.
The testability of the Type III seesaw mechanism at the LHC has 
been investigated in detail in Refs.~\cite{Franceschini:2008pz,Arhrib:2009mz,Li:2009mw}.

We have mentioned the simplest mechanisms at tree level.
Now, if Supersymmetry is realized in nature one has the 
extra possibility to generate neutrino masses through 
the R-parity violating couplings. One could say that in this 
case we use a combination of the different seesaw mechanisms.
We will discuss this possibility in the next sections. 

There are several mechanisms for generating neutrino 
masses at one-loop level. In this case one assumes 
that the mechanisms discussed above are absent and 
only through quantum corrections one generates 
neutrino masses. This possibility is very appealing 
since the neutrino masses are very tiny and the 
seesaw scale can be low.

{\bf {\underline{Zee Model}}}~\cite{Zee}: 
In the so-called Zee model one introduces two extra 
Higgs bosons, $h\sim (1,1,1)$ and $H^{'} \sim (1,2,1/2)$.
In this case the relevant interactions are
\begin{equation}
- {\cal L}_{Zee}= Y \ \ell_L \ h \ \ell_L \ + \ \mu \ H \ H^{'} \ h^\dagger 
\ + \  \sum_{i=1}^2 \ Y_i \ e^c \ H^\dagger_i  \ \ell_L \ + \ \textrm{h.c.},  
\end{equation}
where in general both Higgs doublets couple to the matter fields.
Using these interactions one can generate neutrino masses 
at one-loop level. See Ref.~\cite{Zee} for details. Now, 
it is important to mention that in the simple case where only 
one Higgs doublet couples to the leptons~\cite{Wolfenstein} 
it is not possible to generate neutrino masses in 
agreement with neutrino data. See for example 
Refs.~\cite{He-1,He-2,He-3,He-4,He-5} for details. 

\begin{figure}[t]
\begin{center}
\includegraphics[width=0.49\linewidth]{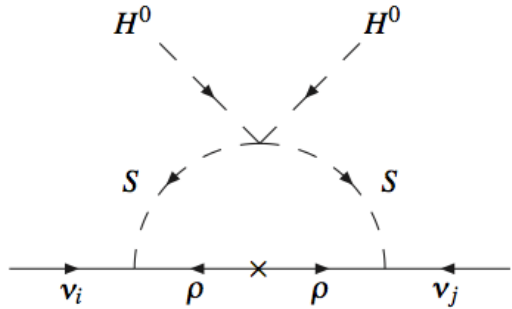}
 \caption{Colored Seesaw Mechanism~\cite{Fileviez-Perez-Wise}. The field $S$ is a colored scalar octet, while $\rho$ is a fermionic colored octet.}
 \label{Colored-Seesaw}
 \end{center}
\end{figure}

{\bf {\underline{Colored Seesaw}}}~\cite{Fileviez-Perez-Wise}:
Now, suppose that one looks for the simplest mechanism for neutrino 
masses at one-loop level where we add only two types of representations, 
a fermionic and a scalar one, and with no extra symmetry. All the possibilities 
were considered in Ref.~\cite{Fileviez-Perez-Wise} where we found that only two cases 
are allowed by cosmology. In this case one has two possible scenarios: 
\begin{enumerate}
\item The extra fields are a fermionic $\rho_1 \sim (8,1,0)$ 
and the scalar $S \sim (8,2,1/2)$. 
\item One adds the fermion $\rho_2 \sim (8,3,0)$ and the scalar field $S \sim (8,2,1/2)$. 
\end{enumerate}
In both cases one generates neutrino masses through the loop in Fig.~\ref{Colored-Seesaw}.
The relevant interactions in this case are given by
\begin{equation}
-{\cal L}_{CS} = Y_2 \ \ell_L \ S \  \rho_1 \ + \ M_{\rho_1} \ {\rm Tr} \ \rho^2_1  
\ + \ \lambda_2 \  {\rm Tr} \left( S^\dagger H \right)^2 \ + \ {\rm h.c.} 
\label{V2}
\end{equation}
Using as input parameters, $M_{\rho_1} = 200$ GeV, $v_0=246$ GeV 
and $M_{S}=2$ TeV we find that in order to get 
the neutrino scale, $\sim 1$ eV, the combination 
of the couplings, $Y_2^2 \lambda_2 \sim 10^{-8}$. 
The experimental bounds on the colored octet masses depend of their decays. 
See Ref.~\cite{FileviezPerez:2010ch} for a detailed discussion.
This mechanism could be easily tested at the LHC. In this case the 
seesaw fields can be produced with very large cross sections through 
the QCD interactions. See Ref.~\cite{FileviezPerez:2010ch} for a detailed analysis of the 
collider signatures in this context. We would like to mention that this mechanism can be 
realized in the context of grand unified theories. 

\section{Neutrinoless double beta decay}
\begin{figure}[h]
\begin{center}
\includegraphics[width=0.3\linewidth]{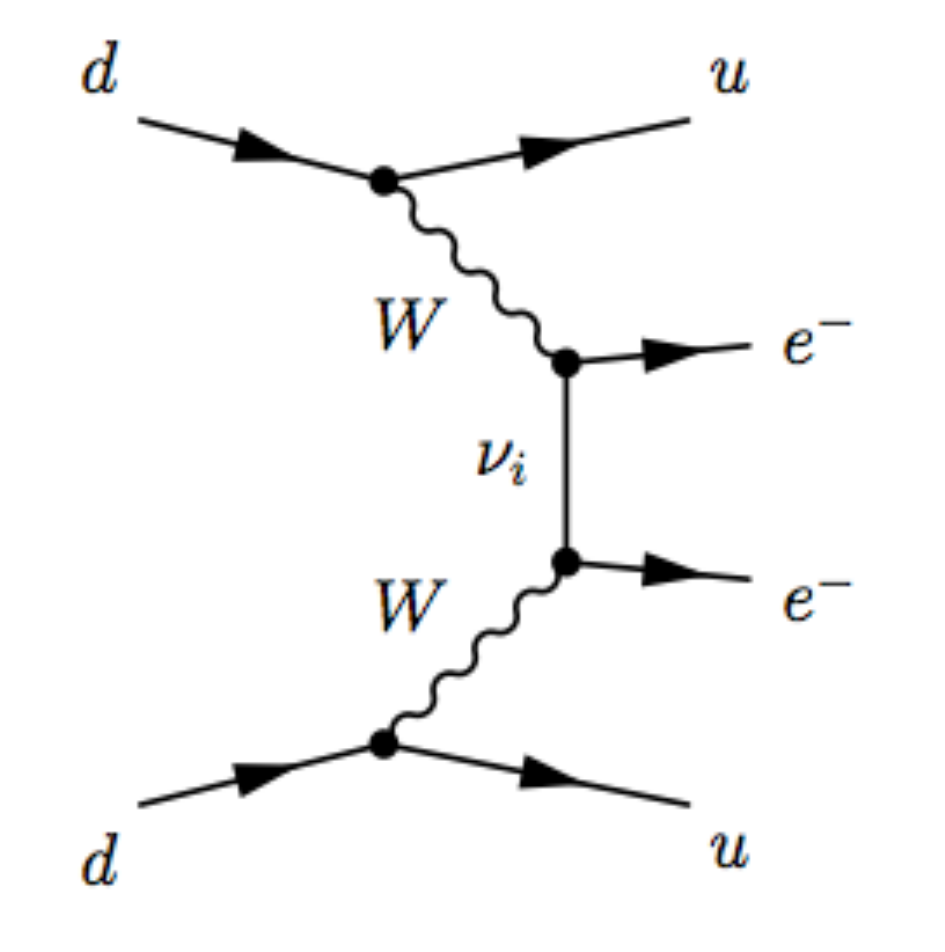}
 \caption{Feynman Graph for neutrinoless double beta decay~\cite{Bilenky:2014uka}. 
 This contribution is present in any model for Majorana neutrino masses.}
 \label{Neutrinoless}
 \end{center}
\end{figure}

The total lepton number can be broken in two units, $\Delta L=2,$ and one can have the exotic process
\begin{equation}
^A_ZX \ \to \  ^A_{Z+2} Y \ + \ 2 e^-,
\end{equation}
which is called neutrinoless double beta decay. Here $^A_ZX$ is a nuclei with atomic number $Z$ 
and atomic mass $A$. For a recent review on neutrinoless double beta decay see 
Ref.~\cite{Bilenky:2014uka}. This process proceeds through the Majorana neutrino mass 
insertion as one can see in Fig.~\ref{Neutrinoless} and the amplitude of these processes is proportional to the quantity
\begin{equation}
m_{\beta \beta}=\sum_i U_{ei}^2 \ m_i,
\end{equation}
where $U$ is the PMNS mixing matrix~\cite{Pontecorvo:1967fh,Maki:1962mu} in the leptonic sector and $m_i$ are the neutrino eigenvalues.
Unfortunately, the neutrino spectrum is unknown and one cannot predict the value of $m_{\beta \beta}$ 
and the lifetime for this process. The most optimistic scenario corresponds to the case when the neutrino spectrum has an inverted 
hierarchy. In this scenario the lower bound on $m_{\beta \beta}$ is about $2 \times 10^{-2}$ eV and 
there is a hope to discover this process at current or future experiments. See Ref.~\cite{Bilenky:2014uka} 
for more details and Refs.~\cite{Weinheimer:2013hya,Schwingenheuer:2012zs,Agostini:2013mzu} for the 
status of the experimental searches. In many extensions of the standard model one can have extra contributions 
for neutrinoless double beta decay. For example, in supersymmetric models with R-parity violation one 
has extra terms mediated by the trilinear and bilinear interactions breaking lepton number. In models with Type II 
seesaw the double charged Higgses give rise to new contributions and the same in other scenarios.
One can say that if neutrinoless double beta decay is discovered one can establish that the neutrino 
is a Majorana particle but we cannot learn too much about the mechanism for neutrino masses. 

\section{Grand Unified Theories}
The grand unified theories (GUTs) have been considered the most appealing extensions of
the Standard Model where one can understand the origin of the Standard Model interactions. In this context 
the gauge interactions of the SM are just different manifestations of the fundamental force defined 
at a different scale. The simplest GUT was proposed in 1974 by H. Georgi and S. Glashow in 
Ref.~\cite{Georgi:1974sy}. This theory is based on the $SU(5)$ gauge symmetry and the matter 
is unified in two representations in agreement with anomaly cancellation
\begin{equation}
\overline{5}=
\left( 
\begin{array}{c}
d^c_1 \\ 
d^c_2 \\
d^c_3 \\
e \\
- \nu
\end{array}
\right)_L,
\hspace{1.0cm} 
10=
\left( 
\begin{array}{ccccc}
0 & u^c_3 & - u^c_2 & - u^1 & - d^1 \\ 
- u^c_3 & 0 & u^1 & - u^2 & - d^2 \\
u^c_2 & -u^c_1 & 0 & - u^3 & - d^3\\
u^1 & u^2 & u^3 & 0 & e^c \\
d^1 & d^2 & d^3 & - e^c & 0
\end{array}
\right)_L.
\end{equation} 
The Higgs sector is composed of only two Higgs fields
\begin{equation}
{5}_H=
\left( 
\begin{array}{c}
T^1 \\ 
T^2 \\
T^3 \\
H^+ \\
H^0
\end{array}
\right),
\hspace{1.0cm} 
24_H=
\left( 
\begin{array}{cc}
\Sigma_8 & \Sigma_{(3,2)} \\ 
 \Sigma_{(3,2)}^* & \Sigma_3
\end{array}
\right)
+ \frac{1}{2\sqrt{15}}
\left( 
\begin{array}{cc}
2 & 0 \\ 
 0 & -3
\end{array}
\right) \Sigma_{24}.
\end{equation} 
The gauge bosons live in the adjoint representation $24_G$ which contains the Standard Model 
gauge bosons and the $V_\mu$ gauge bosons mediating proton decay,
\begin{equation} 
24_G=
\left( 
\begin{array}{cc}
G_\mu & V_\mu \\ 
 V^*_\mu & W_\mu
\end{array}
\right)
+ \frac{1}{2\sqrt{15}}
\left( 
\begin{array}{cc}
2 & 0 \\ 
 0 & -3
\end{array}
\right) B_\mu.
\end{equation} 
The new gauge bosons transform as $V_\mu \sim (3,2,-5/6)$ and mediate the dimension 
six operators discussed in the previous section. Unfortunately, the Georgi-Glashow model 
is not realistic because it fails to explain the values of the gauge couplings 
at the low scale. The idea of defining a simple and realistic grand unified theory based 
on $SU(5)$ has been investigated by many groups. There are two simple scenarios 
which are consistent with the experiment and we discuss their main features.
\begin{itemize}

\item Type II-SU(5)~\cite{Dorsner:2005fq,Dorsner:2005ii,Dorsner:2006hw}:

In this context the neutrino masses are generated through the Type II seesaw 
mechanism and one can have a consistent relation between the charged lepton 
and quark masses using higher-dimensional operators allowed by the gauge symmetry.
The neutrino masses are generated using the following interactions
\begin{equation} 
V \supset Y_\nu \ \overline{5} \ \overline{5} \ 15_H \ + \ \mu \ 5_H^* \ 5_H^* \ 15_H \ + \ {\rm{h.c.}},
\end{equation} 
where the $15_H$ field contains three fields, $\Phi_a \sim (1,3,1)$ is the field needed for Type II seesaw, 
$\Phi_b \sim (3,2,1/6)$ is a scalar leptoquark and $\Phi_c \sim (6,1,-2/3)$. In order to understand the possible predictions in this model, 
the constraints coming from unification have been investigated in detail in Refs.~\cite{Dorsner:2005fq,Dorsner:2005ii,Dorsner:2006hw}.
In Fig.~\ref{triangle} we show the constraints on the spectrum of the theory assuming gauge coupling unification at the one-loop level.
As one can appreciate, in order to find the maximal value of the unification scale the leptoquark $\Phi_b$ must be light. 
Here in order to achieve a proton lifetime consistent with the experiment one needs to use the suppression mechanism proposed 
in Ref.~\cite{Dorsner:2004xa}. 
\begin{figure}[t] 
\begin{center}
	\includegraphics[scale=0.7]{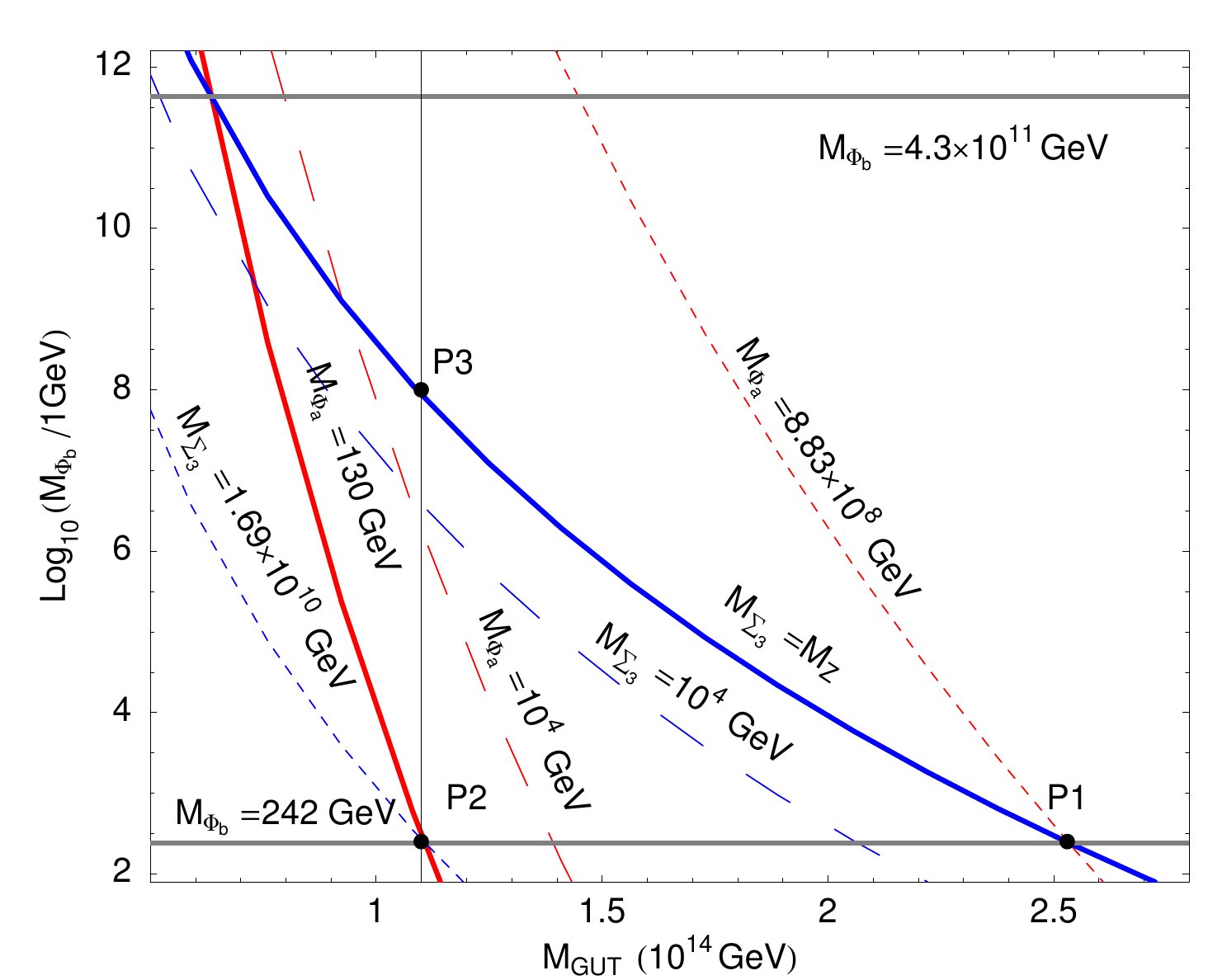}
	\caption{Constraints from gauge coupling unification in Type II-SU(5)~\cite{Dorsner:2006hw}. Here we show the allowed masses for the fields $\Phi_a$, 
	$\Phi_b$, and $\Sigma_3$ in agreement with gauge coupling unification at the one-loop level. The vertical line defines the bound from proton decay. The points P1, P2, 
	and P3 define the allowed area by unification and proton decay.}
	\label{triangle}
\end{center}
\end{figure} 
This model has two main features: a) Predicts the existence of a light lepto-quark which could be discovered at the LHC, 
b) The total lifetime of the proton is $\tau_p \leq 10^{36}$ years. The main phenomenological aspects of the 
leptoquark  $\Phi_b$ were investigated in Ref.~\cite{FileviezPerez:2008dw}. 

\item Type III-SU(5)~\cite{Goran,Dorsner:2006fx,Bajc:2007zf}:

The neutrino masses can be generated through the Type III seesaw mechanism. In this case one needs to include 
an extra fermionic representation in the adjoint representation, $24=(\rho_8, \rho_3, \rho_{(3,2)}, \rho_{(\bar{3},2)},\rho_{24})$.
This mechanism can be realized using the following interactions
\begin{equation} 
V \supset Y_1 \  \bar{5} \ 24 \ 5_H  \ + \  \frac{\bar{5}}{\Lambda} \left(  Y_2 \ 24 \ 24_H \ + \  Y_3 \ 24_H \ 24 \ + \ Y_4 \ \rm{Tr} (24 \ 24_H) \right) 5_H \ + \  {\rm{h.c.}}
\end{equation} 
Notice that $24$ contains two relevant fields for the seesaw mechanism, $\rho_3$ for Type  III seesaw and $\rho_{24}$ for Type I seesaw.  
The unification constraints have been investigated in great detail in this model~\cite{Goran,Dorsner:2006fx,Bajc:2007zf,DiLuzio:2013dda}.
In Fig.~\ref{triangle-24} we show the allowed parameter space in agreement with unification and the values of the gauge couplings at the low scale.
As one can appreciate, the field $\rho_3$ needed for Type III seesaw must be light in the theory. Therefore, one could hope to test the 
Type III seesaw mechanism at the LHC. See Refs.~\cite{Franceschini:2008pz,Arhrib:2009mz,Li:2009mw} for the testability of this mechanism.
This model also predicts a lifetime of the proton equal to $\tau_p < 10^{36-37}$ years~\cite{Goran,Dorsner:2006fx,Bajc:2007zf}. 
\begin{figure}[t] 
\begin{center}
	\includegraphics[scale=0.6]{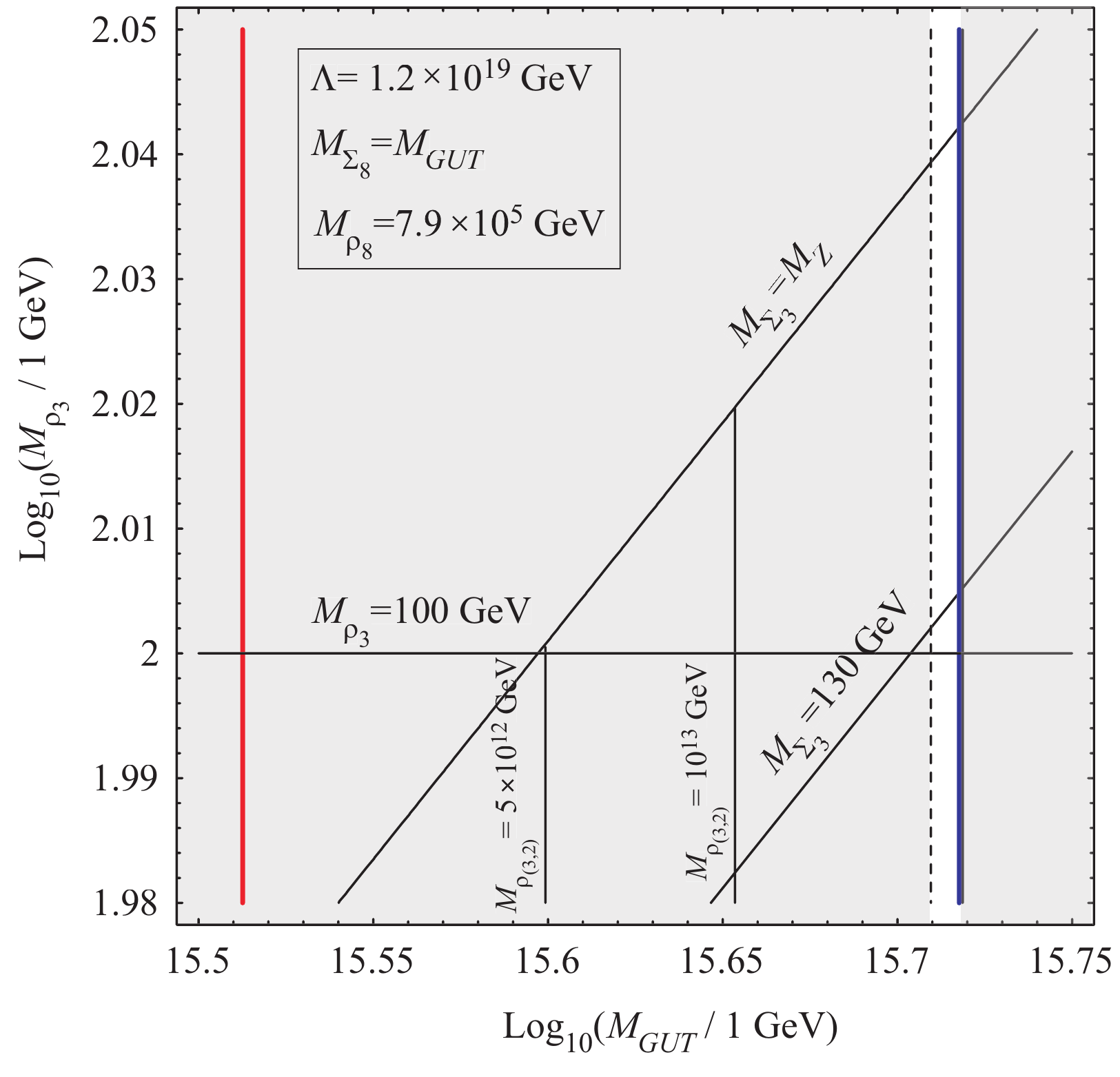}
	\caption{Constraints from gauge coupling unification in Type III-SU(5)~\cite{Dorsner:2006fx}. 
	Here we show the allowed masses for the $\rho_3$, $\Sigma_3$ and $\rho_{(3,2)}$ fields in agreement with gauge coupling unification at one-loop level.
	The dashed and blue vertical lines show the allowed parameter space by perturbativity. See Ref.~\cite{Dorsner:2006fx} for more details.}
	\label{triangle-24}
\end{center}
\end{figure} 
\end{itemize}
As we have discussed above, there are two simple theories based on $SU(5)$ which are consistent with the experiment. 
Theories based on the $SO(10)$ gauge symmetries are very appealing because one has the unification of the Standard Model 
fermions  in only one representation. In the $16$ spinor representation one has SM fermions for each family and the 
right-handed neutrinos. Unfortunately, the Higgs sector of these theories are often very involved and since we have several 
symmetry breaking paths to the Standard Model one cannot make good predictions for proton decay. See 
Ref.~\cite{Senjanovic:2012zt} for a review on grand unified theories.

The supersymmetric version of $SU(5)$ has been proposed by H. Georgi and S. Dimopoulos~\cite{Dimopoulos:1981zb}, 
and independently by N. Sakai~\cite{Sakai:1981gr}. In this context one needs to impose by hand the discrete symmetry 
$R-$parity in order to forbid the dimension four contributions to proton decay coming from the term $\hat{10} \ \hat{\bar{5}} \ \hat{\bar{5}}$.
The dimension five operators are also very important in this context. However, in general it is very difficult to predict the lifetime of the 
proton without knowing the full spectrum of supersymmetric particles. It is well-known, that the minimal renormalizable supersymmetric $SU(5)$ 
is ruled out because the relation $Y_d = Y_e$ is in disagreement with the experiment. Here $Y_d$ and $Y_e$ are the Yukawa matrices 
for down quarks and charged leptons, respectively. The simplest solution to this problem is to include the allowed 
higher-dimensional operators suppressed by the Planck scale. However, in this case one can have even less predictivity for 
proton decay. See Ref.~\cite{Nath:2006ut} for a review on proton decay and 
Refs.~\cite{Bajc:2002bv,EmmanuelCosta:2003pu,Bajc:2002pg,Hisano:2013exa} for a detailed discussion of this issue.
\section{Theories for Local Baryon and Lepton Numbers}
In this section we discuss the simplest realistic theories where the baryon and lepton numbers are defined as local gauge symmetries spontaneously broken 
at the low scale. Here we summarize the main results presented in Refs.~\cite{FileviezPerez:2010gw,Dulaney:2010dj,FileviezPerez:2011pt,FileviezPerez:2011dg,Duerr:2013dza,Duerr:2013opa,Perez:2014qfa,Duerr:2013lka,Perez:2013tea,Perez:2014kfa,Duerr:2014wra}. The original idea related to the spontaneous breaking of local baryon number
with the use of the Higgs mechanism was proposed by A. Pais in Ref.~\cite{Pais:1973mi}. In this article the author did not build a theory based on local baryon number but he proposed to use the Higgs mechanism to understand the spontaneous breaking of this symmetry. See also Refs.~\cite{Rajpoot:1987yg,Foot:1989ts,Carone:1995pu} for previous studies in theories with local baryon number.
 
In the Standard Model the baryon and lepton numbers are accidental global symmetries of the Lagrangian but they are not free of anomalies. 
In order to define a consistent theory where baryon and lepton numbers are local gauge symmetries, all relevant anomalies need to be cancelled. 
Therefore, the SM particle content has to be extended by additional fermions. In our notation, the SM fermionic fields and their 
transformation properties under the gauge group $$SU(3)\otimes SU(2) \otimes U(1)_Y \otimes U(1)_B \otimes U(1)_L$$ are listed in Table~2.
\begin{table}[t]
\begin{center}
  \begin{tabular}{cccccc}
\hline
\hline
~~Field ~~  &~~ $SU(3)$ ~~ &~~ $SU(2)$~~ &~~ $U(1)_Y$~~ &~~ $U(1)_B$~~ &~~ $U(1)_L$
~~ \\
\hline \hline
 $q_L$   & {\bf 3}  & {\bf 2} & $1/6$ & $1/3$ & $0$ \\
 $u_R$   & {\bf 3}  & {\bf 1} & $2/3$ & $1/3$ & $0$ \\
 $d_R$   & {\bf 3}  & {\bf 1} & $-1/3$ & $1/3$ & $0$ \\
 $\ell_L$   & {\bf 1}  & {\bf 2} & $-1/2$ & $0$ & $1$ \\
 $e_R$   & {\bf 1}  & {\bf 1} & $-1$  & $0$ & $1$ \\
 $\nu_R$   & {\bf 1}  & {\bf 1} & $0$  & $0$ & $1$ \\
\hline
\hline
\end{tabular}
\caption{The Standard Model fermionic content and the right-handed neutrinos.}
\end{center}
\label{SMfields}
\end{table}
Here, the right-handed neutrinos are part of the SM fermionic spectrum.
The baryonic anomalies are the following
\begin{eqnarray}
&& \mathcal{A}_1\left(SU(3)^2\otimes U(1)_B\right), \  \mathcal{A}_2\left(SU(2)^2\otimes U(1)_B\right), \mathcal{A}_3\left(U(1)_Y^2\otimes U(1)_B\right), \nonumber \\
&&  \mathcal{A}_4\left(U(1)_Y\otimes U(1)_B^2\right),  \mathcal{A}_5\left(U(1)_B \right), \  \mathcal{A}_6\left(U(1)_B^3\right). \nonumber 
\end{eqnarray}
In the Standard Model the only non-zero anomalies are $\mathcal{A}_2^{SM} = - \mathcal{A}_3^{SM} = 3/2$, while leptonic anomalies which must be cancelled are
\begin{eqnarray}
&& \mathcal{A}_7 \left(SU(3)^2\otimes U(1)_L\right), \  \mathcal{A}_8 \left(SU(2)^2\otimes U(1)_L\right), \  \mathcal{A}_9\left(U(1)_Y^2\otimes U(1)_L\right),  \nonumber \\
&& \mathcal{A}_{10}\left(U(1)_Y\otimes U(1)_L^2\right), \  \mathcal{A}_{11} \left(U(1)_L \right), \  \mathcal{A}_{12} \left(U(1)_L^3\right). \nonumber
\end{eqnarray}
The anomalies which are non-zero in the Standard Model with right-handed neutrinos are $\mathcal{A}_8^{SM}=-\mathcal{A}_9^{SM}=3/2$.
One also has to think about the cancellation of the mixed anomalies for the Abelian symmetries
\begin{eqnarray}
 &&\mathcal{A}_{13} \left(U(1)_B^2\otimes U(1)_L\right),
 \mathcal{A}_{14} \left(U(1)_L^2\otimes U(1)_B\right), \mathcal{A}_{15} \left(U(1)_Y\otimes U(1)_L\otimes U(1)_B\right), \nonumber
\end{eqnarray}
which of course vanish in the Standard Model. Various solutions to the equations which define the cancellation of anomalies were studied 
in Refs.~\cite{FileviezPerez:2010gw,Dulaney:2010dj,FileviezPerez:2011pt,FileviezPerez:2011dg,
Duerr:2013dza,Duerr:2013opa,Perez:2014qfa}. Here we discuss the simplest solutions to understand how one can write down a realistic version 
which could be tested at future experiments.
\begin{itemize}

\item {\textit{Sequential Family}}: In Refs.~\cite{FileviezPerez:2010gw,Dulaney:2010dj,FileviezPerez:2011pt} a sequential (chiral) family was added to the spectrum.
In this case the new quarks have baryon number $-1$ and the leptons have lepton number $-3$. This solution is ruled out today because the new quarks 
are chiral and change the gluon fusion Higgs production by a large factor, approximately a factor 9. Therefore, this is in clear disagreement with the recent experimental results at the Large Hadron Collider.   

\item {\textit{Mirror Family}}: In Refs.~\cite{FileviezPerez:2010gw,Dulaney:2010dj,FileviezPerez:2011pt} the possibility to use mirror fermions was considered.
However, as in the previous case, this scenario is ruled out by the Standard Model Higgs properties.

\item {\textit{Vector-Like Fermions}}: In Ref.~\cite{FileviezPerez:2011pt} the anomalies are cancelled using vector-like fermions. 
In this case, anomaly cancellation requires that the difference between the baryon numbers of the new quarks is equal to $-1$, 
while the difference between the lepton numbers of the new leptons is $-3$. In this scenario the neutrino masses are generated through the 
Type I seesaw mechanism and the new charged leptons acquire masses only from the SM Higgs vacuum expectation value. 
Therefore, the lepton number is broken by two units and one does not have proton decay. 
Unfortunately, the new charged leptons change dramatically the Higgs branching ratio into gamma gamma, reducing 
it by about a factor of 3. Therefore, this scenario is also ruled by the Higgs results at the Large Hadron Collider.         

One can modify this model adding a new Higgs boson with lepton number and generate vector-like masses 
for charged leptons, but one will generate dimension nine operators mediating proton decay, e.g.,
\begin{equation}
{\cal{O}}_9 = \frac{c_9}{\Lambda^5} \left( u_R u_R d_R e_R \right) S_B S_L^\dagger S_L^{'}.
\end{equation}
Here the Higgs bosons transform as

$$S_B \sim (1,1,0,-1,0), S_L \sim (1,1,0,0,-2), \  {\rm{and}}  \ S_L^{'} \sim (1,1,0,0,-3).$$
Now, assuming that $c_9 \sim 1$, and the vacuum expectation values for $S_B, S_L$, and $S_L^{'}$ around a TeV, one finds 
that $\Lambda \geq 10^{7-8}$ GeV. This means that one still has to postulate half of the desert in order to 
satisfy the proton decay bounds or assume a small coupling $c_9$.
\end{itemize}
As one can appreciate the simplest solutions for anomaly cancellation are ruled out by the collider experiments or require an 
extra mechanism to suppress proton decay. 

The possibility to define an anomaly free theory with gauged baryon and lepton numbers using lepto-baryons has been investigated.
In this context lepto-baryons are fermions with baryon and lepton numbers. There are two simple models which are in agreement 
with all bounds from cosmology and collider physics:
\begin{itemize}

\item In the first case adding only six vector-like fields one can define an anomaly free theory. See Table~3 for the particle content and 
Ref.~\cite{Duerr:2013dza} for details. In this context the lightest field could be a neutral Dirac fermion which is a good candidate 
for the cold dark matter in the universe. We will discuss the properties of this model in the next section.

\item There is a second simple model with only four representations. See Table~5 for the particle context and 
Ref.~\cite{Perez:2014qfa} for the detailed discussion. In this case one uses fields in the adjoint and fundamental 
representation of $SU(2)$ to cancel the non-trivial anomalies. The dark matter candidate in this case is a Majorana fermion.

\end{itemize}
%
In order to find realistic scenarios one considers the particle content listed in Table~\ref{fields-I}, where the new fermionic fields 
are singlets or live in the fundamental representation of $SU(2)$. 
\renewcommand{\arraystretch}{2}
\begin{table}[t]
\begin{center}
  \begin{tabular}{cccccc}
\hline
\hline
~~Field ~~  &~~ $SU(3)$ ~~ &~~ $SU(2)$~~ &~~ $U(1)_Y$~~ &~~ $U(1)_B$~~ &~~ $U(1)_L$
~~ \\
\hline \hline
 $\Psi_L$   & {\bf N}  & {\bf 2} & $Y_1$ & $B_1$ & $L_1$ \\
 $\Psi_R$   & {\bf N}  & {\bf 2} & $Y_1$ & $B_2$ & $L_2$ \\
 $\eta_R$   & {\bf N}  & {\bf 1} & $Y_2$  & $B_3=B_1$ & $L_3=L_1$ \\
 $\eta_L$   & {\bf N}  & {\bf 1} & $Y_2$   & $B_4=B_2$ & $L_4=L_2$ \\
 $\chi_R$   & {\bf N}  & {\bf 1} & $Y_3$  & $B_5=B_1$ & $L_5=L_1$ \\
 $\chi_L$   & {\bf N}  & {\bf 1} & $Y_3$   & $B_6=B_2$ & $L_6=L_2$ \\
\hline
\hline
\end{tabular}
\end{center}
 \caption{Fermionic Lepto-baryons in the model proposed in Ref.~\cite{Duerr:2013dza}.}
 \label{fields-I}
\end{table}
The $\mathcal{A}_2\left(SU(2)^2\otimes U(1)_B\right)$ anomaly can be cancelled if one imposes the condition
\begin{equation}
 B_1 - B_2 = -\frac{3}{N}.
\end{equation}
Now, to cancel the $\mathcal{A}_1\left(SU(3)^2\otimes U(1)_B\right)$  anomaly when $N \neq 1$ one needs to impose the relation
\begin{equation}
 2(B_1 - B_2)-(B_3 - B_4) - (B_5- B_6) = 0,
\end{equation}
which can be cancelled using
\begin{equation}
 B_1 = B_3=B_5, \text{ and } B_2 = B_4=B_6.
\end{equation}
See Table~\ref{fields-I} for more details.
There is a ``duality" between the cancellation the baryonic and leptonic anomalies. 
Both types of anomalies can be cancelled in the same way.
The anomaly $\mathcal{A}_2\left(SU(2)^2\otimes U(1)_L\right)$ can be cancelled using the condition
\begin{equation}
 L_1 - L_2 = -\frac{3}{N},
\end{equation}
for the leptonic charges and the others imposing the relation
\begin{equation}
 L_1 = L_3=L_5, \text{and } L_2 = L_4 = L_6 \, .
\end{equation}
Finally, one has to think about the anomalies with weak hypercharge. Once one uses the above assignment of baryon and lepton numbers, $\mathcal{A}_4$ and $\mathcal{A}_{10}$ 
are always cancelled and do not provide a condition for the hypercharges. In order to cancel the anomaly $U(1)_Y^2 \otimes U(1)_B$, one needs
\begin{equation}
 Y_2^2 + Y_3^2 - 2 Y_1^2 = \frac{1}{2}\, .
\end{equation}
A simple set of solutions for this equation is given by
\begin{equation}
 (Y_1,Y_2,Y_3) \subset \left\{ (\pm \frac{1}{2}, \pm 1,0), (\pm \frac{1}{6}, \pm \frac{2}{3}, \pm \frac{1}{3}), (0,\pm \frac{1}{2},\pm \frac{1}{2} )\right\}.
\end{equation}
It is important to mention that since the new particles are vector-like with respect to the Standard Model gauge group, the SM anomalies 
do not pose a problem. One can prove that the $U(1)_L \otimes U(1)_B^2 $, $U(1)_L^2 \otimes U(1)_B $ and $U(1)_Y\otimes U(1)_L\otimes U(1)_B$
anomalies are cancelled automatically. 

Now, one needs to discuss the possibility to satisfy all cosmological and collider constraints. 
In these scenarios one must avoid a stable electrically charged or colored field.
Therefore, the new fields should have a direct coupling to the SM fermions or the lightest 
particle in the new sector must be stable and neutral. Here we discuss the possible scenarios for different values of $N$.
${N=1}$: In this case the new fields do not feel the strong interaction, and the only solution which allows for a stable field in the new sector 
is the one with $Y_1=\pm 1/2$, $Y_2=\pm 1$, and $Y_3=0$. Then, when the lightest field is neutral one can have a dark 
matter candidate. The stability of the dark matter candidate in these models is an automatic consequence coming from symmetry breaking.
This is a very appealing solution because as a bonus one has a candidate to describe the cold dark matter in the Universe.

${N=3}$: In this scenario one can use the weak hypercharges  $Y_1=\pm 1/6$, $Y_2=\pm 2/3$, and $Y_3=\pm 1/3$, and 
a stable colored field can be avoided. In order to generate vector-like masses for the new fields one needs a scalar $S_{BL} \sim (1,1,0,-1,-1)$, 
and one generates dimension seven operators such as $q_L q_L q_L \ell_L S_{BL}/\Lambda^3$ mediating proton decay. Then, in this case the great desert must be postulated.

${N=8}$: This scenario could be interesting but in order to couple the new lepto-baryons to the SM fermions we need to include extra colored scalar fields. 
One way is to add color octet scalars that let the new fermions couple to leptons. The new colored scalars can decay at one loop to a pair of gluons after spontaneous symmetry breaking. 
This scenario is possible but we will focus our discussions on the simplest case which corresponds to $N=1$. 

\begin{table}[t]
\begin{center}
\begin{tabular}{cccccc}
\hline\hline
~~Fields ~~  &~~ $SU(3)$ ~~ &~~ $SU(2)$~~ &~~ $U(1)_Y$~~ &~~ $U(1)_B$~~ &~~ $U(1)_L$ ~~ \\
\hline 
\hline
  $\Psi_L$ & {\bf 1} & {\bf 2} & - $\frac{1}{2}$ & $B_1$ & $L_1$ \\ 
   $\Psi_R$ & {\bf 1} & {\bf 2} & - $\frac{1}{2}$ & $B_2$ & $L_2$ \\ 
   $\eta_R$ & {\bf 1} & {\bf 1} & - $1$ & $B_1$ & $L_1$ \\ 
   $\eta_L$ & {\bf 1} & {\bf 1} & - ${1}$ & $B_2$ & $L_2$ \\ 
   $\chi_R$ & {\bf 1} & {\bf 1} & 0 & $B_1$ & $L_1$ \\   
   $\chi_L$ & {\bf 1} & {\bf 1} & 0 & $B_2$ & $L_2$ \\ 
\hline \hline
\end{tabular}
 \caption{Lepto-baryons in the model presented in Ref.~\cite{Duerr:2013dza}.}
  \label{ModelI}
  \end{center}
\end{table}

\subsection{Theoretical Framework}
The main goal here is to discuss a realistic anomaly free theory based on the gauge group~\cite{FileviezPerez:2010gw} $$SU(3) \otimes SU(2) \otimes U(1)_Y \otimes U(1)_B \otimes U(1)_L.$$
The simplest solution discussed in the last section is the one with colorless fermions. 
The new fermion fields of this model are given in Table~4 assuming $N=1$, $Y_1=\pm 1/2$, $Y_2=\pm 1$, and $Y_3=0$.
For simplicity, in this section we discuss only the baryonic sector of the theory. 
In this case the relevant Lagrangian is given by
\begin{equation}
 \mathcal{L} = \mathcal{L}_\text{SM} + \mathcal{L}_\text{B},
\end{equation} 
where $ \mathcal{L}_\text{SM}$ is the Lagrangian for the fields present in the SM. Since all quarks have baryon number their kinetic term is modified. 
In the above equation $\mathcal{L}_\text{B}$ reads as 
\begin{equation}
\label{eq:LB}
 \mathcal{L}_B = - \frac{1}{4} B_{\mu \nu}^B B^{\mu \nu, B}  - \frac{\epsilon_{B}}{2} B_{\mu \nu}^B B^{\mu \nu} +  \mathcal{L}_f +   \mathcal{L}_{S_B},
\end{equation} 
where
$B_{\mu \nu}= \partial_\mu B_\nu - \partial_\nu B_\mu$ is the $U(1)_Y$ strength tensor, $B_{\mu \nu}^B= \partial_\mu B_\nu^B - \partial_\nu B_\mu^B$ is 
the $U(1)_B$ strength tensor, and $\mathcal{L}_f$ is
\begin{align}
 \mathcal{L}_f   & =  i \overline{\Psi}_L \slashed{D} \Psi_L + i \overline{\Psi}_R \slashed{D} \Psi_R + i \overline{\eta}_L \slashed{D} \eta_L +  i \overline{\eta}_R \slashed{D} \eta_R + i \overline{\chi}_L \slashed{D} \chi_L + i \overline{\chi}_R \slashed{D} \chi_R \nonumber \\
&  -  Y_1 \overline{\Psi}_L H \eta_R - Y_2 \overline{\Psi}_L \tilde{H} \chi_R  
 - Y_3 \overline{\Psi}_R H \eta_L - Y_4 \overline{\Psi}_R \tilde{H} \chi_L \nonumber \\
& -  \lambda_{\Psi} \overline{\Psi}_L \Psi_R S_{B} - \lambda_{\eta} \overline{\eta}_R \eta_L S_{B} 
 - \lambda_{\chi} \overline{\chi}_R \chi_L S_{B} \ + \ {\rm h.c.}
\end{align}
The term $\mathcal{L}_{S_B}$ is defined by 
\begin{equation}
\mathcal{L}_{S_B} =   \left( D_\mu S_B\right)^\dagger D^\mu S_B - m_B^2 S_B^\dagger S_B  - \lambda_B (S_B^\dagger S_B)^2  - \lambda_{HB} (H^\dagger H)(S_B^\dagger S_B) .
\end{equation} 
As we have discussed above the baryon numbers $B_1$ and $B_2$ of the new fermions are constrained 
by the condition
\begin{equation}
 B_1 - B_2 = -3.
\end{equation}
In this context the scalar sector is composed of the SM Higgs and the new boson $S_B$ with the following quantum numbers
\begin{equation}
 H \sim (\mathbf{1},\mathbf{2},1/2,0), \hspace{0.5cm} {\text{and}}  \hspace{0.5cm} S_{B} \sim (\mathbf{1},\mathbf{1},0,-3). 
\end{equation}
Notice that the need to generate vector-like masses for the new fermions define the baryon number of the new scalar.
Once $S_B$ gets a vacuum expectation value we will have only $|\Delta B|=3$ interactions and proton decay never occurs.
Therefore, the great desert is not needed. For the application of this idea in low energy supersymmetry see Ref.~\cite{Arnold:2013qja}.
%
\begin{table}[t]
\begin{center}
\begin{tabular}{cccccc}
\hline\hline
~~Fields ~~  &~~ $SU(3)$ ~~ &~~ $SU(2)$~~ &~~ $U(1)_Y$~~ &~~ $U(1)_B$~~ &~~ $U(1)_L$ ~~ \\
\hline 
\hline
  $\Psi_L$ & {\bf 1} & {\bf 2} & $\frac{1}{2}$ & $\frac{3}{2}$ & $\frac{3}{2}$ \\ 
   $\Psi_R$ & {\bf 1} & {\bf 2} & $\frac{1}{2}$ & -$\frac{3}{2}$ & -$\frac{3}{2}$ \\ 
   $\Sigma_L$ & {\bf 1} & {\bf 3} & $0$ & -$\frac{3}{2}$ & -$\frac{3}{2}$ \\ 
   $\chi_L$ & {\bf 1} & {\bf 1} & 0 & -$\frac{3}{2}$ & -$\frac{3}{2}$ \\ 
\hline \hline
\end{tabular}
 \caption{Lepto-baryons in the model presented in Ref.~\cite{Perez:2014qfa}.} 
 \end{center}
  \label{modelwith4fields}
\end{table}
\subsection{Phenomenological Aspects}
The class of theories discussed above predict the existence of a new neutral gauge boson associated to the local baryon number. 
The interactions of the lepto-phobic gauge boson are given by
\begin{equation}
{\cal L}_B \supset - g_B \bar{\chi} \left( B_1 P_L \ + \ B_2 P_R \right)   \gamma^\mu  \chi \  Z_\mu^B + \frac{1}{2} M_{Z_B}^2 Z_\mu^B Z^{B, \mu} - \frac{1}{3} g_B 
\sum _{i} \bar{q}_i \gamma^\mu q_i Z_\mu^B,
\end{equation}    
neglecting the kinetic mixing between the two Abelian symmetries, the term $\epsilon_B$ in Eq.~(\ref{eq:LB}), such that $Z_\mu^B = B_\mu^B$. Here $P_L=(1-\gamma^5)/2$ and $P_R=(1+\gamma^5)/2$ are the usual projection operators, while $B_1$ and $B_2$ are the baryon numbers of the dark matter candidate, see Table 4 for the quantum numbers.
The mass of the leptophobic gauge boson is given by $M_{Z_B}=3 g_B v_B$, where $v_B$ is the vacuum expectation value of the $S_B$ boson. 
The Dirac spinor, $\chi=\chi_L+\chi_R$, is stable and it is the cold dark matter candidate.
\begin{figure}[t]
\begin{center}
\includegraphics[width=0.49\linewidth]{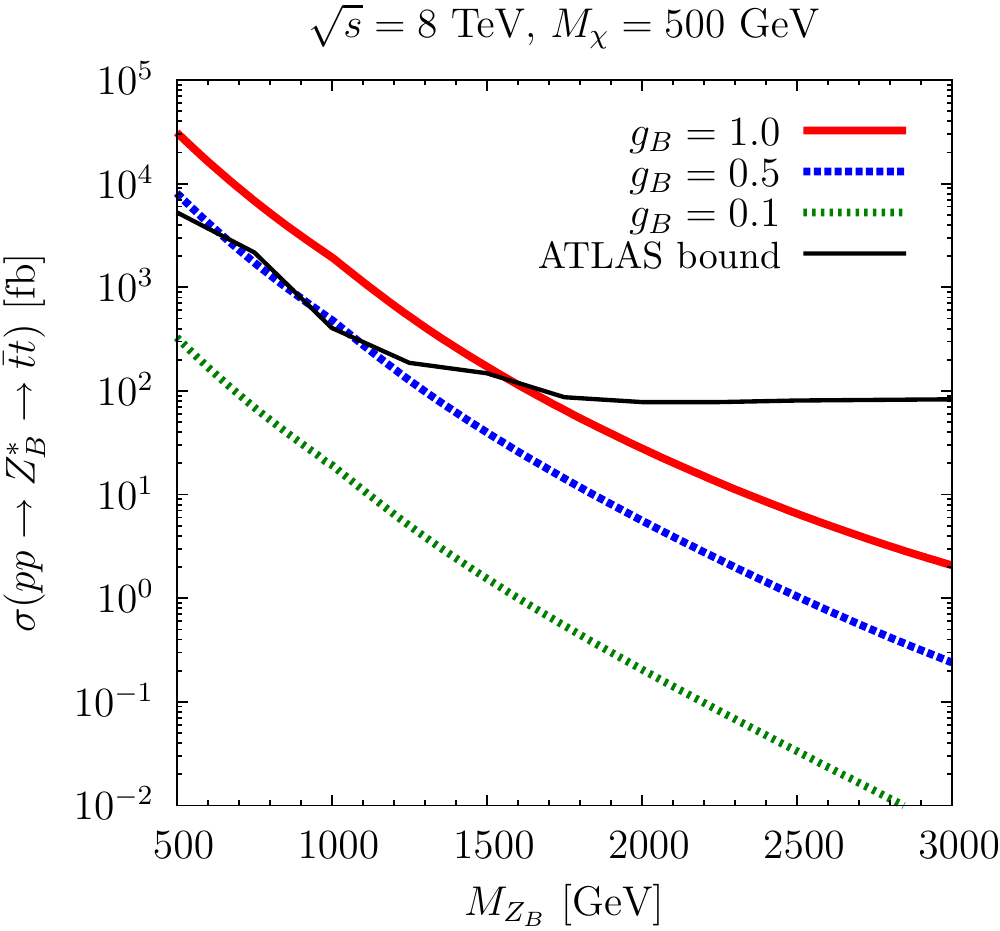}
 \caption{Decay of the leptophobic gauge boson $Z_B$ into two top quarks. Here we show the experimental bounds from the ATLAS collaboration~\cite{ATLAS} (solid black) and the theoretical predictions for different values of the gauge couplings ($g_B=1$ in red, $g_B=0.5$ in blue, and $g_B=0.1$ in green) when $\sqrt{s}=8$ TeV. For more details see Ref.~\cite{Duerr:2014wra}.}
 \label{fig:ZBbounds}
\end{center}
\end{figure}

\begin{figure}[t]
\begin{center}
\includegraphics[width=0.6\linewidth]{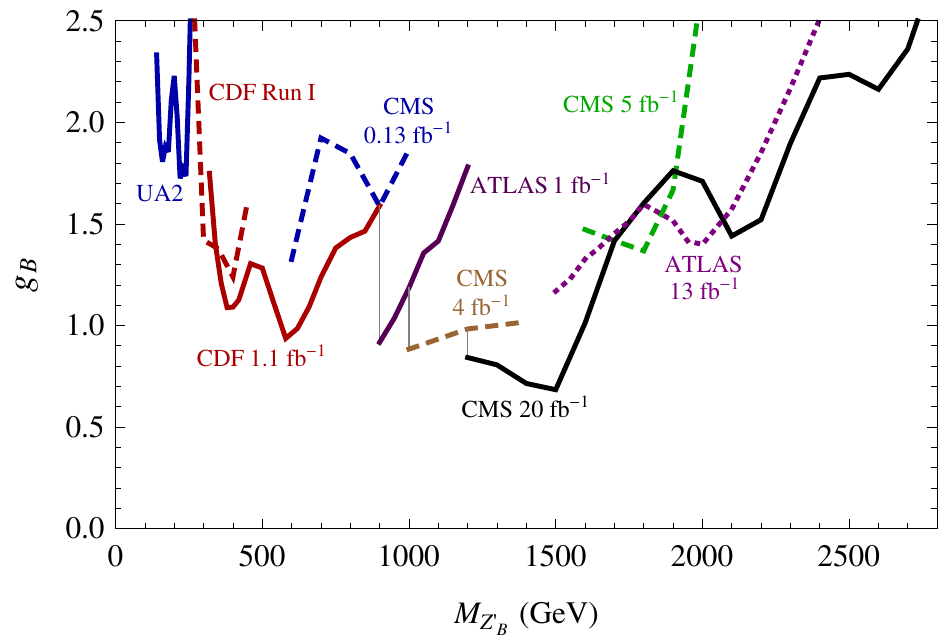}
 \caption{Experimental bounds on the properties of a leptophobic gauge boson in the plane gauge coupling--gauge boson mass. 
 For more details see Ref.~\cite{Dobrescu:2013cmh}.}
 \label{ZBbounds}
\end{center}
\end{figure}

Using the decay of $Z_B$ into two top quarks the ATLAS collaboration has set bounds on this type of gauge bosons, 
see Ref.~\cite{ATLAS}. Here the relevant production channel is $pp \ \to \ Z_B^* \ \to \ \bar{t} t$. The numerical results for the cross section are 
shown in Fig.~\ref{fig:ZBbounds} for different values of the gauge coupling $g_B$, we use the values $0.1$, $0.5$, and $1.0$. 
The black curve in Fig.~\ref{fig:ZBbounds} shows the experimental bounds from the ATLAS collaboration~\cite{ATLAS}. 
Then, since the area above the black curve is ruled out by the experiment one can say that the gauge coupling must be smaller 
than 0.5 to be consistent with the experiment in most of the parameter space. In Fig.~\ref{ZBbounds} we show the bounds 
on the gauge couplings and the gauge boson mass in a large range. As one can appreciate the gauge coupling can be large in the low mass region.
See Refs.~\cite{Dobrescu:2013cmh,An:2012ue,Tulin:2014tya} for a detailed discussion of the experimental bounds.
The properties of the Higgs sector of this models have been investigated in detail 
in Ref.~\cite{Duerr:2014wra}. Here the new physical Higgs boson can decay with a large branching ratio 
into dark matter when the mixing between the two physical Higgses is small. 
This scenario is preferred by the recent discovery of a SM-like Higgs boson.  

\subsection{Cosmological Aspects}
We have mentioned above that in the context of the simplest theories with gauged baryon and lepton numbers 
one has a candidate for the cold dark matter in the Universe. In the model discussed in the previous 
section the dark matter candidate is a Dirac fermion, $\chi=\chi_L+\chi_R$. Since the local baryon number 
symmetry is spontaneously broken at the low scale one must understand the possibility to have a consistent 
scenario for baryogenesis and explain the possible relation between the baryon asymmetry and the 
cold dark matter density. This issue has been investigated in Ref.~\cite{Perez:2013tea} and we outline the main results.
See Ref.~\cite{Zurek:2013wia} for a review of asymmetric dark matter mechanisms.

{\textit{Baryon and Dark Matter Asymmetries}}:
The model discussed in the previous section enjoys three anomaly-free symmetries~\cite{Perez:2013tea}:
\begin{itemize}

\item $B-L$ in the Standard Model sector is conserved in the usual way. We will refer to this symmetry as $(B-L)_{SM}$.  

\item The accidental $\eta$ global symmetry in the new sector. Under this symmetry the new fields transform in the following way:
\begin{align*}
\Psi_{L,R} \to e^{i \eta}  \Psi_{L,R}, \
 \eta_{L,R} \to e^{i \eta} \eta_{L,R}, \
 \chi_{L,R} \to e^{i \eta} \chi_{L,R}. 
\end{align*}

\item Above the symmetry breaking scale, total baryon number $B_T$ carried by particles in both, the standard model and 
the new sectors is also conserved. Here we will assume that the symmetry breaking scale for baryon number is low, close to the electroweak scale. 

\end{itemize}
We recall the thermodynamic relationship between the number density asymmetry $n_+ - n_-$ and the chemical potential
in the limit $\mu \ll T$,
\begin{equation}
\frac{\Delta n}{\rm{s}}=\frac{n_+ - n_-}{\rm{s}}=\frac{15 \ g}{2\pi^2 g_{*} \ \xi} \frac{\mu}{T},
\end{equation}
where $g$ counts the internal degrees of freedom, ${\rm{s}}$ is the entropy density, and $g_{*}$ 
is the total number of relativistic degrees of freedom. Here $\xi=2$ for fermions and $\xi=1$ for bosons.

In order to derive the relationship between the baryon asymmetry and the dark matter relic density, one defines the densities 
associated with the conserved charges in this theory. Following the notation of Ref.~\cite{Harvey}, the $B-L$ asymmetry 
in the SM sector is defined in the usual way
\begin{eqnarray}
\Delta (B-L)_{SM} &=& \frac{15}{ 4 \pi^2 g_{*} T}   3 ( \mu_{u_L} +  \mu_{u_R} +  \mu_{d_L}  + \mu_{d_R} -  \mu_{\nu_L} -  \mu_{e_L} -  \mu_{e_R}),
\end{eqnarray}
and the $\eta$ charge density is given by
\begin{equation}
\begin{aligned}
\Delta \eta  &= \frac{15}{ 4 \pi^2 g_{*} T}   \big( 2 \mu_{\Psi_L} +  2 \mu_{\Psi_R} +  \mu_{\chi_L}  + \mu_{\chi_R}  +  \mu_{\eta_L} +  \mu_{\eta_R} \big).
\end{aligned}
\end{equation}
Isospin conservation $T_3=0$ implies
\begin{equation}
\mu_{u_L}=\mu_{d_L},\qquad \mu_{e_L}=\mu_{\nu_L}, \qquad \mu_+ = \mu_0\,.
\end{equation}
Standard Model interactions and the Yukawa interactions present in the theory give us other equilibrium conditions for the chemical potentials.
The conservation of electric charge $Q_{em}=0$ in this model implies
\begin{multline}
6 (\mu_{u_L} + \mu_{u_R}) - 3 (\mu_{d_L} + \mu_{d_R}) - 3 (\mu_{e_L} + \mu_{e_R}) +
2 \mu_0 -  \mu_{\Psi_L}- \mu_{\Psi_R} -  \mu_{\eta_L}- \mu_{\eta_R}=0. 
\end{multline}
The electroweak sphalerons in this context must satisfy the conservation of total baryon number, and the associated `t Hooft operator is
\begin{equation}
(q_L q_L q_L \ell_L)^3 \overline{\Psi}_R \Psi_L\,.
\end{equation}
Therefore, the sphalerons give us an additional equilibrium condition between the standard model particles and new degrees of freedom
\begin{equation}
3 ( 3 \mu_{u_L} + \mu_{e_L}) + \mu_{\Psi_L} - \mu_{\Psi_R}=0. 
\end{equation}
This result is very unique. It is important to emphasize that the baryon number in the SM sector is broken but the total baryon number is conserved.
Notice that in this model the sphalerons play the crucial role of transferring the asymmetry from the standard model sector to the dark matter sector.

The total baryon number is conserved above the symmetry breaking scale. Therefore the requirement of $B_T=0$ gives
\begin{align}\nonumber
B_T &= \frac{15}{ 4 \pi^2 g_{*} T} \Big[ 3 ( \mu_{u_L} +  \mu_{u_R} +  \mu_{d_L}  + \mu_{d_R}) + B_1 ( 2  \mu_{\Psi_L} +  \mu_{\eta_R} + \mu_{\chi_R}  ) +  B_2 (  2  \mu_{\Psi_R} +  \mu_{\eta_L} + \mu_{\chi_L} ) \nonumber \\
&\qquad - 6  \mu_{S_B}) \Big]=0.
\end{align}
Using all conditions on the chemical potential in the theory, one can write the final baryon asymmetry in the SM sector 
in terms of $\Delta(B-L)_{\text{SM}}$ and $\Delta\eta$~\cite{Perez:2013tea},
\begin{eqnarray}\nonumber
B^\text{SM}_f &\equiv& \frac{15}{ 4 \pi^2 g_{*} T}    \left( 12  \mu_{u_L} \right) = C_1\,\Delta (B-L)_\text{SM} + C_2\,\Delta \eta,
\end{eqnarray}
with
\begin{equation}
C_1=\frac{32}{99} \qquad {}\text{and} \qquad C_2 = \frac{(15 - 14 B_2)}{198}.
\end{equation}
Requiring that the dark matter asymmetry is bounded from above by the observed dark matter density, 
one finds the following upper bound on the dark matter mass
\begin{equation}\label{eq:bound}
M_\chi \leq \frac{\Omega_{DM} \ C_2 \ M_p}{ | \Omega_B - C_1 \ \Omega_{B-L}|}.
\end{equation}
Here $\Omega_{B-L}={\rm{s}} \ \Delta (B-L)_{SM} M_p/\rho_c$, $M_p$ is the proton mass, $\rho_c$ is the critical density and ${\rm{s}}$ is the entropy.
As one can appreciate, this bound is a function of the baryon number $B_2$, and the $B-L$ asymmetry generated through a mechanism 
such as leptogenesis. Therefore, in general one could say that there is a consistent relation between 
the baryon and dark matter asymmetries.  In this model one does not have a mechanism to explain the primordial asymmetry 
in the dark matter sector. Therefore, using $\Delta \eta=0$ we find the following relation
\begin{equation}\label{eq:bound}
\Omega_{B} = \frac{32}{99} \Omega_{B-L},
\end{equation}
where $\Omega_{B-L}$ is generated through a mechanism such as leptogenesis.
In this case one has only thermal dark matter relic density and a simple relation between the $B-L$ asymmetry 
and the final baryon asymmetry.  

\begin{figure}[t]
\begin{center}
\includegraphics[width=0.6\linewidth]{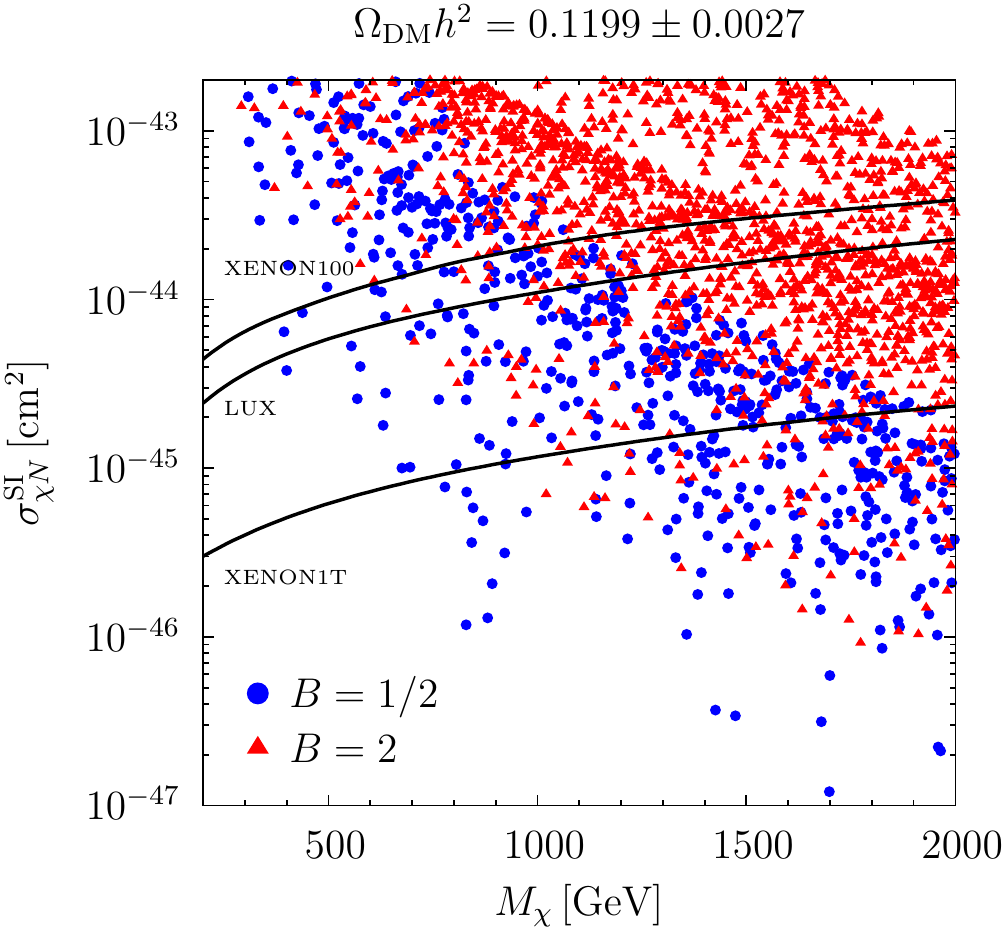}
\caption{Prospects for DM direct detection, assuming the value of the DM relic density $\Omega_\text{DM} h^2 = 0.1199 \pm 0.0027$ measured by Planck~\cite{Ade:2013zuv}. The plot shows the spin-independent elastic DM--nucleon cross section $\sigma_{\chi N}^\text{SI}$ as a function of the DM mass $M_\chi$. The exclusion limits of XENON100~\cite{Aprile:2012nq} and LUX~\cite{Akerib:2013tjd} are given, as well as the projected limit for XENON1T~\cite{Aprile:2012zx}. The gauge coupling is varied inside $g_B \in [0.1,0.5]$, and the gauge boson mass is varied inside $M_{Z_B}= 0.5-5.0$ TeV. 
Blue dots are for $B=1/2$, red triangles are for $B=2$. See Ref.~\cite{Duerr:2014wra} for more details.}
\label{fig:directdetection}
\end{center}
\end{figure}

{\textit{Thermal Baryonic Dark Matter}}:
In Refs.~\cite{Duerr:2013lka,Duerr:2014wra} the possibility to 
achieve the correct relic density and the constraints from direct detection have been investigated.
Assuming that the total cold dark matter relic density is explained with the Dirac spinor $\chi$, 
the predictions for direct detection are shown in Fig.~\ref{fig:directdetection}.
One can see that there are many allowed scenarios when one can satisfy the relic density and direct detection constraints from dark matter experiments.
In this case the dark matter annihilation cross section into SM particles and the nucleon--dark matter elastic cross section proceed through the leptophobic $Z_B$.
A nice feature of this model is that the direct detection cross section is independent of the matrix element because baryon number is a conserved current.
See Refs.~\cite{Duerr:2013lka,Duerr:2014wra}  for more details.

{\textit{Upper Bound on the Symmetry Breaking Scale:}}
The existence of a non-zero relic density can be used to find an upper bound on the symmetry breaking scale on 
models with gauged baryon number when the dark matter candidate is a Dirac fermion.
Here we summarize the results presented in Ref.~\cite{Duerr:2014wra}. Using the upper bound on the 
dark matter relic density, $\Omega_{DM} h^2 \leq 0.12$, and the fact that the dark matter and the gauge boson masses are generated 
through the same Higgs mechanism, i.e. $M_\chi = \lambda_\chi v_B/\sqrt{2}$ and $M_{Z_B}=3 g_B v_B$ in these models, 
one finds an upper bound on the symmetry breaking scale given by
\begin{equation}
v_B^2 \leq \frac{g_B^4 \ \lambda_\chi^2 \ (B_1+B_2)^2 \ 1.77 \times 10^{9} \ {\rm{GeV}^2}}{168 \pi \ \left( (2 \lambda_\chi^2 - 9 g_B^2)^2 + \frac{9}{4 \pi^2} g_B^8 \right) x_f},
\end{equation}
for a given value of the freeze-out temperature $x_f$. Therefore, it is possible to find an upper bound on the gauge boson mass using the above equation
\begin{equation}
M_{Z_B} \leq 316.1 \  \frac{(B_1+B_2)}{\sqrt{x_f}} {\rm{TeV}}.
\end{equation}
Now, using $x_f=20$ and $B_1+B_2=1/2$ as example, the upper bound on the gauge boson mass reads as
\begin{equation}
M_{Z_B} \leq 35.3 \ {\rm{TeV}},
\end{equation} 
and $M_\chi \leq 17.7$ TeV. Notice that this bound is true when $Z_B$ is heavier than the dark matter candidate.
This result is very interesting because one can say that there is a hope to test or rule out this model at the current or future collider experiments.
Of course, the LHC cannot test the theory if the new gauge boson has a mass close to the upper bound. However, 
one could rule out this type of spectrum at future 100 TeV colliders. 
Notice that this bound is much smaller than the one coming from unitarity constraints~\cite{Griest:1989wd}.

\subsection{Towards Low Scale Unification}
The theories with local baryon number open up a new possibility for physics beyond the Standard Model. 
Since the great desert is not needed in this picture one can imagine that the unification of gauge forces 
can be realized at the low scale. Using a bottom-up approach we will discuss the possibility to have the unification 
of the gauge interactions at the low scale~\cite{Perez:2014kfa}.
\begin{figure}[t]
	\centering
		\includegraphics[width=0.6\textwidth]{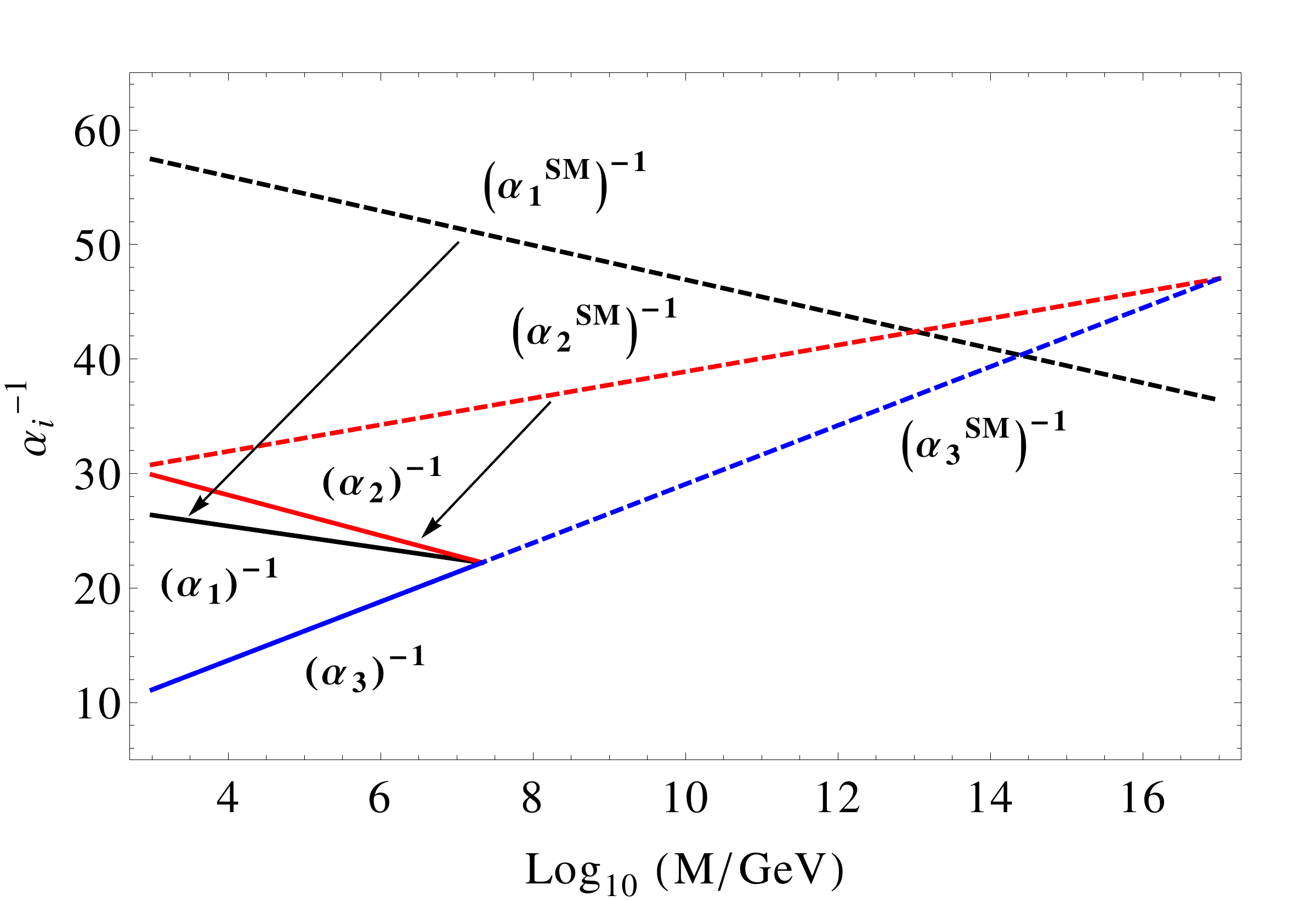}
		\caption{Evolution of the gauge couplings in the Standard Model represented by dashed lines
		and in the new model where the unification scale is realized at $10^4$ TeV. See Ref.~\cite{Perez:2014kfa} for details.}
		\label{runing-SM}
\end{figure}
In order to investigate the unification of gauge interactions at the low scale it is important to make sure that the proton 
is stable or the baryon number violating operators are highly suppressed.  
This theory based on the gauge group $$SU(3) \otimes SU(2) \otimes U(1)_Y \otimes U(1)_B \otimes U(1)_L,$$ has been 
discussed in the previous sections. In one of the models the new fermions called ``lepto-baryons" needed for anomaly cancellation are
\begin{align}
\Psi_L \sim (1 , 2 , 1/2 , B, L ), \ \Psi_R \sim (1, 2, 1/2, -B, -L ), \nonumber \\
\Sigma_L \sim (1, 3, 0, -B, -L),   \  {\rm and } \  \chi_L \sim (1, 1, 0, -B, -L),     \nonumber
\end{align}
where $B=L=3/ (2 n_F)$, and $n_F$ is the number of copies. Notice that these fields change 
only the evolution of the $SU(2)$ and $U(1)_Y$ gauge couplings. The Higgs sector of this theory is composed of the fields $S_B \sim (1, 1, 0, 2B, 2L)$ 
and $S_L \sim (1, 1, 0, 0, 2)$. When $U(1)_B$ is spontaneously broken 
and $n_F \neq 3n$ ($n$ is an integer number) the proton is stable because one only has $\Delta B=\pm 3/n_F$ interactions.

The evolution of the Standard Model gauge couplings at one-loop level is described by the equation
\begin{equation}
k_i\alpha^{-1}_i(M) = \alpha^{-1}_i(M_Z) - \frac{B_i}{2 \pi} \ {\rm Log}\left(\frac{M}{M_Z}\right),
\end{equation}
where $k_i = (k_1, k_2, k_3)$ are the normalization factors. In the simplest grand unified theory based on $SU(5)$ 
one has $k_i=(5/3,1,1)$. However, in general the $k_1$ value depends of the embedding in a given theory. 
The coefficients
\begin{equation}
B_i =  b^\text{SM}_i  + \theta (M-M_F) \ n_F \ b^{\rm{new}}_i  \times \frac{{\rm Log}(M/M_F)}{{\rm Log}(M/M_Z)},
\end{equation}
contain all possible contributions from the low scale, $M_Z$, to the unification scale, $M_U$.
Here, $M_F$ is the mass scale of the new fermions, $\theta (x)$ is the step function, and 
$b^{\rm{SM}}_i = (41/6, -19/6, -7)$ are the coefficients in the Standard Model.
In the model discussed above the new coefficients are $b_i^{\rm{new}}=(2/3,2,0)$.  

In Fig.~\ref{runing-SM} one has the evolution of the gauge couplings in the Standard Model and in the model with lepto-baryons. 
The lepto-baryons at the low scale change the evolution of $\alpha_2$ and $\alpha_1$ dramatically without affecting $\alpha_3$. 
Here one assumes four copies of lepto-baryons with baryon and lepton numbers equal to $3/8$. Since they have different baryon and 
lepton numbers than the Standard Model fields one never induces new sources of flavor violation and the unification scale 
$M_U$ is larger than $10^4$ TeV in order to suppress the flavor violating effective operators. The evolution of the new couplings is defined by the equation
\begin{align}
\alpha^{-1}_X (M_Z) &= k_X \alpha^{-1}_X (M_U) + \frac{B_X}{2 \pi} \ {\rm Log}\left(\frac{M_U}{M_Z}\right),
\end{align}
where $X=B,L$. The $B_X$ coefficients are given by
\begin{align}
B_B = b^\text{SM}_B + \theta (M_U-M_F) \ b^{\rm{new}}_B \times \frac{{\rm Log}(M_U/M_F)}{{\rm Log}(M_U/M_Z)},
\end{align}
and
\begin{align}
B_L = b^\text{SM}_L + \theta (M_U-M_F) \ b^{\rm{new}}_L \times  \frac{{\rm Log}(M_U/M_F)}{ {\rm Log}(M_U/M_Z)} \,.
\end{align}
Notice that the values of these gauge couplings at different scales and the $k_X$ are barely constrained. 
However, one can envision that it is possible to define a theory where all gauge couplings are unified, the Standard Model couplings, 
$\alpha_1$, $\alpha_2$, $\alpha_3$, and the new couplings $\alpha_B$ and $\alpha_L$.Therefore, assuming unification at the 
$M_U$ scale, one can find the values of $\alpha_B$ and $\alpha_L$ at the $M_Z$ scale for given values of $k_B$ and $k_L$.
The contribution of the Standard Model fields to the running of the baryonic and 
leptonic couplings is given by the coefficients $b^\text{SM}_B = 8/3$ and $b^\text{SM}_L = 6$.
The new fields contribute as follows
\begin{align}
b^{\rm{new}}_B &= \left(\frac{16}{3}n_F + \frac{4}{3}\right)B^2={3} \frac{(4 n_F + 1)}{n_F^2}, \\
b^{\rm{new}}_L &=  b^{\rm{new}}_B + \frac{10}{3} \,.
\end{align}
\begin{figure}[t]
	\centering
		\includegraphics[width=0.6\textwidth]{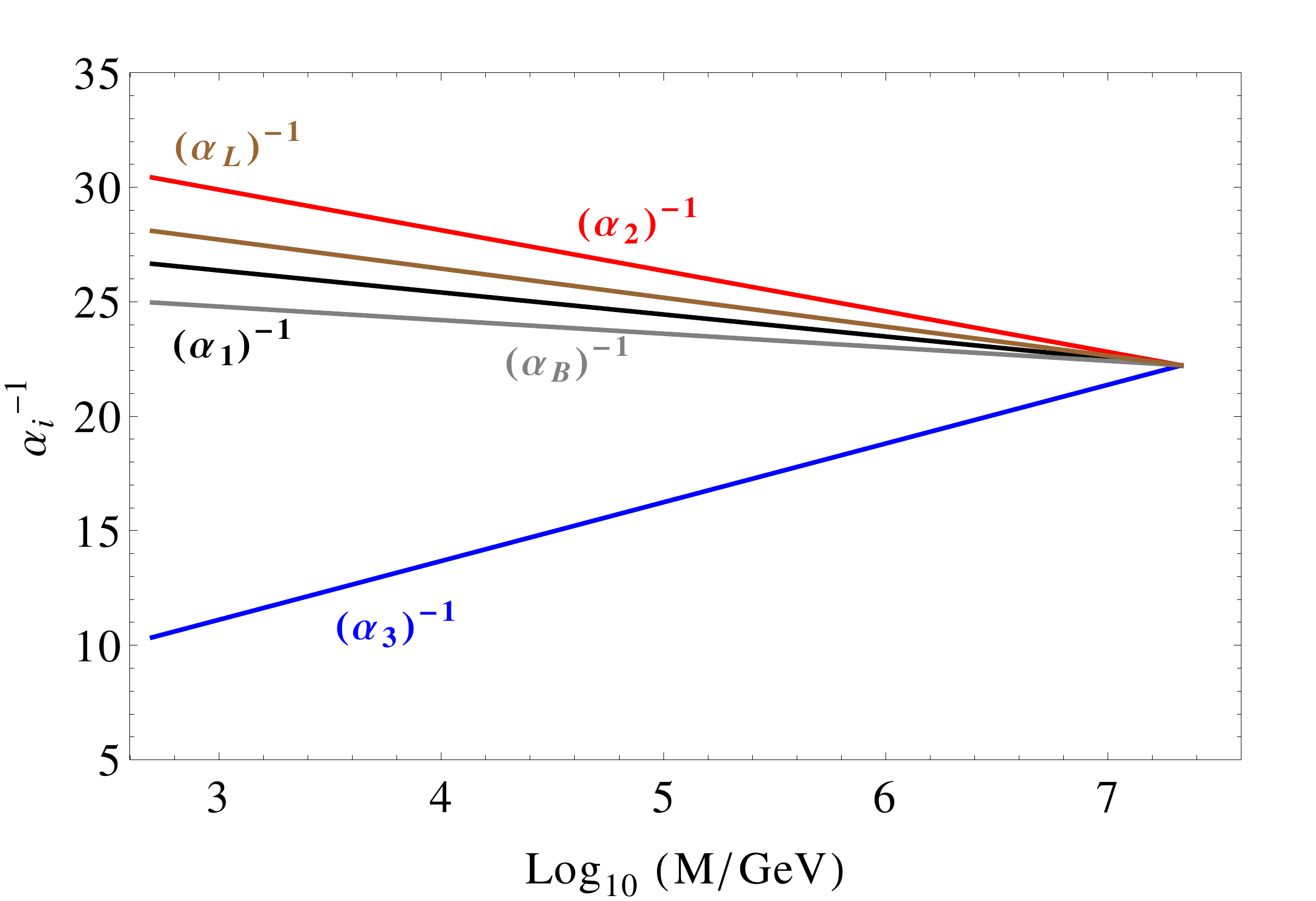}
	\caption{Evolution of the gauge couplings with lepto-baryons where the unification scale is  $M_U \sim 2 \times 10^4$ TeV. 
	Assuming for simplicity $k_B=k_L=k_1$ for the running of the $\alpha_B$ and $\alpha_L$. At the $M_Z$ scale one
	finds $\alpha^{-1}_B (M_Z) = 25.17$ and $\alpha^{-1}_L (M_Z) = 28.69$~\cite{Perez:2014kfa}.}
	\label{all-running}
\end{figure}
In Fig.~\ref{all-running} one shows the numerical results for the running of all gauge couplings. 
 As one can appreciate, these results tell us that it is possible to have consistent unification of 
 the gauge interactions using a bottom-up approach. Of course, the embedding of this model in 
 a grand unified theory is very important.
 
 Theories with local baryon number and left-right symmetry have been investigated in Ref.~\cite{Duerr:2013opa}.
 These theories are based on the gauge group $SU(3) \otimes SU(2)_L \otimes SU(2)_R \otimes U(1)_L \otimes U(1)_B$ 
 and define a possible new path towards the unification of gauge interactions. 

\section{Baryon and Lepton Number Violation in Supersymmetry}
\begin{figure}[t] 
\begin{center}
	\includegraphics[scale=0.75]{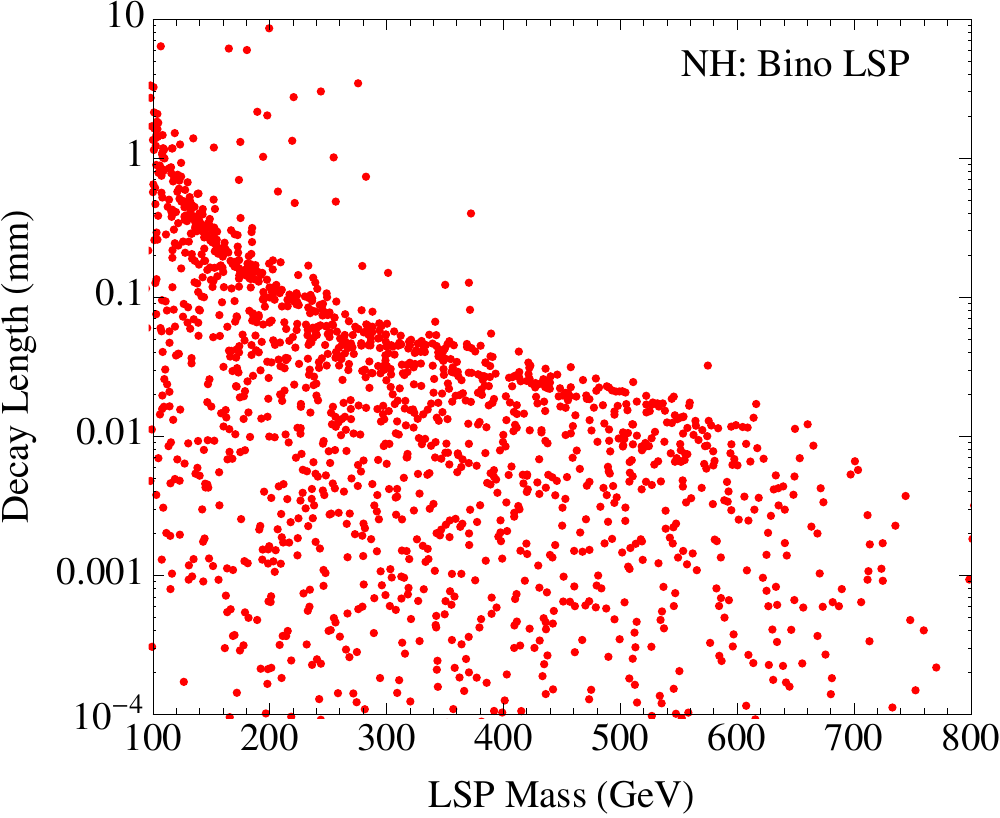}
	\put(-40,-4){(a)}
	\includegraphics[scale=0.75]{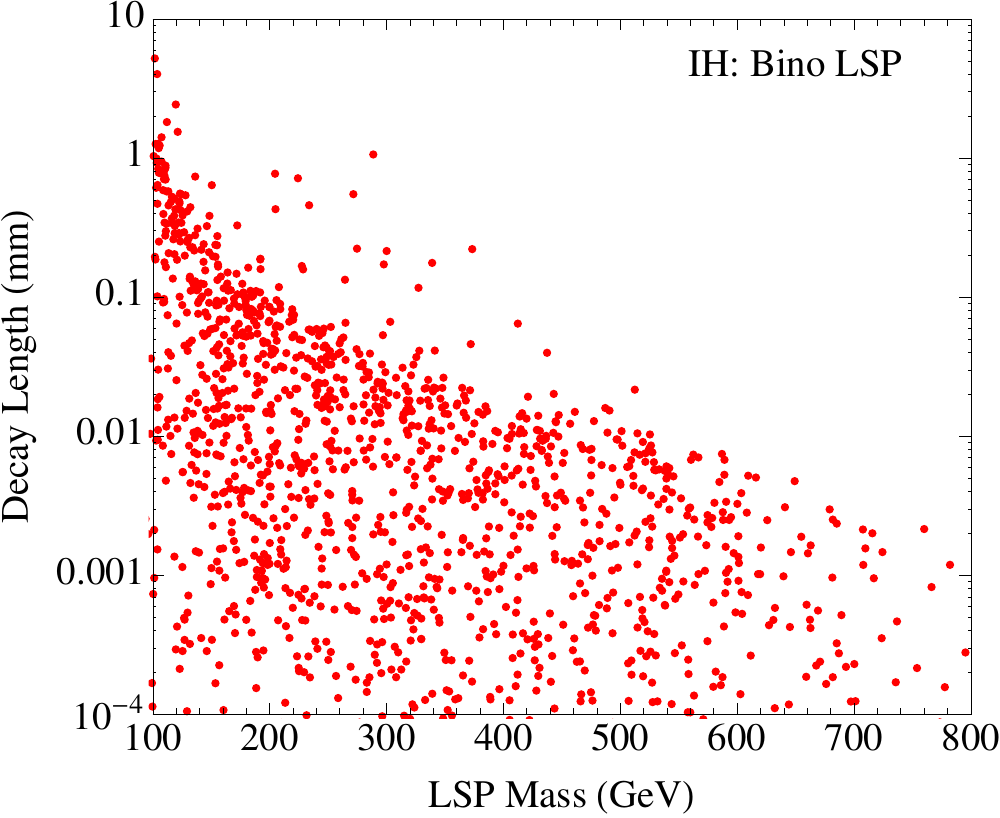}
	\put(-40,-4){(b)}
\caption{
Decay length in millimeters versus LSP mass for a dominantly bino LSP in (a) for a NH and in (b) for an IH. See Ref.~\cite{FileviezPerez:2012mj} for details.}
\label{LSP.DL.Bino}
\end{center}
\end{figure}	
\begin{figure}[t]
\begin{center}
	\includegraphics[scale=0.75]{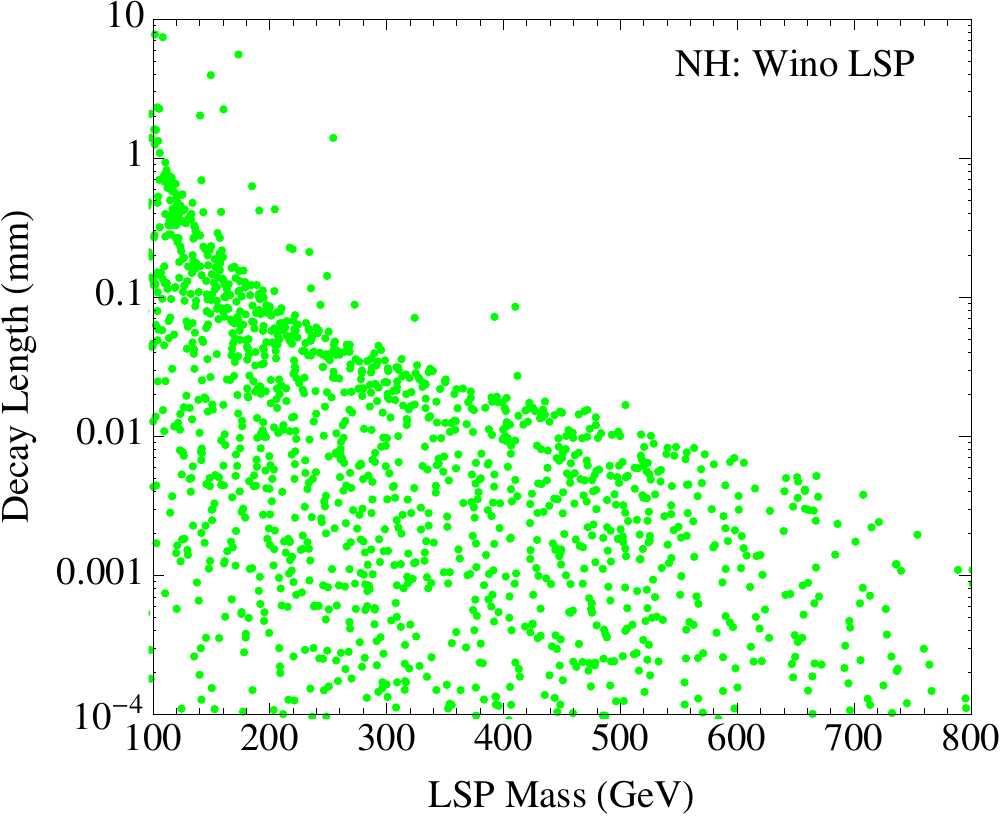}
	\put(-40,-4){(a)}
	\includegraphics[scale=0.75]{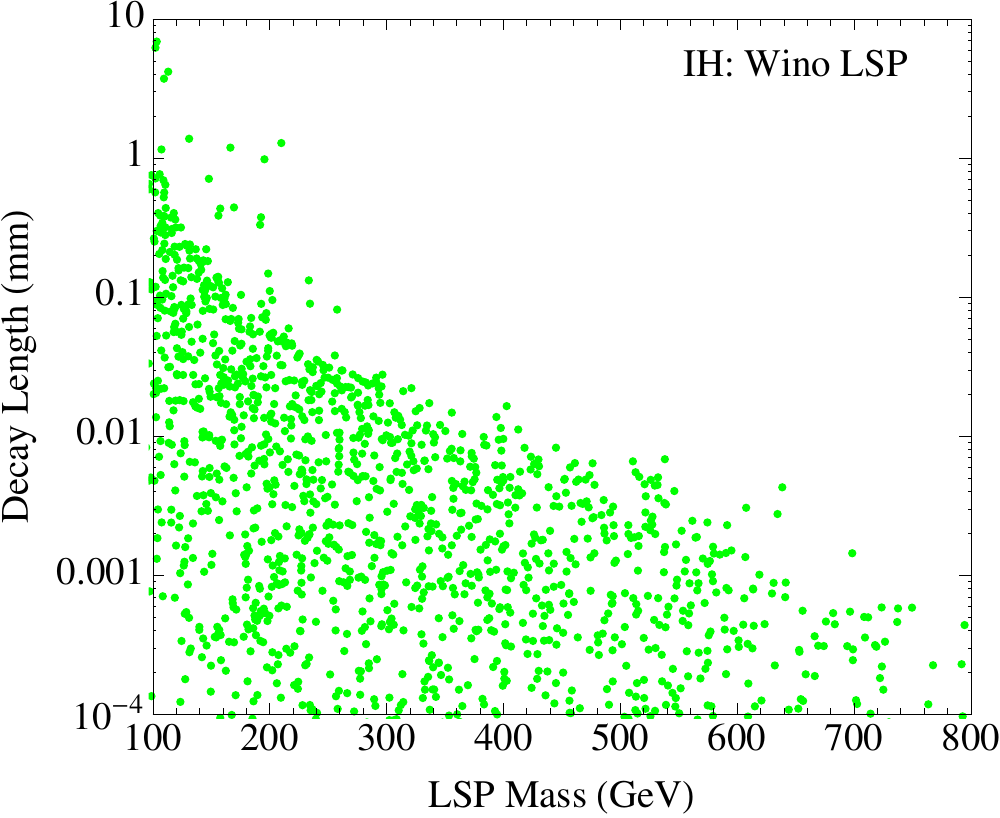}
	\put(-40,-4){(b)}
\caption{
Decay length in millimeters versus LSP mass for a dominantly wino LSP in (a) for a NH and in (b) for an IH. See Ref.~\cite{FileviezPerez:2012mj} for details.}
\label{LSP.DL.Wino}
\end{center}
\end{figure}
\begin{figure}[t]
\begin{center}
	\includegraphics[scale=0.75]{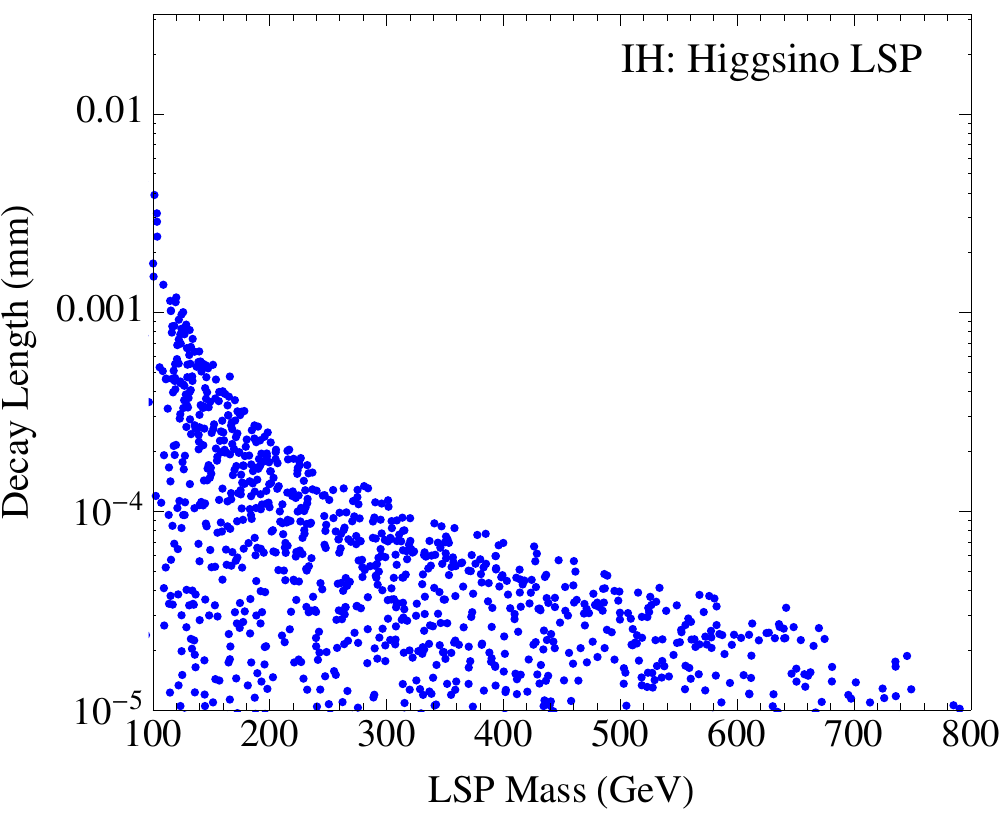}
	\put(-40,-4){(a)}
	\includegraphics[scale=0.775]{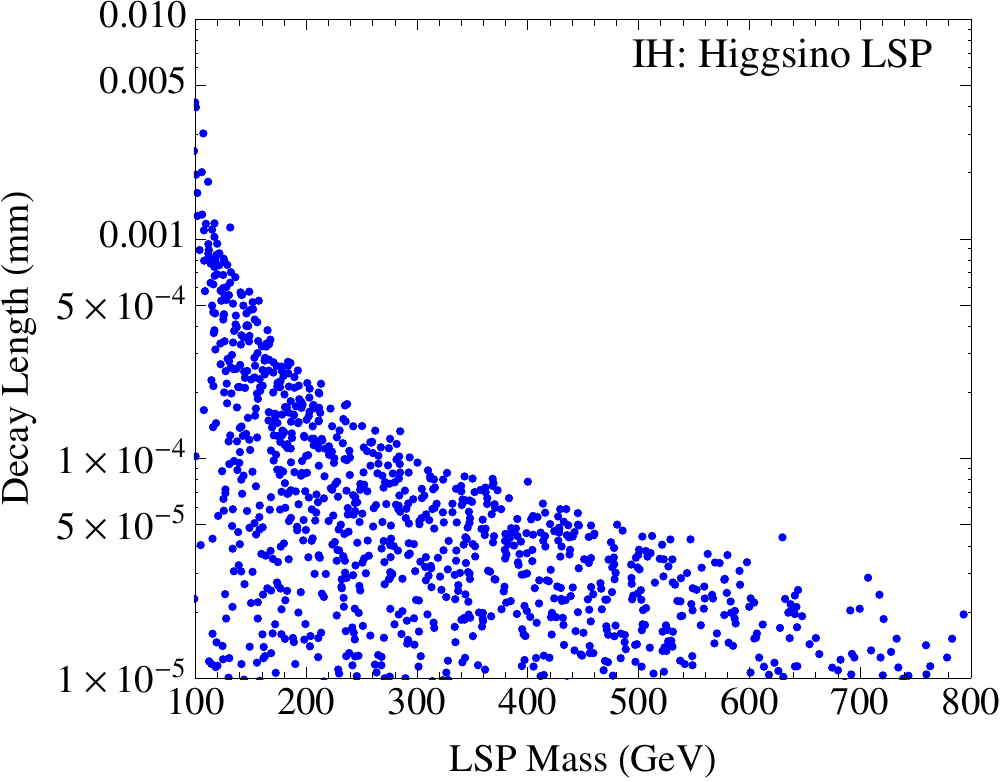}
	\put(-40,-4){(b)}
\caption
{Decay length in millimeters versus LSP mass for a dominantly Higgsino LSP in (a) for a NH and in (b) for an IH. See Ref.~\cite{FileviezPerez:2012mj} for details.}
\label{LSP.DL.Higgsino}
\end{center}
\end{figure}
\begin{figure}[t] 
\begin{center}
	\includegraphics[scale=0.75]{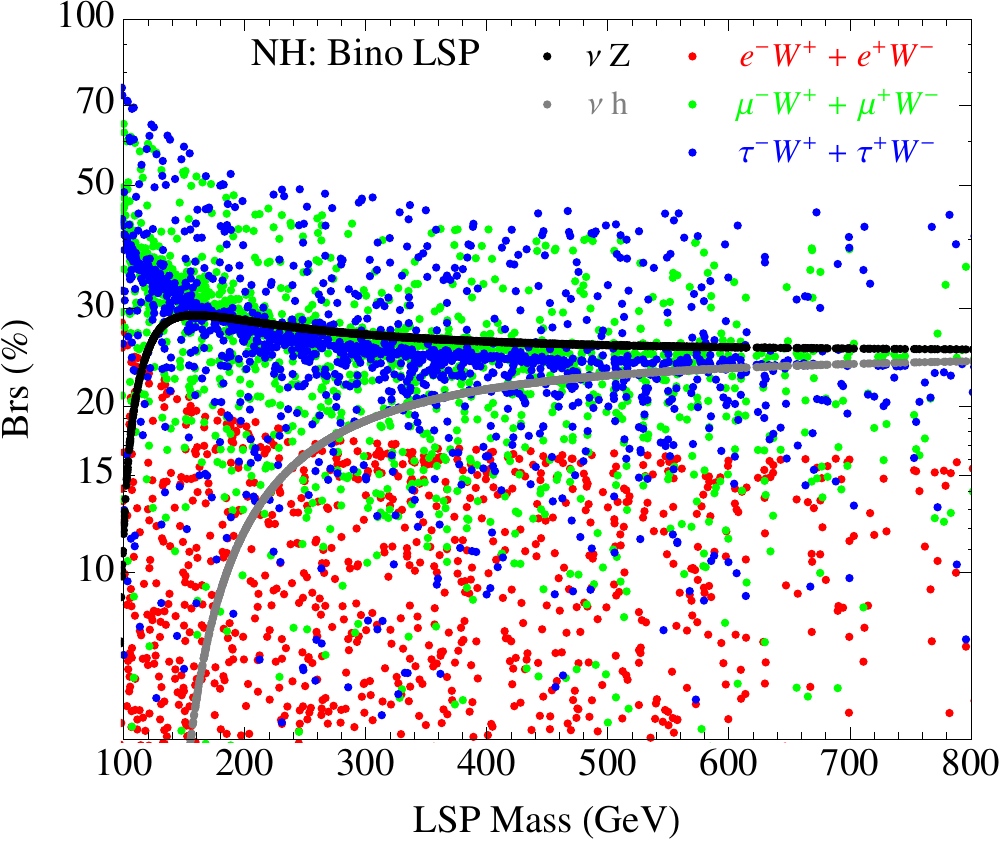}
	\put(-40,-4){(a)}
	\includegraphics[scale=0.75]{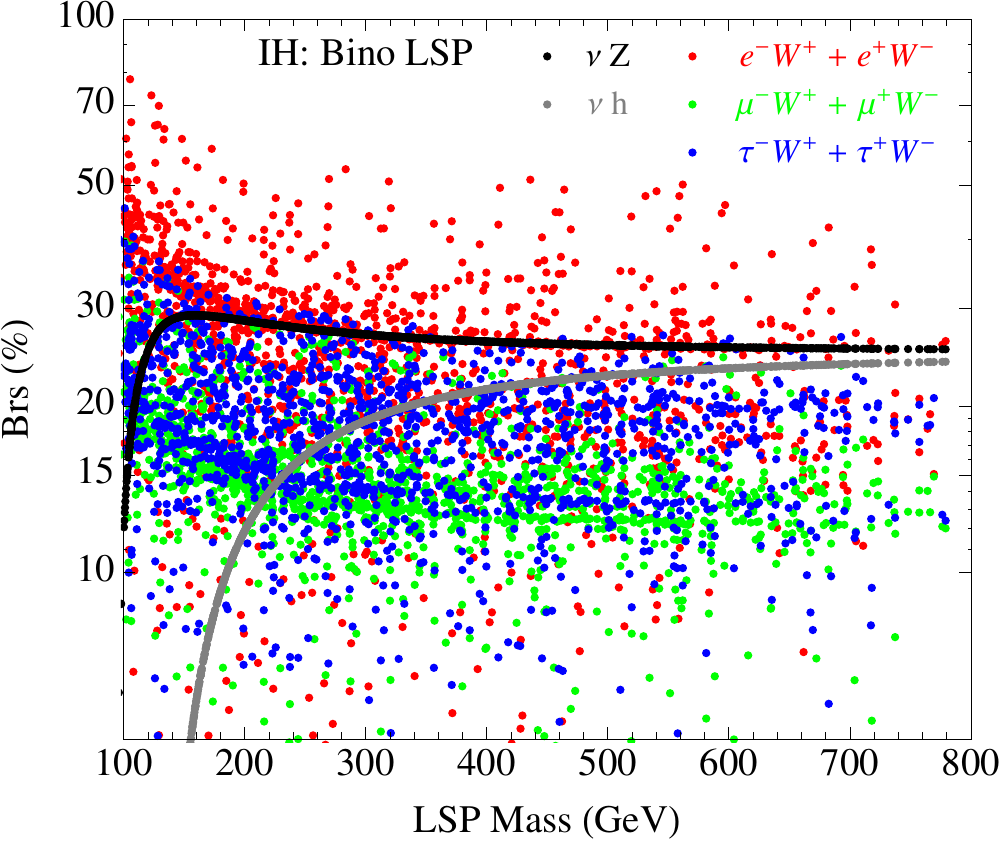}
	\put(-40,-4){(b)}
\caption{LSP branching ratios versus LSP mass for a dominantly bino LSP in (a) for a NH and in (b) for an IH.  
See Ref.~\cite{FileviezPerez:2012mj} for details.}
\label{LSP.BR.Bino}
\end{center}
\end{figure}
\begin{figure}[t]
\begin{center}
	\includegraphics[scale=0.75]{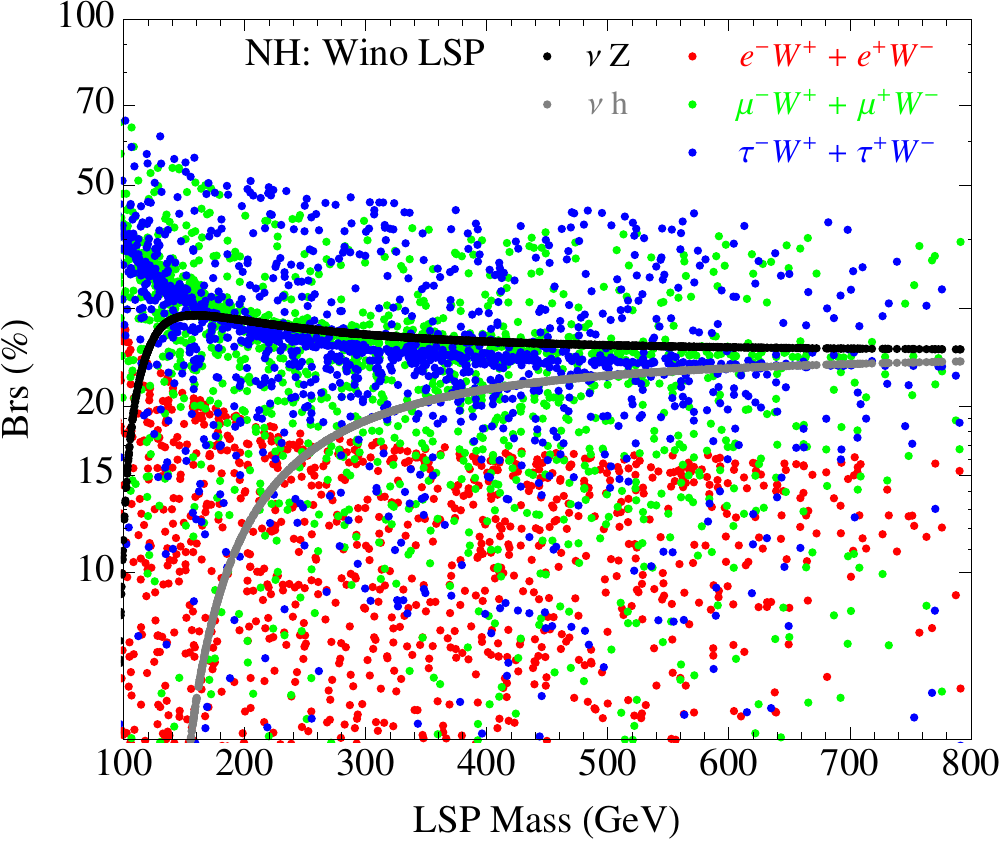}
	\put(-40,-4){(a)}
	\includegraphics[scale=0.75]{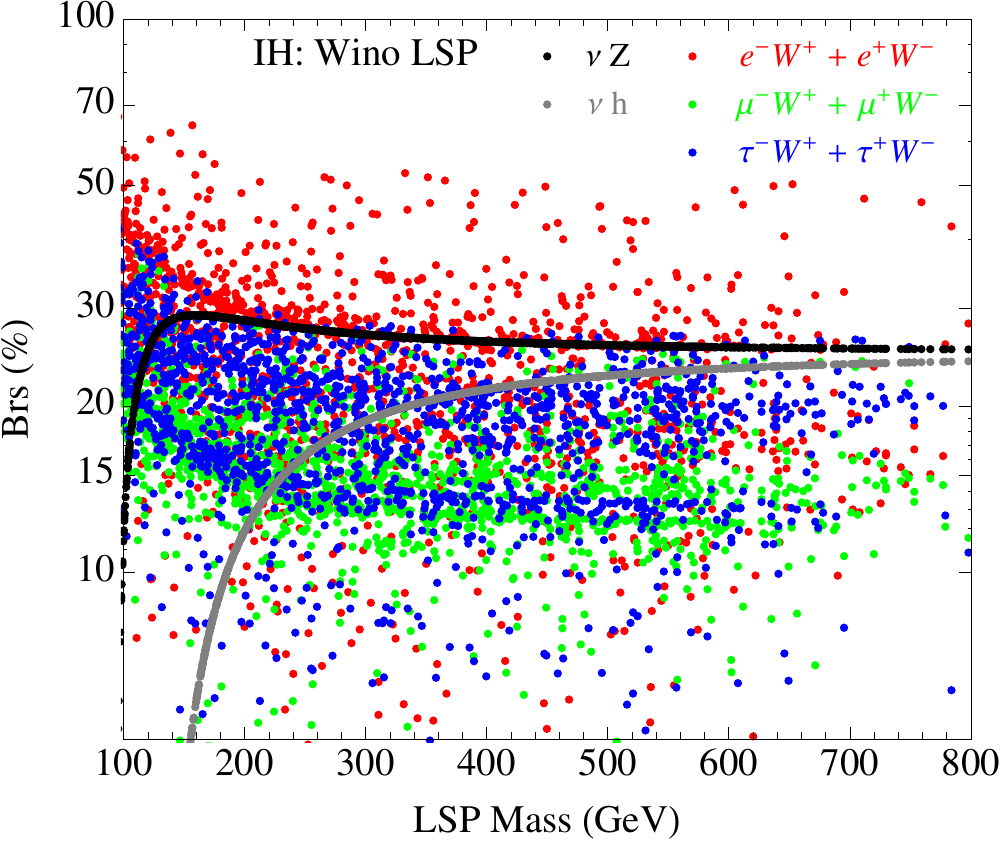}
	\put(-40,-4){(b)}
\caption{LSP branching ratios versus LSP mass for a dominantly wino  LSP in (a) for a NH and in (b) for an IH. 
See Ref.~\cite{FileviezPerez:2012mj} for details.}
\label{LSP.BR.Wino}
\end{center}
\end{figure}
\begin{figure}[t]
\begin{center}
	\includegraphics[scale=0.75]{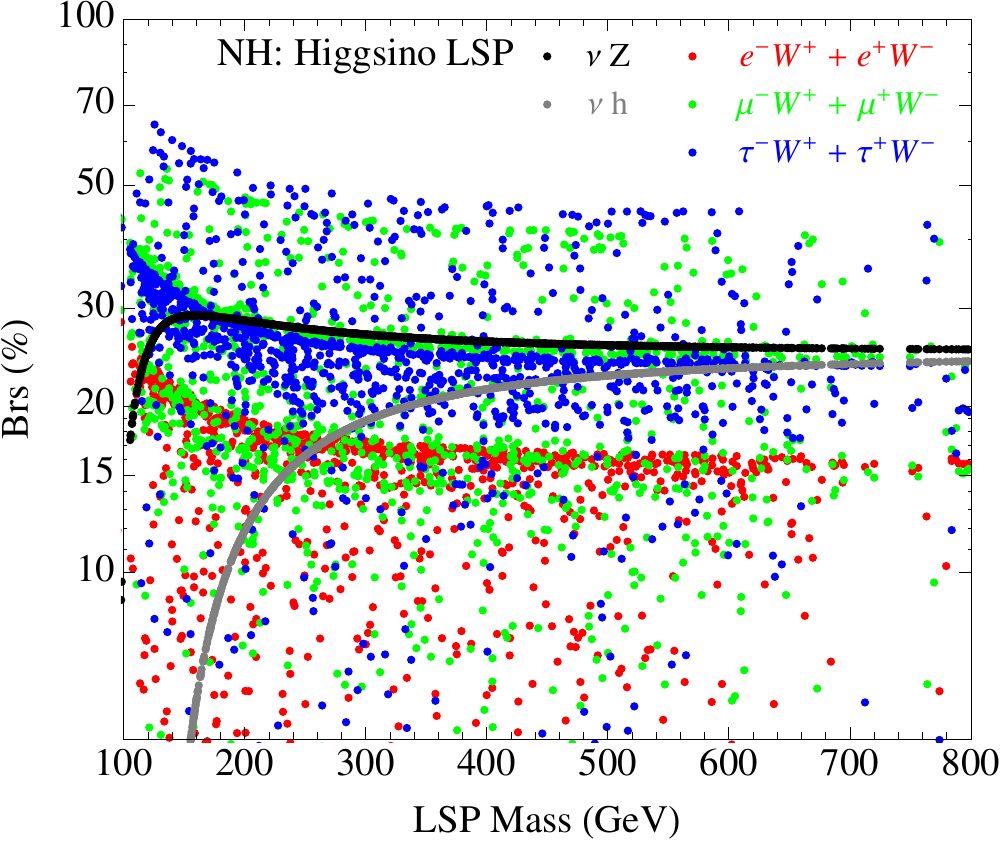}
	\put(-40,-4){(a)}
	\includegraphics[scale=0.75]{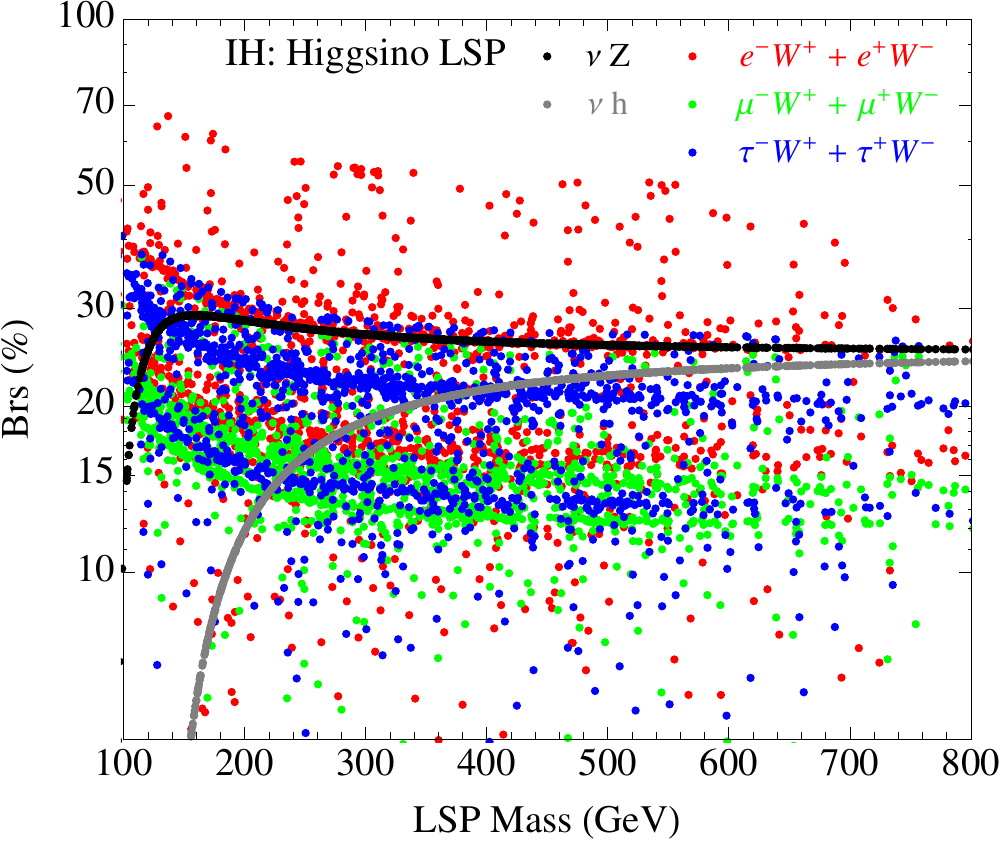}
	\put(-40,-4){(b)}
\caption
{LSP branching ratios versus LSP mass for a dominantly Higgsino LSP in (a) for a NH and in (b) for an IH.  
See Ref.~\cite{FileviezPerez:2012mj} for details.}
\label{LSP.BR.Higgsino}
\end{center}
\end{figure}

In the minimal supersymmetric Standard Model~\cite{MSSM1,MSSM2,MSSM3} one has interactions violating the total lepton and baryon numbers 
at the renormalizable level. These interactions are given by
\begin{equation}
{\cal{W}}_{RpV}=\epsilon \hat{L} \hat{H}_u \ + \  \lambda \hat{L} \hat{L} \hat{e}^c \ + \   \lambda^{'} \hat{Q} \hat{L} \hat{d}^c \ + \  \lambda^{''} \hat{u}^c \hat{d}^c \hat{d}^c,
\end{equation}
breaking $B-L$ and the discrete symmetry called matter parity, $M=(-1)^{3(B-L)}$. There is a simple relation between $R$-parity and 
$M$-parity, $R=(-1)^{2S}M$~\cite{R1,R2} where $S$ is the spin. Now, even if the conservation of $M$-parity is imposed by hand there are also 
dimension five operators which give rise to proton decay. These operators 
 are 
\begin{equation}
{\cal{W}}_{RpC}^5=  \frac{\lambda_L}{\Lambda} \hat{Q} \hat{Q} \hat{Q} \hat{L} \ + \  \frac{\lambda_R}{\Lambda} \hat{u}^c \hat{d}^c \hat{u}^c \hat{e}^c 
\ + \  \frac{\lambda_{\nu^c}}{\Lambda} \hat{u}^c \hat{d}^c \hat{d}^c \hat{\nu}^c.
\end{equation}
Notice that these interactions conserve $B-L$ and in order to satisfy the experimental bounds on proton decay one needs to assume a large cutoff scale, i.e. $\Lambda > 10^{17}$ GeV 
when we assume low-energy supersymmetry and the coefficients are of order one. The simplest way to understand the origin of the $B$
and $L$ violating interactions in the MSSM is to consider a theory where $B-L$ is inside the algebra. The relation between $M$-parity 
and $B-L$ has been investigated by many groups~\cite{Hayashi,Martin,AM,Martin2,Masiero,Goran1,Goran2,Feldman}. It is important to say that
most of the people in the SUSY community assume the conservation of matter parity because the lightest neutralino can describe the cold dark matter of the
Universe. Recently, this issue was investigated in Refs.~\cite{FileviezPerez:2008sx,Barger:2008wn,FileviezPerez:2009gr,Everett:2009vy,FileviezPerez:2010ek} 
and it has been shown that the minimal theory based on $B-L$ predicts that $R-$parity must be spontaneously broken. These results are interesting because 
the minimal theory for spontaneous $R-$parity violation could tells us what we should expect at the Large Hadron Collider if low-energy supersymmetry is realized in nature. 
In this section we discuss in great detail the minimal theory for spontaneous $R-$parity violation and the testability at the Large Hadron Collider. 

\subsection{The Minimal Theory for Spontaneous $R$-parity Violation}
The simplest gauge theory for spontaneous R-parity breaking was proposed in Ref.~\cite{Barger:2008wn}. In this context one can understand dynamically the origin of the 
R-parity violating terms in the MSSM. Here we discuss the structure of the theory and the full spectrum.
This theory is based on the gauge group $$SU(3) \otimes SU(2) \otimes U(1)_Y \otimes U(1)_{B-L},$$ 
and the different matter chiral superfields are given by  
\begin{equation}
\hat{Q} = \left(
\begin{array} {c}
\hat{u} \\ \hat{d}
\end{array}
\right) \ \sim \ (2,1/6,1/3),
\ \ 
\hat{L} = \left(
\begin{array} {c}
 \hat{\nu} \\ \hat{e}
\end{array}
\right) \ \sim \ (2,-1/2,-1),
\end{equation}
\begin{equation}
\hat{u}^c \ \sim \ (1,-2/3,-1/3),
\ \ 
\hat{d}^c \ \sim \ (1,1/3,-1/3),
\ \ 
\hat{e}^c \ \sim \ (1,1,1).
\end{equation}
In order to cancel the $B-L$ anomalies one introduces three chiral superfields for the right-handed neutrinos,
\begin{equation}
\hat{\nu}^c \ \sim \ (1,0,1).
\end{equation}
\\
The Higgs sector is composed of two Higgs chiral superfields as in the MSSM 
\begin{equation}
\hat{H}_u = \left(
\begin{array} {c}
	\hat{H}_u^+
\\
	\hat{H}_u^0
\end{array}
\right) \ \sim \ (2, 1/2, 0),
\ \ \
\hat{H}_d = \left(
\begin{array} {c}
	\hat{H}_d^0
\\
	\hat{H}_d^-
\end{array}
\right) \ \sim \ (2, -1/2, 0).
\end{equation}
\\
In this content the superpotential reads as
\begin{equation}
{\cal W}_{BL}={\cal W}_{MSSM} \ + \ Y_\nu \ \hat{L} \ \hat{H}_u \ \hat{\nu}^c,
\end{equation}
where
\begin{eqnarray}
	{\cal W}_{MSSM} &=& Y_u \ \hat{Q} \ \hat{H}_u \ \hat{u}^c
\ + \ Y_d \ \hat{Q} \ \hat{H}_d \ \hat{d}^c
\ + \ Y_e \ \hat{L} \ \hat{H}_d \  \hat{e}^c
\ + \ \mu \ \hat{H}_u \  \hat{H}_d. 
\end{eqnarray}
In addition to the superpotential, the model is also specified by the soft terms
\begin{eqnarray}
V_{soft} & = & m_{\tilde \nu^c}^2 \abs{\tilde{\nu}^c}^2 \ + \ m_{\tilde L}^2 \ \abs{\tilde L}^2 \ + \ m_{\tilde e^c}^2 \ \abs{\tilde e^c}^2
\ + \ m_{H_u}^2 \abs{H_u}^2 + m_{H_d}^2 \abs{H_d}^2 \nonumber \\
& + &  \left( \frac{1}{2} M_{BL} \tilde{B^{'}} \tilde{B^{'}}	+  A_\nu \ \tilde{L} \ H_u \ \tilde{\nu}^c  \  + \  B\mu \ H_u H_d
		\ + \  \mathrm{h.c.} \right) \ + \ V_{soft}^{MSSM},
\label{soft}
\end{eqnarray}
where the terms not shown here correspond to terms in the soft MSSM potential.
Since we have a new gauge symmetry in the theory we need to modify the kinetic terms for all MSSM matter superfields, 
and include the kinetic term for right-handed neutrino superfields
\begin{equation}
{\cal{L}}_{Kin} (\nu^c) = \int d^2 \theta d^2 \bar{\theta}  \   (\hat{\nu}^c)^\dagger  e^{g_{BL} \hat{V}_{BL}}  \hat{\nu}^c.
\end{equation}   
Here $\hat{V}_{BL}$ is the $B-L$ vector superfield. Using these interactions we can study the full spectrum of the theory.
As in the MSSM, electroweak symmetry is broken by the vevs of $H_u^0$ and $H_d^0$, while $U(1)_{B-L}$ 
is broken due to the vev of right-handed sneutrinos. Notice that this is the only field which can break local $B-L$ and 
give mass to the new neutral gauge boson in the theory. Therefore, the theory predicts spontaneous R-parity violation.
It is important to mention that the $B-L$ and R-parity breaking scales are determined by the soft supersymmetric breaking scale, and one must 
expect lepton number violation at the LHC. 

The neutral fields are defined as
\begin{eqnarray}
H_u^0 &=& \frac{1}{\sqrt{2}} \left(  v_u \ + \ h_u \right) \ + \ \frac{i}{\sqrt{2}} A_u, \\
H_d^0 &=& \frac{1}{\sqrt{2}} \left(  v_d \ + \ h_d \right) \ + \ \frac{i}{\sqrt{2}} A_d, \\
\tilde{\nu}^i &=& \frac{1}{\sqrt{2}} \left(  v_{L}^{i} \ + \  h_L^{i}   \right) \ + \ \frac{i}{\sqrt{2}} A_L^{i} , \\
\tilde{\nu}^c_i &=& \frac{1}{\sqrt{2}} \left(  v_{R}^{i} \ + \  h_R^{i}   \right) \ + \ \frac{i} {\sqrt{2}} A_R^{i}, 
\end{eqnarray}
and the relevant scalar potential reads as
\begin{eqnarray}
V &=& V_F \ + \ V_D \ + \ V_{soft}, 
\end{eqnarray}
with
\begin{eqnarray}
V_F &=&  |\mu|^2 |H_u^0|^2 \ + \  | - \mu H_d^0 + \tilde{\nu}_i Y_\nu^{ij} \tilde{\nu}^c_j |^2 \ + \ \sum_{i}  | Y_\nu^{ij} \tilde{\nu}^c_j|^2 |H_u^0|^2  \ + \ \sum_{j}  |\tilde{\nu}_i   Y_\nu^{ij}|^2 |H_u^0|^2, \\
V_D &=& \frac{(g_1^2 + g_2^2)}{8} \left(  |H_u^0|^2 - |H_d^0|^2 - \sum_{i} |\tilde{\nu}_i|^2 \right)^2 \ + \ \frac{g_{BL}^2}{8} \left( \sum_{i} ( |\tilde{\nu}^c_i|^2 - |\tilde{\nu}_i|^2 ) \right)^2, \\
V_{soft} &=& (\tilde{\nu}^c_i)^\dagger m_{\tilde{\nu}^c_{ij}}^2 \tilde{\nu}^c_j \ + \  \tilde{\nu}_i^\dagger m_{\tilde{L}_{ij}}^2 \tilde{\nu}_j \ + \  m_{H_u}^2 |H_u^0|^2 \ + \ m_{H_d}^2 |H_d^0|^2 
\ + \ \left( \tilde{\nu}_i  a_\nu^{ij} \tilde{\nu}_j^c H_u^0 -  B \mu H_u^0 H_d^0 \ + \  \rm{h.c.}\right). \nonumber \\
\end{eqnarray}
In order to have phenomenologically allowed solutions the $v_L^i$ have to be small, and the $v_R^i$ have 
to be much larger than $v_u, v_d$ and $v_L^i$. These vacuum expectation values must be small in order to have small 
neutrino masses generated through the R-parity violating terms.
Up to negligibly small terms the right-handed sneutrino acquire 
a vev in only one family. A possible solution is $v_R^i=(0,0,v_R)$. In this case
\begin{eqnarray}
\label{vR.sln}
(v_R)^2 &\approx& -\frac{8 (m_{\tilde{\nu}^c}^2)_{33}}{g_{BL}^2}, \\
v_L^k &\approx & \frac{v_R}{\sqrt{2}} \frac{\left( \mu v_d Y_\nu^{k3} - a_\nu^{k3} v_u \right)}{\left[ (m_{\tilde{L}}^2)_{kk} - \frac{(g_1^2 + g_2^2)}{8} (v_u^2 - v_d^2) - \frac{g_{BL}^2}{8} (v_R)^2\right]}.
\end{eqnarray}
Notice that the vacuum expectation value for the right-handed sneutrino is determined by the soft term. In this way we understand why the $B-L$ and $R$-parity violating scales are defined by the SUSY breaking scale. 
In the MSSM, the large top Yukawa coupling drives the up-type soft Higgs mass squared parameter to negative values for generic boundary conditions leading to radiative electroweak symmetry breaking~\cite{Ibanez:1982fr,Ibanez:1982fr-2}; a celebrated success of the MSSM. A valid question is then if the same success is possible in achieving a tachyonic right-handed sneutrino mass in this $B-L$ model as required by Eq.~(\ref{vR.sln}).  Unfortunately, this is not possible through a large Yukawa coupling since the Yukawa couplings of the right-handed neutrino are all dictated to be small by neutrino masses.  However, there is an alternate possibility whereby a positive mass squared parameter for the right-handed sneutrino at the high scale will run to a tachyonic value at the low scale. This is due to the presence of the so-called $S$-term (due to $D$-term contributions to the RGE) in the soft mass RGE, as discussed for this $B-L$ model in Refs.~\cite{Ambroso:2009jd, Ambroso:2009sc, Ambroso:2010pe}. For the implementation of the radiative mechanism in the non-minimal model where the right-handed neutrino masses are generated through the Higgs mechanism see Ref.~\cite{FileviezPerez:2010ek}.
%
\subsection{$R$-Parity Violating Interactions and Mass Spectrum}
After symmetry breaking lepton number is spontaneously broken in the form of bilinear R-parity violating interactions. There are no trilinear R-parity violating interactions at the renormalizable level. These bilinear interactions mix the leptons with the Higgsinos and gauginos
	$$\frac{1}{2} g_{BL} v_{R} ( \nu_3^c \tilde B^{'} ), \quad 
	\frac{1}{2} g_{2} v_L^i ( \nu_i \tilde W^0 ), \quad
	\frac{1}{\sqrt{2}} g_{2} v_L^i (e_i \tilde W^+ ),$$
	$$\frac{1}{2} g_{1} v_L ^i( \nu_i \tilde B ), \quad
	\frac{1}{\sqrt{2}} Y_\nu^{i3} v_R ( L^T_i i\sigma_2 \ \tilde{H}_u ), \quad
	\frac{1}{\sqrt{2}} Y_\nu^{ij} v_L^i ( \tilde{H}_u^0 \ \nu^c_j  ), \quad
	\frac{1}{\sqrt{2}} Y_e^i v_L^i ( \tilde{H}_d^- \ e^c_i ).$$
The first term is new and is the only term not suppressed by neutrino
masses. The fifth term corresponds to the so-called $\epsilon$ term,
and second, third and fourth terms are small but important for the decay
of neutralinos and charginos. There are also lepton number violating interactions coming from the soft terms and the $B-L$ D-term.
From $V_{soft}$ one gets
$$ A_\nu^{i3} \frac{v_R}{\sqrt{2}} \tilde{L}_i^T \ i \sigma_2 \ H_u,$$
while from the D-term one finds
$$g_{BL}^2 v_R \ \tilde{\nu}^c \left( \tilde{q}^\dagger \frac{1}{6} \tilde{q} \ - \  \tilde{l}^\dagger \frac{1}{2} \tilde{l} \right).$$
As one can expect these terms are important to understand the scalar sector of the theory.

The neutral gauge boson associated to the $B-L$ gauge group is $Z_{BL}$.
Using the covariant derivative for the right-handed sneutrinos, $D_\mu \tilde{\nu}^c = \partial_\mu \tilde{\nu}^c + \frac{i}{2} g_{BL} B_\mu^{'} \tilde{\nu}^c$, 
the mass term for $Z_{BL}$ is
\begin{equation}
	M_{Z_{BL}}=\frac{g_{BL}}{2} v_R.
\end{equation}
Now, using the experimental collider constraint~\cite{Carena:2004xs}
\begin{equation}
\label{ZBL.const}
	\frac{M_{Z_{BL}}}{g_{BL}} \geq 3 \text{ TeV},
\end{equation}
and Eq. (21) one finds the condition
\begin{equation}
	| (m_{\tilde{\nu}^c)_{33}} | > 2.12 \ g_{BL} \ \rm{TeV}.
\end{equation}
Then, if $g_{BL} =0.1$ the soft mass above has to be larger than 200 GeV.  
This condition can be easily satisfied without assuming a very heavy spectrum for the supersymmetric particles. 

As in any supersymmetric theory where $R$-parity is broken all the fermions with the same quantum numbers mix and 
form physical states which are linear combinations of the original fields. The neutralinos in this theory are a linear combination 
of the  fields, $$\left(\nu_i, \ \nu^c_j, \ \tilde B', \ \tilde B, \ \tilde W^0, \ \tilde H_d^0, \ \tilde H_u^0\right).$$ Then, the neutralino mass matrix is given by
\begin{equation}
	{\cal M}_{N} =
	\begin{pmatrix}
			0
		&
			\frac{1}{\sqrt{2}} \ Y_\nu^{ij} v_u
		&
			-\frac{1}{2} g_{BL} \ v_L^i
		&
			-\frac{1}{2} g_1 \ v_L^i
		&
			\frac{1}{2} g_2 \ v_L^i
		&
			0
		&
			\frac{1}{\sqrt{2}} \ Y_\nu^{ij} \ v_R^j
	\\
			\frac{1}{\sqrt{2}} \ Y_\nu^{ij} \ v_u
		&
			0
		&
			\frac{1}{2} g_{BL} \ v_R^j
		&
			0
		&
			0
		&
			0
		&
			\frac{1}{\sqrt{2}} \ Y_\nu^{ij} \ v_L^i
	\\
			-\frac{1}{2} g_{BL} \ v_L^i
		&
			\frac{1}{2} g_{BL} \ v_R^j
		&
			M_{BL}
		&
			0
		&
			0
		&
			0
		&
			0
	\\
			-\frac{1}{2} g_1 \ v_L^i
		&
			0
		&
			0
		&
			M_1
		&
			0
		&
			-\frac{1}{2} g_1 v_d
		&
			\frac{1}{2} g_1 v_u
	\\
			\frac{1}{2} g_2 \ v_L^i
		&
			0
		&
			0
		&
			0
		&
			M_2
		&
			\frac{1}{2} g_2 v_d
		&
			-\frac{1}{2} g_2 v_u
	\\
			0
		&
			0
		&
			0
		&
			-\frac{1}{2} g_1 v_d
		&
			\frac{1}{2} g_2 v_d
		&
			0
		&
			-\mu
	\\
			\frac{1}{\sqrt{2}} \ Y_\nu^{ij} \ v_R^j
		&
			\frac{1}{\sqrt{2}} \ Y_\nu^{ij} \ v_L^i
		&
			0
		&
			\frac{1}{2} g_1 v_u
		&
			-\frac{1}{2} g_2 v_u
		&
			-\mu
		&
			0
	\end{pmatrix}.
\label{neutralino}
\end{equation}
We have discussed above that only one right-handed sneutrinos get a vev, $v_R^i=(0,0,v_R)$.
Now, integrating out the neutralinos one can find the mass matrix for the light neutrinos. In this case 
one has three active neutrinos and two sterile neutrinos, and the mass matrix in the basis 
$(\nu_e, \nu_\mu, \nu_\tau, \nu^c_e, \nu^c_\mu)$ is given by
%
\begin{equation}
\label{Mnu}
M_\nu =
\begin{pmatrix}
	A \ v_{L}^i v_{L}^j
	+ B \ \left[Y_\nu^{i3} v_{L}^j + Y_\nu^{j3} v_{L}^i \right]
	+ C \ Y_\nu^{i3} Y_\nu^{j3} 
	&
	\frac{1}{\sqrt{2}} v_u Y_\nu^{i \beta} 
\\
	\frac{1}{\sqrt{2}} v_u Y_\nu^{\alpha j}
	&
	O_{2\times2}
\end{pmatrix},
\end{equation}
where
\begin{align}
	A & = \frac{2 \mu^2}{\tilde m^3}, \ \ \
	B = \left(\frac{v_u}{\sqrt{2} v_R}  + \frac{\sqrt{2} \mu v_d v_R}{\tilde m^3}\right),
	\ \ \
	C = \left(\frac{2 M_{BL} v_u^2}{g_{BL}^2 v_R^2} + \frac{v_d^2 v_R^2}{\tilde m^3}\right), 
\end{align}
\begin{align}	
	\tilde m^{3} & = \frac
		{
			4
			\left[
				\mu v_u v_d \left(g_1^2 M_2 + g_2^2 M_1 \right)
				- 2 M_1 M_2 \mu^2
			\right]
		}
		{g_1^2 M_2 + g_2^2 M_1}
		.
\end{align}
%
Here $\alpha$ and $\beta$ take only the values 1 and 2. From the experimental limits on active neutrino masses we obtain $(Y_\nu)_{i \alpha} \lesssim 10^{-12}$.
This can be compared to $(Y_\nu)_{i 3} \lesssim 10^{-5}$, which is less constrained because of the TeV scale
seesaw suppression. It has been pointed out 
in Refs.~\cite{Barger:2010iv, Ghosh:2010hy} (and earlier in a different context~\cite{Mohapatra}) that this theory predicts the existence of two light sterile neutrinos which 
are degenerate or lighter than the active neutrinos, a so-called $3+2$ neutrino model. 

In this theory the chargino mass matrix, in the basis $\left(e^c_j, \ \tilde W^+, \ \tilde H_u^+\right)$ and $\left(e_i, \ \tilde W^-, \ \tilde H_d^-\right)$, is given by
\begin{equation}
\label{chargino}
{\cal M}_{{\tilde \chi}^{\pm}}	=	
	\left(
	\begin{array}{cc}
		0
	&
		M_C
	\\
		M_C^T
	&	0
	\end{array}
	\right),
\end{equation}
with
\begin{equation}
	M_C =
	\begin{pmatrix}
			-\frac{1}{\sqrt{2}} Y_e^{ij} v_d
		&
			0
		&
			\frac{1}{\sqrt{2}} Y_e^{ij}  v_L^j
	\\
			\frac{1}{\sqrt{2}} \ g_2  v_L^i
		&
			M_2
		&
			\frac{1}{\sqrt{2}} g_2 v_d
	\\
			-\frac{1}{\sqrt{2}} Y_\nu^{ij} v_R^j
		&
			\frac{1}{\sqrt{2}} g_2 v_u
		&
			\mu
	\end{pmatrix}.
\end{equation}
\\
In the sfermion sector, the mass matrices $\mathcal{M}_{\tilde u}^2$, and $\mathcal{M}_{\tilde d}^2$ for squarks, 
and $\mathcal{M}_{\tilde e}^2$ for charged sleptons, in the basis $\left(\tilde f, \  {\tilde f}^{c*} \right)$, are given by
\begin{eqnarray}
\mathcal{M}_{\tilde u}^2&=&\left(\begin{array}{cc}
		m_{\tilde Q}^2
		\ + \ m_{u}^2
		\ + \  \left(\frac{1}{2} \ - \  \frac{2}{3} s_W^2 \right) \ M_Z^2 \ c_{2\beta} 
		+  \frac{1}{3} D_{BL}
		&
		\frac{1}{\sqrt 2} \left(a_u \ v_u - Y_u \ \mu \ v_d\right)
	\\
		\frac{1}{\sqrt 2} \left(a_u \ v_u - Y_u \ \mu \ v_d\right)
		&
		m_{\tilde u^c}^2
		\ + \ m_{u}^2
		\ + \ \frac{2}{3} M_Z^2 \ c_{2\beta}\ s_W^2 
		 -  \frac{1}{3} D_{BL}
	\end{array}\right),
	\nonumber \\
\end{eqnarray}	
\begin{eqnarray}	
\mathcal{M}_{\tilde d}^2 &=& \left(\begin{array}{cc}
		m_{\tilde Q}^2
		\ + \ m_{d}^2
		\ - \  \left(\frac{1}{2}  \ - \ \frac{1}{3} \  s^2_W \right) M_Z^2 \ c_{2 \beta}
		 +  \frac{1}{3} D_{BL}
		&
		\frac{1}{\sqrt 2} \left(Y_d \ \mu \ v_u - a_d \ v_d\right)
	\\
		\frac{1}{\sqrt 2} \left(Y_d \ \mu \ v_u - a_d \ v_d\right)
		&
		m_{\tilde d^c}^2
		\ + \ m_{d}^2
		\ - \  \frac{1}{3} \ M_Z^2  \ c_{2\beta} \ s^2_W
		-  \frac{1}{3} D_{BL}
	\end{array}\right),
	\nonumber \\
	\\
\mathcal{M}_{\tilde e}^2 &=& \left(\begin{array}{cc}
\label{Selectron.Mass}
		m_{\tilde L}^2
		\ + \ m_{e}^2
		\ - \ \left( \frac{1}{2} \ - s_W^2 \right) M_Z^2 \ c_{2\beta} 
		-  D_{BL}
		&
		\frac{1}{\sqrt 2} \left(Y_e \ \mu \ v_u - a_e \ v_d\right)
	\\
		\frac{1}{\sqrt 2} \left(Y_e \ \mu \ v_u - a_e \ v_d\right)
		&
		m_{\tilde e^c}^2
		\ + \ m_{e}^2
		\ - \  M_Z^2 \ c_{2\beta}\ s_W^2
		+  D_{BL}
	\end{array}\right), \nonumber \\ 
\end{eqnarray}
where $c_{2\beta} = \cos 2\beta$, $s_W = \sin\theta_W$ and
\begin{equation}
D_{BL} = \frac{1}{8} \ g_{BL}^2  v^2_R= \frac{1}{2} M_{Z_{BL}}^2.
\end{equation}
$m_u, \ m_d$ and $m_e$ are the respective fermion masses and $a_u, \ a_d$ and $a_e$
are the trilinear $a$-terms corresponding to the Yukawa couplings $Y_u, \ Y_d$ and $Y_e$. 
Regardless, the physical states are related to the gauge states by
\begin{align}
\label{eq_squarkmixing}
	\begin{pmatrix}
		\tilde f_1
		\\
		\tilde f_2
	\end{pmatrix}
	& =
	\begin{pmatrix}
		\cos \theta_{\tilde f}
		&
		\sin \theta_{\tilde f}
		\\
		- \sin \theta_{\tilde f}
		&
		\cos \theta_{\tilde f}
	\end{pmatrix}
	\begin{pmatrix}
		\tilde f
		\\
		\tilde f^{c*}
	\end{pmatrix}.
\end{align}
The masses in the sneutrino sector are given by
\begin{eqnarray}
M_{\tilde{\nu}_i}^2 &=& m_{\tilde{L}_i}^2 \ + \ \frac{1}{2} M_{Z}^2 \cos 2 \beta - \frac{1}{2} M_{Z_{BL}}^2, \\
M_{\tilde{\nu}_3^c}^2 &=& M_{Z_{BL}}^2, \\
M_{\tilde{\nu}_\alpha^c}^2 &=& m_{\tilde{\nu}^c_\alpha}^2 \ + \ D_{BL},
\end{eqnarray}
and $\alpha=1..2$. For simplicity we listed the mass matrices in 
the limit $v_L^i, a_\nu, Y_\nu \to 0$. It is important to mention that all sfermion masses are modified due to the existence of the $B-L$ D-term. 
See Ref.~\cite{FileviezPerez:2012mj} for a detailed study of the spectrum in this model.

At first glance, finding a dark matter in $R$-parity violating theories seems hopeless. But while the traditional neutralino LSP case is no longer valid, the situation is not lost. As first discussed in~\cite{Takayama:2000uz,Takayama:2000uz-2}, such models can have an unstable LSP gravitino, with a lifetime longer than the age of the universe.  The strong suppression on its lifetime is due to both the Planck mass suppression associated with its interaction strength and bilinear $R$-parity violation which is small due to neutrino masses and must facilitate the decay of the LSP. In the mass insertion approximation, this can be understood as the gravitino going into a photon and neutralino which then has some small mixing with the neutrino due to $R$-parity violation ($m_{\chi \, \nu}$), thereby allowing $\tilde G \to \gamma \nu$. Adopting approximations made in~\cite{Takayama:2000uz}, the lifetime for the gravitino decaying into a photon and neutrino (in years) is about
\begin{equation}
	\tau(\tilde G \to \gamma \nu) \sim 2 \times 10^{10}
	\left(\frac{m_{3/2}}{100 \text{ GeV}} \right)^{-3}
	\left(\frac{m_{\chi \nu}/m_\chi}{10^{-6}} \right)^{-2}
	\text{ years},
\end{equation}
which for appropriate values of the gravitino mass leads to a long enough life time. Unlike in $R$-parity conserving models with a gravitino LSP, there are no issues with big bang nucleosynthesis from slow NLSP decay since the NLSP decays more promptly through $R$-parity violating interactions. Several interesting studies have been conducted on the signatures and constraints of unstable gravitino dark matter, see for example~\cite{Takayama:2000uz,Buchmuller:2007ui,Buchmuller:2007ui-2}.

\subsection{Lepton Number Violation and Decays}
\begin{figure}[t] 
\begin{center}
	\includegraphics[scale=1.25]{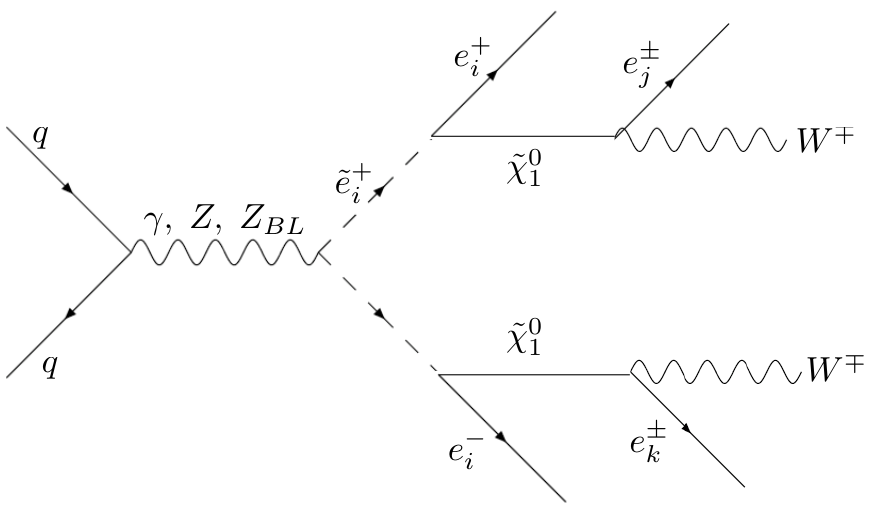}
	\put(-67,20){\Large $\tilde e^-_i$}
\caption
{
Topology of the signals with multi-leptons~\cite{FileviezPerez:2012mj}.
}
\label{Signal}
\end{center}
\end{figure}
\begin{figure}[t] 
\begin{center}
	\includegraphics[scale=0.8]{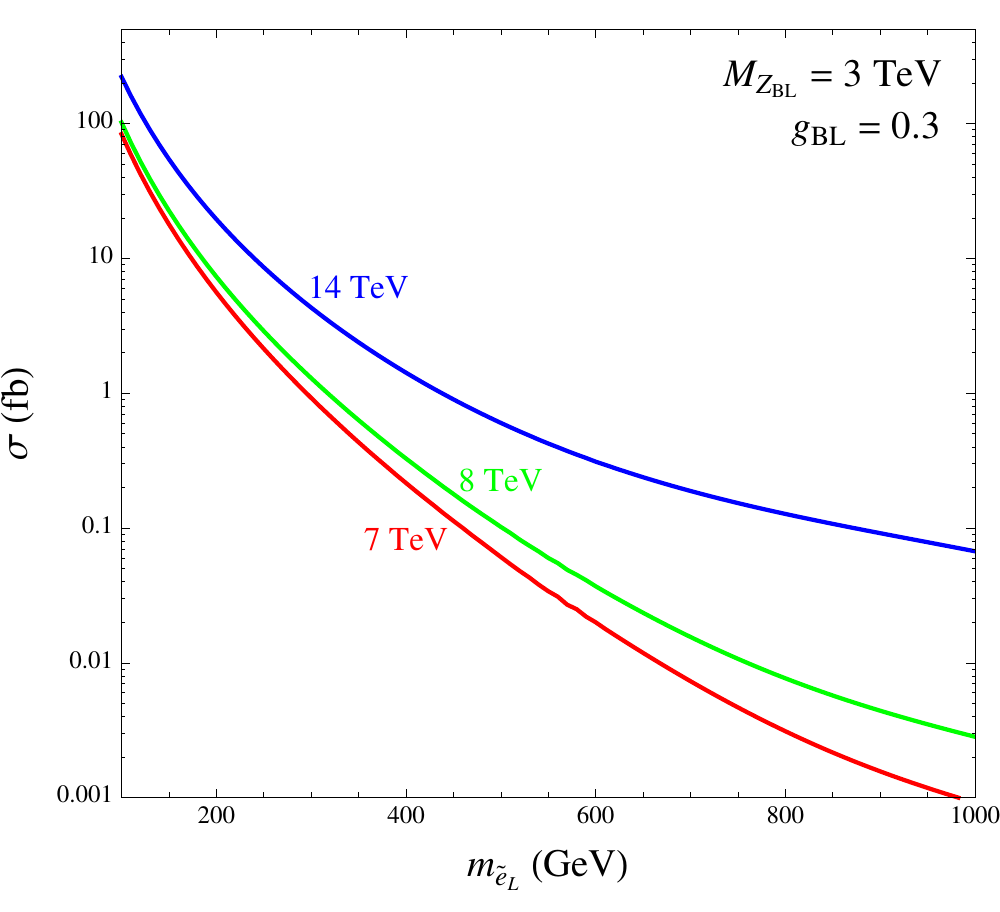}
	\caption{Cross sections for $p\, p \to \gamma, Z,Z_{BL} \to \tilde{e}^{\pm}_i \tilde{e}^{\mp}_i$ with LHC center of mass energies of 7 TeV (red), 8 TeV (green) and 14 TeV (blue), assuming $M_{Z_{BL}} = 3$ TeV and $g_{BL}=0.3$ versus the left-handed slepton mass. See Ref.~\cite{Perez:2013kla} for details.}
\label{cs}
\end{center}
\end{figure}
\begin{figure}[t]
\begin{center}
	\includegraphics[scale=0.7]{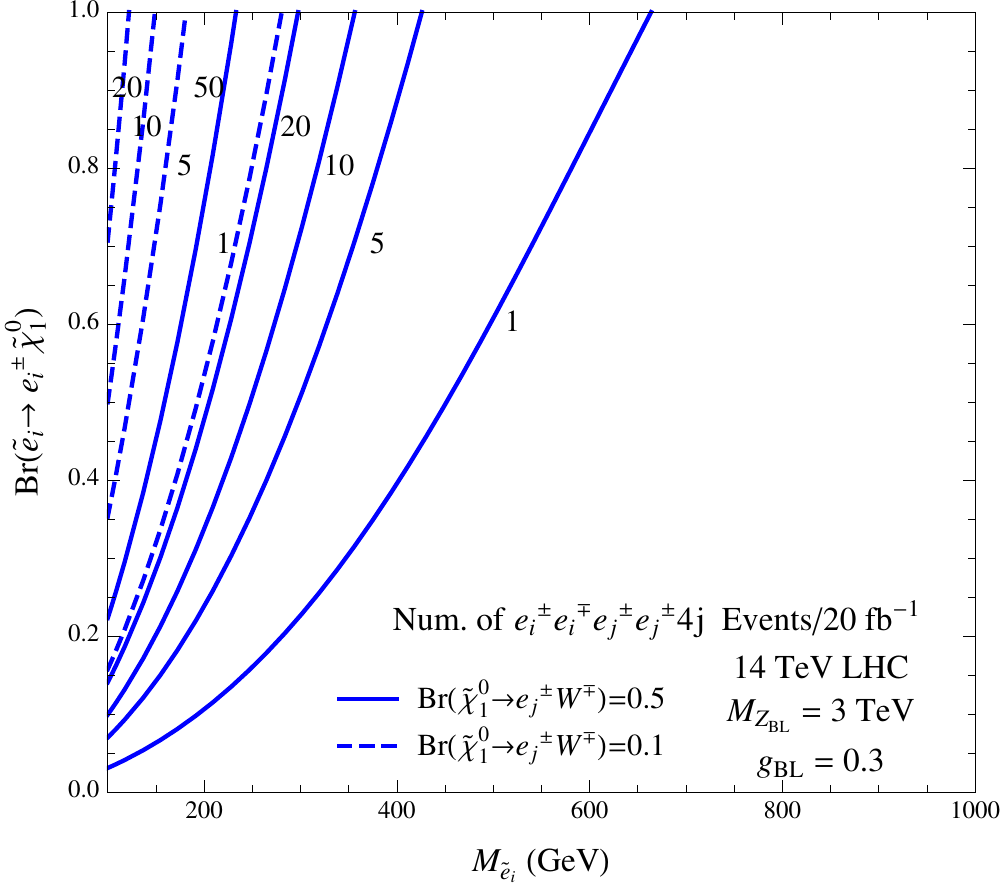}
	\includegraphics[scale=0.7]{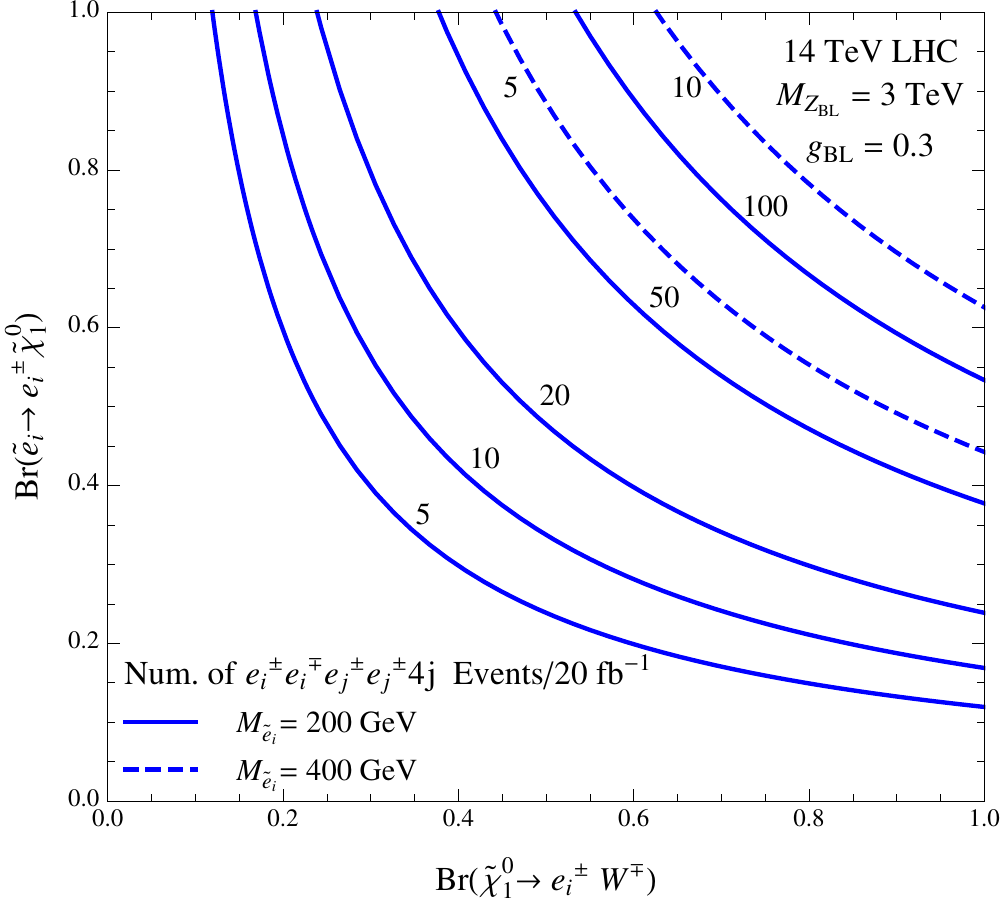}
	\caption{Number of events in the ${\rm{Br}} (\tilde{e}^{\pm}_i \to e^{\pm}_i \tilde{\chi}_1^0)$ vs $M_{\tilde{e}_i}$ plane in the left panel assuming 
	$\sqrt{s}=14$ TeV, ${\rm{Br}} (\tilde{\chi}^0_1 \to e^{\pm}_k W^{\mp})=0.1 (0.5)$, when $M_{Z_{BL}}=3$ TeV and ${\cal{L}}=20 \ {\rm{fb}}^{-1}$. 
	In the right panel the number of events are plotted in the plane ${\rm{Br}} (\tilde{e}^{\pm}_i \to e^{\pm}_i \tilde{\chi}_1^0)$-${\rm{Br}} (\tilde{\chi}^0_1 \to e^{\pm}_k W^{\mp})$~\cite{Perez:2013kla}.}
\label{14TeV}
\end{center}
\end{figure}

At this point, the relevant pieces of this model have been laid out and the question of interesting signals can now be tackled. Since lepton number is violated, same-sign dileptons and multijets are possible final states. Such signatures are interesting since they have no SM background. However, the final states depend critically on the nature of the LSP and since R-parity is violated the possibilities are more numerous than normal, \textit{i.e.} colored and charged fields. In this section we will discuss only the decays of the lightest neutralino through the $R$-parity violating couplings.
\\
{\textit{Neutralino Decays}:}
The leading decay channels for the lightest neutralino, $\tilde{\chi}_1^0$, include
\begin{align}
\label{chi.decays}
	\tilde{\chi}_1^0 \to e^{\pm}_i W^{\mp},
	\  \
	\tilde{\chi}_1^0 \to \nu_i Z,
	\  \
	\tilde{\chi}_1^0  \to  \nu_i h_k,
	\ \
	\tilde{\chi}_1^0 \to e_i^\pm H^\mp.
\end{align}
The amplitudes for the two first channels are proportional to the mixing between the leptons and neutralinos, while the last one is proportional to the Dirac-like Yukawa terms. While decays to all the MSSM 
Higgses are possible, typically, only the lightest MSSM Higgs, $h$ ($k = 1$), is light enough for the scenario we consider 
here and so we will only take it into account. A naive estimation of the decay width yields
\begin{equation}
\Gamma (\tilde{\chi}_1^0) \sim \frac{g_2^2}{32 \pi} |V_{\nu \chi }|^2 M_{\chi},
\end{equation}
where $V_{\chi \nu}$ is the mixing between the neutralino and neutrino which is proportional to $\sqrt{m_\nu/M_{\chi}}$.
Assuming that $m_\nu < 0.1$ eV for the decay length one finds $L(\tilde{\chi}_1^0) \gg 0.6$ mm.
Therefore, even without making a detailed analysis of the decays of the lightest neutralino one expects signals with lepton 
number violation and displaced vertices in part of the parameter space.

In Figs.~\ref{LSP.DL.Bino}-\ref{LSP.DL.Higgsino} we show the decay lengths versus LSP mass resulting from a scan over all the possible values of $\epsilon_1$ and $\epsilon_2$ and over the parameters and ranges specified in Ref.~\cite{FileviezPerez:2012mj}. The points are divided according to the largest component of the LSP and the neutrino hierarchy with a dominantly bino, wino and Higgsino LSP in the case when the neutrino spectrum has a Normal Hierarchy (NH) shown in (a) and for an Inverted Hierarchy (IH) in (b), respectively. The relevant decay lengths can be understood by studying the mixings in Eq.~(\ref{neutralino}).  Since the higgsino-neutrino decay strength is the largest, $\sim Y_\nu v_R$, the Higgsino LSP has the shortest decay length.  It is followed by the wino LSP with mixing $\sim g_2 v_L$ and finally the bino with coupling $\sim g_1 v_L$ and therefore the largest possible decay lengths.  Displaced vertices associated with the lifetime of the LSP will only be discernible in a very limited part of the parameter space.

The LSP branching ratios into the various possible channels versus the LSP mass are displayed in Figs.~\ref{LSP.BR.Bino}-\ref{LSP.BR.Higgsino}, plotting the dominantly bino, wino and Higgsino LSP in (a) for a NH and in (b) for an IH.  Although it is not obvious from the Figs.~\ref{LSP.BR.Bino}-\ref{LSP.BR.Higgsino}, the branching ratio to the electron $W^\pm$ channel is always smaller then either the $\mu^\mp W^\pm$ or the $\tau^\mp W^\pm$ in the NH. 
%
\subsection{Signatures at the LHC}

We have mentioned in the previous section that the minimal supersymmetric $B-L$ theory predicts lepton number violation at the LHC 
and the relevant couplings are small. Therefore, the production mechanisms for the supersymmetric particles at the LHC are 
not modified but the LSP is not stable and decays via the lepton number violating couplings. In this case one could have 
the productions of several SM particles together with two LSPs
\begin{displaymath}
p \ p \  \to \  \Psi_1 \ldots \Psi_n \ {\rm{LSP}} \ {\rm{LSP}} \  \to \   \Psi_1 \ldots \Psi_n \  \Psi_i \ \Psi_j,
\end{displaymath}
where $\Psi_i$ is a SM particle, and $n=2,4,6,..$. One example is the scenario where the stop is the LSP and decays as a leptoquark:
\begin{displaymath}
p \ p \  \to \  \tilde{t}_1^* \  \tilde{t}_1 \to \   b \ \bar{b} \ \tau^+ \ \tau^-.
\end{displaymath}
In Table~\ref{tbl.decays} we list the possible LSP scenarios and their main decays, focusing only on light third generation sfermions.
\begin{table}[h]
\begin{center}
\begin{tabular}{|c|c|c|c|}
\hline
~~~~~~~~ LSP  Scenario~~~~~~~~&~~~~ Decays ~~~~\\ \hline \hline
$\tilde{t}_1$ & $t \ \bar{\nu}, \  j \ \bar{\nu}, \ b \ e^+_i, \ j \ e^+_i$ \\ \hline
$\tilde{b}_1$ & $b \ \bar{\nu}, \  j \ \bar{\nu}, \ t \ e^-_i, \ j \ e^-_i$ \\ \hline
$\tilde{\chi}^0_1$ & $ e^{\pm}_i \ W, \  \nu \  Z, \  e^{\pm}_i \ H^{\mp}, \  \nu \ H_i^0$ \\ \hline
$\tilde{\chi}^{\pm}_1$ & $ e^{\pm}_i \ Z, \  \nu \  W^\pm, \  e^{\pm}_i \ H_i^0, \  \nu \ H_i^0$ \\ \hline
$\tilde{\tau}^{\pm}$ & $ e^{\pm}_i \nu, \  \bar{q} \ q', \ h W^\pm $ \\ \hline
$\tilde{\nu}_3$ & $ \bar{q} q, \ \bar{e}_i e_j, \ WW, \ ZZ, hh, HH $ \\ \hline
\end{tabular}
\caption{Lepton number violating LSP decays.}
\label{tbl.decays}
\end{center}
\end{table}
As discussed in Ref.~\cite{FileviezPerez:2012mj}, the most exotic signals in this theory correspond to a neutralino LSP. 
In this case we can produce two charged sleptons through the photon, the $Z$ and the $B-L$ gauge boson:
\begin{equation}
p \ p \  \to \ \gamma, Z, Z_{B-L} \  \to \ \tilde{e}_i^+ \tilde{e}_i^-.  
\end{equation}  
The sleptons subsequently decay as $\tilde{e}_i  \to e_i \  \tilde{\chi}^0_1$ and finally the neutralinos decay through 
$\tilde{\chi}^0_1 \to e^{\pm}_j W^{\mp}$. In Ref.~\cite{FileviezPerez:2012mj} has been shown that both of these branching ratios are typically large. Therefore, one can expect a significant number of events with four leptons and two W's.
When the $W$ gauge bosons decay hadronically, one is left with the unambiguous lepton number violating final states:
\begin{eqnarray}
e_i^{\pm} \ e^{\mp}_i \ e^{\pm}_j \ e^{\pm}_k \ 4 j.
\end{eqnarray}
where $e_i=e,\mu,\tau$. See Fig.~\ref{Signal} for the topology of these events.

The number of events for these channels with multileptons is given by
\begin{equation}
N_{ijk}={\cal{L}} \times \sigma (p p \to \tilde{e}^{\pm}_i \tilde{e}^\mp_i) \times C_{jk},
\end{equation}
where ${\cal{L}}$ is the integrated luminosity, and $C_{jk}$ is given by
\begin{equation}
C_{jk}= 2 (2-\delta_{jk}) \times {\rm{Br}} (\tilde{e}^{\pm} \to e^{\pm} \tilde{\chi}_1^0)^2 \times {\rm{Br}} (\tilde{\chi}^0_1 \to e^{\pm}_j W^{\mp}) \times {\rm{Br}} (\tilde{\chi}^0_1 \to e^{\pm}_k W^{\mp}) 
\times {\rm{Br}} (W \to jj)^2. 
\end{equation} 
In order to analyze the testability of the channels with multi-leptons we need to estimate the cross section using the production mechanism mentioned above and focusing on left-handed sleptons, as motivated above. 
This study was done in Ref.~\cite{Perez:2013kla} and here we discuss the main results.

In Fig.~\ref{cs} we show the production cross section for left-handed sleptons assuming that the mass of the $B-L$ gauge boson is $M_{Z_{BL}}=3$ TeV and the corresponding coupling 
is $g_{BL}=0.3$. In this figure we show the numerical results for $\sigma (p p \to \tilde{e}^{\pm}_i \tilde{e}^\mp_i)$ when the center mass energy is $\sqrt{s}=7$ TeV,  
$\sqrt{s}=8$ TeV, and $\sqrt{s}=14$ TeV. It is important to mention that when $\sqrt{s}=14$ TeV the cross section is above 1 fb when the selectron mass is below 400 GeV. For selectron masses in the TeV range, the signals discussed above will be very difficult to test.

The number of events with four leptons and four jets are estimated for the scenario: $\sqrt{s}=14$ TeV and 20 fb$^{-1}$ in Fig.~\ref{14TeV}. 
The left panel in Fig.~\ref{14TeV} shows the number of events in the ${\rm{Br}} (\tilde{e}^{\pm}_i \to e^{\pm}_i \tilde{\chi}_1^0)$--$M_{\tilde{e}_i}$ 
plane assuming an optimistic (pessimistic) branching ratio for $\tilde \chi^0_1 \to e_i^\pm W^\mp$ of 0.5 (0.1). In the left panel, the number of events 
is shown in the ${\rm{Br}} (\tilde{e}^{\pm}_i \to e^{\pm}_i \tilde{\chi}_1^0)$--${\rm{Br}} (\tilde{\chi}_1 \to e^{\pm}_i W^{\mp})$ for a light (heavy) selectron mass of 200 GeV (400 GeV) for the 14 TeV run. 
Here one assumes $M_{Z_{BL}}=3$ TeV and $g_{BL}=0.3$. As one can see, only when the sleptons are light with mass below 300 GeV we expect a large number of events. See Ref.~\cite{Perez:2013kla} for a more detailed study. For recent phenomenological studies in these theories see 
Refs.~\cite{Marshall:2014kea,Marshall:2014cwa,Ovrut:2014rba}.

\section{Final Discussions}
In this review we have discussed the desert hypothesis in particle physics and 
its role in physics beyond the Standard Model. As we have mentioned, the main reason to
assume the great desert between the weak and Planck scales is the proton stability. 
The simplest grand unified theories predict proton decay and in order to satisfy the 
experimental bounds on the proton decay lifetimes one needs to assume that these 
theories describe physics at the very high scale. This approach has been accepted by the particle physics 
community even if there is no way to test these theories. The desert hypothesis is very naive 
and we have shown in this review the possibility to define the simplest theories where the great desert is not needed.

In the first part of the review we have discussed simple theories where the baryon and lepton numbers are defined as local symmetries.
In this context one can define an anomaly free theory adding new fermions with lepton and baryon number which we call ``lepto-baryons".
After symmetry breaking one finds that the baryon number is broken in three units. Therefore, the proton is stable and there is no need to 
assume the great desert. We have shown that the simplest theories with local B and L contain a candidate for the cold dark matter in the Universe.
Using the relic density constraints we found an upper bound on the symmetry breaking scale. Therefore, one can test or rule out these theories in current or future colliders. We have discussed the relation between the baryon asymmetry and cold dark matter density in these theories, showing that these theories 
are consistent with cosmology. We have shown that these theories open the possibility of having unification at the low scale. 
These theories are appealing extensions of the Standard Model of particle physics.

In the second part of the review we have presented the simplest supersymmetric gauge theory which predicts that $R-$parity must be spontaneously 
broken at the supersymmetric breaking scale. As we have discussed, in the context of supersymmetric theories one needs to assume the 
great desert to suppress the dimension five and six operators mediating proton decay. At the same time the unification of gauge couplings 
is realized at the high scale. We have discussed the main features of the simplest gauge theory for spontaneous $R-$parity violation.
This theory predicts lepton number violating signatures at collider. We have investigated the spectrum and the signatures with multi-leptons which could 
help us to test the theory at the Large Collider Collider. The most important result in this section is that the simplest theory for $R-$parity 
predicts that this symmetry must be broken. Maybe this is the key element needed to discover supersymmetry at colliders. As it has been 
discussed, in this case one does not expect signatures with missing energy and one should look for signals with multi-leptons with 
the same electric charge and multi-jets.  
 
We would like to emphasize that the origin of baryon and lepton number (non)conservation provides a unique guide to understand the 
different theories for physics beyond the Standard Model. We have discussed two main type of these theories which could be 
tested in current and future experiments. In our opinion the theories with local baryon and lepton numbers define a new way to 
think about the unification of gauge interactions and cosmology.

\section*{Acknowledgments}
I would like to thank my collaborators Jonathan M. Arnold, Borut Bajc, 
Vernon Barger, Ilja Dorsner, Michael Duerr, Bartosz Fornal, Manfred Lindner, 
Pran Nath, Sebastian Ohmer, Hiren H. Patel, German Rodrigo, Goran Senjanovic, Sogee Spinner 
and Mark B. Wise for great collaborations in different projects related to this review. 
I would like to thank the Max Planck Institut fuer Kernphysik (MPIK) and 
the California Institute of Technology (Caltech) for supporting my work.
This work has been partly supported by the Gordon and Betty Moore Foundation through 
Grant 776 to the Caltech Moore Center for Theoretical Cosmology and Physics. 

\newpage 
 

\end{document}